\documentclass[aps,pra,twocolumn,floats,showpacs,letterpaper,eqsecnum,superscriptaddress]{revtex4}
\usepackage{amssymb}
\usepackage{amsbsy}
\usepackage{amsmath}
\usepackage{graphicx}
\usepackage{graphics}
\usepackage{setspace}
\usepackage{array}
\usepackage{color}
\usepackage{fontenc}
\usepackage{textcomp}
\usepackage{rotating}
\usepackage{bm}

\let\a=\alpha \let\b=\beta \let\g=\gamma  
   \let\k=\kappa
\let\l=\lambda     
\let\s=\sigma   

\let\w=\omega    
   
 \let\W=\Omega    
  \let\re=\ref

\def\nn{\nonumber}

\let\p=\partial
\def\bpm{\begin{pmatrix}}
\def\epm{\end{pmatrix}}
\def\be{\begin{equation}}
\def\ee{\end{equation}}
\def\bea{\begin{eqnarray}}
\def\eea{\end{eqnarray}}
\def\ba{\begin{array}}
\def\ea{\end{array}}
\def\del{\delta}

\def\td{\tilde}
\def\wtd{\widetilde}

\def\ep{{\epsilon}}
\def\vep{{\varepsilon}}

\def\re{\text{Re}}

\def\bfr{\mathbf{r}}

\def\Tr{\text{Tr}}

\def\rmd{\mathrm{d}}

\newcommand{\bk}{{\bf k}}

\newcommand{\bp}{{\bf p}}
\newcommand{\bq}{{\bf q}}

\newcommand{\trasp}{\mathsf{T}}

\newcommand{\St}{\mathfrak{St}}
\newcommand{\vf}{\mathsf{v}_\mathrm{F}}
\newcommand{\va}{\mathsf{v}_\mathrm{a}}

\newcommand{\sfp}{\mathsf{p}}

\newcommand{\ef}{\mathbf{E}}
\newcommand{\velf}{\mathbf{v}_\mathrm{F}}
\newcommand{\wA}{\W_{A^\prime}}
\newcommand{\what}{\widehat}
\newcommand{\mathsym}[1]{{}}
\newcommand{\unicode}[1]{{}}

\newcommand{\sff}{\mathsf{f}}
\newcommand{\sech}{\mathrm{sech}}

\newcommand{\vex}[1]{\bm{\mathrm{#1}}}
\newcommand{\msf}[1]{\mathsf{#1}}

\begin{document}

\title{Transport coefficients of graphene: Interplay of impurity scattering, Coulomb interaction, and optical phonons}
\author{Hong-Yi~Xie} \email{hongyi.xie@rice.edu}
\affiliation{Department of Physics and Astronomy, Rice University, Houston, Texas 77005, USA}
\author{Matthew~S.~Foster}
\affiliation{Department of Physics and Astronomy, Rice University, Houston, Texas 77005, USA}
\affiliation{Rice Center for Quantum Materials, Rice University, Houston, Texas 77005, USA}
\date{\today\\}

\begin{abstract}
We study the electric and thermal 
transport
of the Dirac 
carriers
in monolayer graphene using the Boltzmann-equation approach. 
Motivated by recent thermopower measurements 
[F. Ghahari, H.-Y.~Xie, T. Taniguchi, K. Watanabe, M.~S.~Foster, and P.~Kim, Phys.\ Rev.\ Lett.\ {\bf 116}, 136802 (2016)], we
consider the effects of quenched disorder, Coulomb interactions, and 
electron--optical-phonon scattering. 
Via an unbiased numerical solution to the Boltzmann equation we calculate the electrical conductivity, thermopower, and 
electronic component of the
thermal conductivity, 
and discuss the 
validity
of Mott's formula and of the Wiedemann-Franz law. 
An analytical solution for the disorder-only case shows that 
screened
Coulomb impurity scattering, although elastic, violates the Wiedemann-Franz law even at low temperature. 
For the combination of carrier-carrier Coulomb and short-ranged impurity scattering,
we observe the crossover from the interaction-limited (hydrodynamic) regime to the disorder-limited (Fermi-liquid) regime. 
In the former, the thermopower and the thermal conductivity follow the results anticipated by the relativistic hydrodynamic theory. 
On the other hand, we find that optical phonons become nonnegligible at relatively low temperatures
and that the induced electron thermopower violates Mott's formula. 
Combining all of these scattering mechanisms, we obtain the thermopower that quantitatively coincides with the experimental data.        
\end{abstract}

\maketitle

\tableofcontents

\section{Introduction}

Electric and thermal transports in monolayer graphene are influenced by various scattering mechanisms, such as quenched impurities, interparticle interactions, and phonons~\cite{castro, dars11rmp, NomuraMac0607,AERF0608,AHGDS2007,OGM07,kashsmin,lars2007,MarkusSub2008,matt08,mueller2008,matt09,Adam08, hwang2009,jang08, hartnoll2007, mirlin2011,narozhny15,hwang2008,kim2010,sohier2014,Kubakaddi09,Bao,zuev09, Wei2009}. 
In
weakly disordered
graphene, interaction effects can become dominant 
at non-zero temperature; 
in the vicinity of zero doping the 
Coulomb-interacting massless Dirac carriers form a relativistic electron-hole plasma.
In this 
interaction-limited 
regime, hydrodynamic theory \cite{mueller2008,matt09,narozhny15} predicts intriguing non-Fermi-liquid transport properties. 
First, the electron-hole fluid exhibits a finite and nonvanishing dc electrical conductivity 
at the Dirac point
even in the absence of impurities, due entirely to inelastic electron-hole collisions. 
Moreover, Mott's formula \cite{MottDavis} and the Wiedemann-Franz law~\cite{AMbook} are violated. 
In a Fermi liquid, these respectively determine the thermoelectric power and the electronic component of the thermal conductivity 
from the electrical conductivity.
Instead, 
for graphene in the hydrodynamic regime
the thermopower at non-zero doping approaches the 
thermodynamic
entropy per charge, 
and 
the thermal conductivity 
at the Dirac point diverges as the impurity concentration vanishes. 

The theory predicts upper bounds for the thermoelectric power 
and electronic component of the thermal conductivity, limited only by disorder.
While violating classical relations between thermoelectric coefficients, 
the latter are strongly constrained and interrelated by the relativistic hydrodynamics \cite{mueller2008,matt09}. 
We emphasize that this violation of the Mott and Wiedemann-Franz relations 
is different from the usual physics of narrow-gap/gapless
semiconductors, for example, the bipolar diffusion process~\cite{HJGbook}, where separated electron and hole currents are assumed.
Strong inelastic electron-hole scattering in ultraclean graphene 
implies that a composite electron-hole fluid emerges \cite{matt09}, which cannot be decomposed into
valence and conduction band components.

Three very recent experiments \cite{Crossno15,Bandurin15,GXFKexp}
have provided substantial evidence
for interaction-limited transport in graphene. 
Measurements of the electronic component of the 
thermal conductivity near charge neutrality \cite{Crossno15} showed large violations of the Wiedemann-Franz law \cite{Lucas15,Principi15}.
Non-local transport in doped graphene \cite{Bandurin15} was used to probe the viscosity of the electron fluid \cite{Mueller09,Levitov15,Polini15,Mirlin15}. 
Finally, thermoelectric power measurements \cite{GXFKexp} showed a substantial deviation from the Mott formula. 
In this work, we model the experiment in \cite{GXFKexp} using the Boltzmann equation to incorporate carrier-impurity, 
carrier-carrier, and carrier-optical phonon scattering mechanisms. 

The thermopower measurements in Ref.~\onlinecite{GXFKexp} 
were performed on high-mobility graphene 
encapsulated by
hexagonal-boron-nitride.
The experiment was done 
over a large span of dopings, with charge-carrier density $n \equiv \rho/(-e)$ ranging from zero to 
$\pm \,3.0 \times 10^{12}\,\mathrm{cm}^{-2}$
[$\rho$ is the charge density and $e > 0$ is the elementary charge].
The measurements were performed at 
relatively high temperatures ($130 \, \mathrm{K} \lesssim T \lesssim 350 \, \mathrm{K}$) in order to fulfill the non-degenerate condition 
$k_B T \gtrsim \mu$ over much of the doping span, while avoiding 
the electron-hole puddle regime at low temperatures near charge neutrality~\cite{RNpuddles,Martin2008,dars11rmp}. 
Here $\mu$ denotes the chemical potential, determined by the temperature and the fixed charge-carrier density $n$.
In this regime the measured thermopower is consistently larger than that predicted by Mott's formula~\cite{MottDavis}, 
but saturates below the ideal hydrodynamic prediction. This novel feature suggests that in order to quantitatively 
characterize the thermoelectric transport in graphene, one should consider 
additional \emph{inelastic} scattering mechanisms. 

We exclude acoustic phonons, since at low doping the 
electron--acoustic-phonon
scattering~\cite{hwang2008,kim2010} is quasi-elastic 
and incapable of producing large violations of
Mott's formula. 
As discussed in Ref.~\onlinecite{kim2010}, the Bloch-Gr\"uneisen temperature 
$T_\mathrm{BG} \equiv 2  \hbar \va k_\mathrm{F}/k_\mathrm{B}$ plays the key role, where $\va$ 
is the acoustic phonon velocity and $k_\mathrm{F}$ the Fermi wavevector. Assuming the acoustic phonon 
velocity equal to $2.6 \times 10^6$ cm/s, one can estimate the Bloch-Gr\"uneisen temperature as $T_\mathrm{BG} \approx 70 \sqrt{n}$ K, 
where the density $n$ is measured in units of $10^{12}$ $\mathrm{cm}^{-2}$. The experiment in Ref.~\onlinecite{GXFKexp} 
is performed in the regime $T \gtrsim T_\mathrm{BG}$ where the acoustic-phonon--scattering is quasi-elastic.
In addition we disregard the effects of external magnetic fields or spin-flip mechanisms.   

In this paper we consider the inelastic \emph{optical-phonon} scattering and model 
graphene by the Hamiltonian
\be  \label{total-ham}
	H = 
	H_\mathrm{0} 
	+ 
	H_\mathrm{oph} 
	+ 
	V_{\mathrm{imp}}^{(\mathrm{s})} 
	+ 
	V_{\mathrm{imp}}^{(\mathrm{l})}  
	+ 
	V_{\mathrm{int}} 
	+ 
	V_{\text{e-oph}},
\ee
where 
$H_\mathrm{0}$ describes the free Dirac fermions, 
$H_\mathrm{oph}$ the optical-phonon bath, 
$V_{\mathrm{imp}}^{(s)}$ and $V_{\mathrm{imp}}^{(\mathrm{l})}$ the quenched short-ranged and long-ranged 
(Coulomb impurity)
disorder
potentials, 
respectively, 
$V_\mathrm{int}$ the interparticle Coulomb interactions, 
and 
$V_\text{e-oph}$ the electron--optical-phonon coupling. 
We assume that both the time-reversal symmetry and spin SU(2) rotation symmetry 
are preserved 
in the presence of disorder and interactions.
We also presume that the particle-hole symmetry as well as the honeycomb 
lattice space group symmetries 
(translations, rotations, and reflections) are preserved under disorder 
average~\cite{matt08}. 
Concretely, each term in the Hamiltonian (\ref{total-ham}) is constructed as follows.

The short-ranged impurity Hamiltonian 
$V_{\mathrm{imp}}^{(\mathrm{s})}$ 
takes the form as introduced in Ref.~\onlinecite{matt08}, which incorporates 
five types of time-reversal-symmetric disorder, all assumed to be zero-mean, short-ranged, and Gaussian-correlated. 
Five independent parameters $\{ g_u, g_A, g_{A3}, g_m, g_v \}$ characterize their statistical fluctuations. 
In the Boltzmann equation these parameters appear effectively in certain combinations 
[$G_{0,\mathrm{f},\mathrm{b}}$ in Eq.~(\ref{imp-col})]. 
The term 
$V_{\mathrm{imp}}^{(\mathrm{l})}$
gives the scalar potential due to Coulomb impurities. 
These are subject to the temperature and density-dependent static screening by the electron-hole plasma~\cite{Adam08, hwang2009}.

\begin{figure}
\centering
\includegraphics[width=0.4\textwidth]{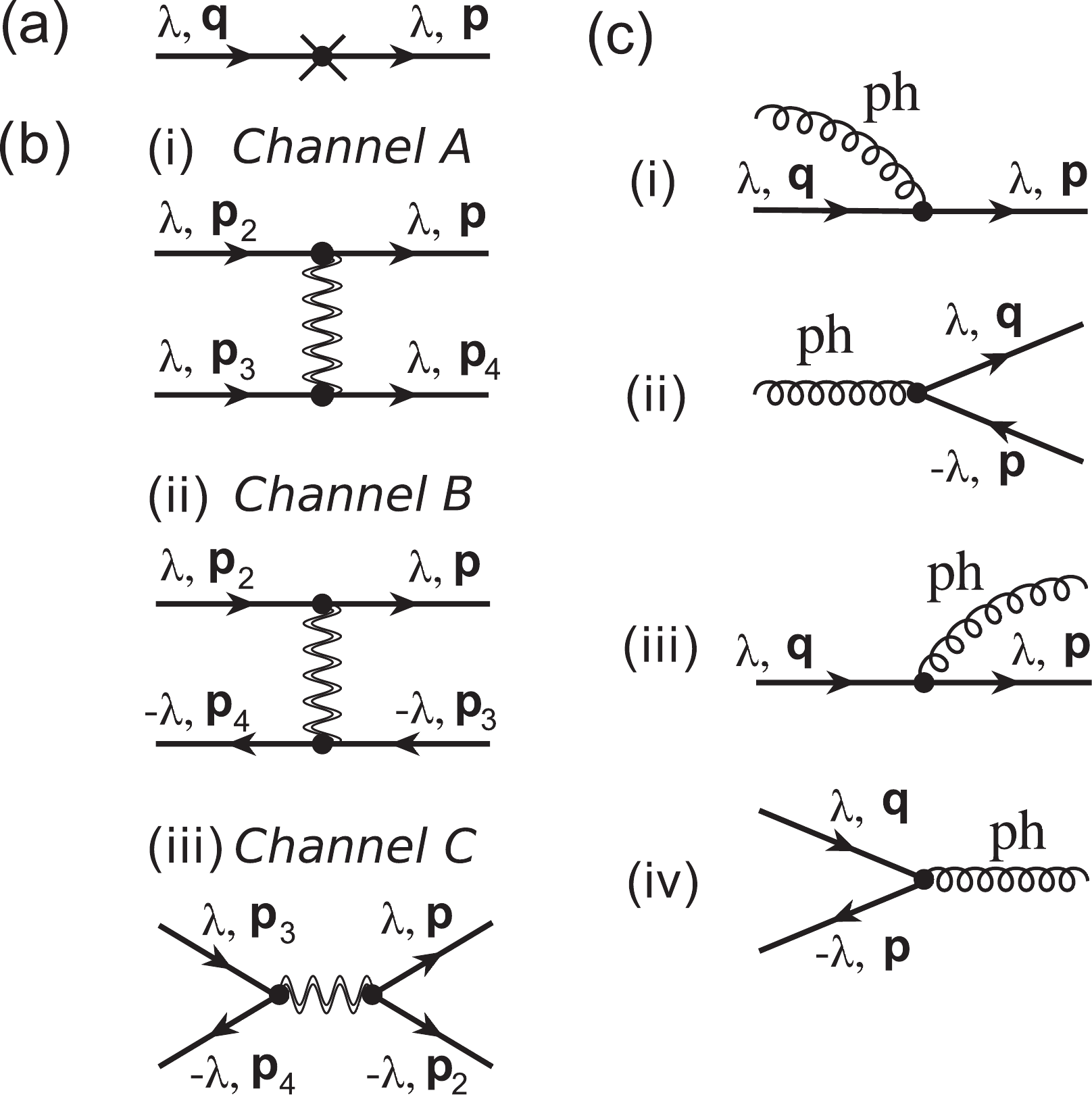}  
\caption{Diagrams representing the collision integrals in Eq.~(\ref{collisions}). 
The arrows indicate the flow of electric charge only and the wave vector labels correspond to incoming and 
outgoing fermions on the left and right of the scattering vertices, respectively. 
(a) Static (both short- and long-ranged) impurity scattering. 
(b) Coulomb collision processes that preserve electron and hole numbers separately. 
The label $\lambda \in \pm 1$ denotes electrons ($+1$) or holes ($-1$).
(c) Carrier--optical-phonon scattering: (i) and (ii) Phonon absorption; (iii) and (iv) Phonon emission.}  
\label{scats}
\end{figure}

Coulomb interactions between carriers are encoded in
$V_{\mathrm{int}}$. 
Dynamical screening is treated within
the random phase approximation~\cite{wunsch06, darsarma07, macdonald2009, mirlin2011}.
Screening is crucial both in the low-temperature degenerate Fermi liquid phase, 
but also in the high-temperature, non-degenerate regime of primary interest here. 
Different from a single component plasma, graphene ultimately screens \emph{better} at higher temperatures, 
due to the proliferation of thermally-excited electron-hole pairs. 
At intermediate temperatures and finite charge density, the Thomas-Fermi length reaches a maximum. 
The interaction strength is encoded in the 
fine structure constant $\a_\mathrm{int}$ that depends on the dielectric constants of the substrates~\cite{jang08}. 
In the kinetic theory, 
dynamical screening 
suppresses 
the ``collinear'' singularity of the Coulomb collision integral, which is due to 
the linear dispersion of Dirac fermions and the energy conservation (see Appendix~\ref{collinear-singularity}). 
Note that for simplicity we only consider 
two-body
collision processes that preserve the population of electrons and holes separately.
We leave the effects of 
(three-body or impurity-assisted)
electron-hole 
Auger imbalance relaxation
processes~\cite{matt09} to future study.         

Three types of in-plane optical-phonon modes~\cite{why-inplane} allowed by time-reversal 
symmetry~\cite{manes07,basko2008} couple to electrons. For simplicity we consider only the $A_1^\prime$ modes that correspond 
to the ``Kekul\'e'' vibration of the honeycomb lattice and couple the electrons between $K$ and $K^\prime$ valleys~\cite{manes07,basko2008}. 
The $A_1^\prime$ phonons have been suggested to be the most relevant optical-phonon branch for 
influencing electrical
transport at relatively 
low temperature~\cite{sohier2014}, 
possessing
the lowest excitation energy and the strongest coupling to electrons. 
We note that in the context of the Boltzmann equation the collision integral for $A_1^\prime$ phonons can also qualitatively 
describe the effect of the other optical-phonon branches. Similar 
to the case of short-ranged impurity scattering,
the collision integrals for different optical-phonon branches are distinguished 
only by the factors $(1 \pm \hat{\bp} \cdot \hat{\bq})/2$ that 
enhance the electron forward ($+$) or backward ($-$) scattering. 
Furthermore, we use the single-mode Einstein 
model (dispersionless) $H_\mathrm{opt}$ to describe the $A_1^\prime$ phonons; the electron-phonon coupling $V_{\text{e-opt}}$ 
takes the form introduced in Refs.~\onlinecite{sohier2014,manes07,basko2008}. Two parameters are present: The $A_1^\prime$-phonon 
frequency $\w_{A^\prime}$ and electron-phonon coupling $\b_{A^\prime}$ (doping and temperature dependent, see the discussion in Sec.~\ref{comp-exp}). 
We in addition assume that the phonons are in thermal equilibrium, that is, the phonon kinetics are not involved 
(no drag effect on electrons) since the optical-phonon dispersion is weak~\cite{ph-drag}. 
 
All of the scattering mechanisms that we consider are 
depicted in Fig.~\ref{scats}. In particular, the Coulomb interaction
mediates three scattering channels that we label A, B, and C. Channel A describes intraband carrier-carrier scattering. 
Channels B and C encode interband conduction electron-valence hole (``electron-hole'') scattering. 
These involve different kinematic regions of frequency $\omega$ and momentum $\vex{q}$ transfer across the Coulomb line,
as channels A and B have $|\omega| \leq v_F q$ (``quasi-static''), while channel C has $|\omega| \geq v_F q$ (``optical''). 
Plasmons appear in channel C.

This paper is organized as follows. In Sec.~\ref{sec-results} we present the main results of our calculations and 
interpret the experimental data in Ref.~\onlinecite{GXFKexp}. In Sec.~\ref{tech-sec} we transcribe the Boltzmann equation 
that is derived via the Schwinger-Keldysh formalism, with the collision integrals for the impurity scattering, Coulomb interaction, 
and 
electron--optical-phonon scattering corresponding to the Feynman diagrams 
depicted in Fig.~\ref{scats}. 
Then we introduce the orthogonal-polynomial method for solving 
the linearized Boltzmann equation. 
Results for impurity-only and interaction-limited transport are discussed in more detail in Sec.~\ref{Sec: Transport Coeffs}. 
The collinear singularity of the Coulomb collision integrals and the RPA dynamical screening are discussed in 
the Appendix.


\section{Main results} \label{sec-results}

In general one has the linear response relations~\cite{mahanbook}
\begin{subequations} \label{linear-resp}
\begin{align}
& \mathbf{J} = \s \boldsymbol{\mathcal{E}} + \s \a_\infty (-\nabla_\bfr T), \label{Jlin-resp}\\
& \mathbf{J}_\mathrm{Q} = T \s \a_\infty \boldsymbol{\mathcal{E}} +\left( \k_\infty + T \s \a_\infty^2 \right) (-\nabla_\bfr T),
\end{align}
where 
$\mathbf{J}$ is the charge current, 
$\mathbf{J}_\mathrm{Q}$ the heat current, 
$\boldsymbol{\mathcal{E}}$ the electrochemical field, 
$\nabla_\bfr T$ the temperature gradient, and 
$\s$, $\k_\infty$, and $\a_\infty$ are the electrical conductivity, thermal conductivity, and thermoelectric power, respectively. 
We use the subscript ``$\infty$'' to indicate bulk thermoelectric transport coefficients. In a finite (mesoscopic) sample, 
slow imbalance relaxation 
(recombination-generation)
can give rise to different transport coefficients and/or a spatially inhomogeneous response \cite{matt09},
but we do not consider this possibility here.
The Lorenz ratio is	
\be 
L \equiv \frac{\k_\infty}{\s \, T},
\ee  
\end{subequations}
for which we discuss the validity of the Wiedemann-Franz law $L_0 =\pi^2 k_\mathrm{B}^2/(3e^2)$~\cite{AMbook}. 
Solving the linearized Boltzmann equation (\ref{l-bolt-eq}), inserting the distribution function solution into 
Eq.~(\ref{currents}), and comparing the result to Eq.~(\ref{linear-resp}), we obtain the transport coefficients.

\subsection{Quantum kinetic equation}

We derive the quantum kinetic equation for electron ($\l=+1$) and hole ($\l=-1$) 
distribution 
functions $f_{\l} \left( \bp,\bfr,t \right)$ 
via the Schwinger-Keldysh technique~\cite{keldysh}, where $\bp$ is the quasiparticle wave vector, $\bfr$ the position, and $t$ the time. 
In the presence of an electric driven field, the stationary Boltzmann equation takes the form
\be  \label{f-bolt-eq}
	\left[ \velf \cdot \nabla_\bfr - 
	\frac{\l \, e}{\hbar}
	\ef \cdot \nabla_\bp \right] f_{\l} \left( \bp,\bfr \right) = \St_\l[\{f_{\l^\prime}\}], 
\ee
where $\velf$ is the Fermi velocity parallel to the wave vector $\bp$, 
$e > 0$ is the elementary charge, and
$\ef$ is the total electric field. 
The collision integral $\St_\l[\{f_{\l^\prime}\}]$ incorporates the three scattering mechanisms in the Hamiltonian (\ref{total-ham}) (see also Fig.~\ref{scats}),
\be  \label{collisions}
	\St_\l[\{ f_{\l^\prime} \}] = \St_{\mathrm{imp}, \l} [f_\l]  + \St_{\mathrm{int}, \l} [\{ f_{\l^\prime} \}]  + \St_{\text{oph}, \l}[\{f_{\l^\prime}\}],
\ee 
where 
$\St_{\mathrm{imp}, \l} [f_\l]$ describes elastic scattering induced by impurities [Fig.~\ref{scats}(a)], 
$\St_{\mathrm{int}, \l} [\{ f_{\l^\prime} \}]$ the inelastic Coulomb 
scattering 
between quasiparticles [Fig.~\ref{scats}(b)], and  
$\St_{\text{oph}, \l}[\{f_{\l^\prime} \}]$ the inelastic scattering 
of carriers
by optical phonons [Fig.~\ref{scats}(c)]. 

Assuming that the distribution functions $f_{\l=\pm 1}(\bp,\bfr)$ are diagonal in valley and spin space~\cite{negoffdiag}, 
we present the explicit expressions for the collision integrals in Eq.~(\ref{collisions}) in Sec.~\ref{c-integrals}. 
In the hydrodynamic regime the response to static fields is dominated by the zero modes
of the inelastic carrier-carrier collision integrals, associated to energy and momentum conservation
\cite{mueller2008,matt09}.
For this reason we also neglect the weak off-diagonal components in electron-hole space,
which do not directly contribute to the dc response \cite{lars2007}.

\subsection{Benchmark: Impurity-only transport}  \label{dis-only}

In the presence of only elastic scattering 
[see Fig.~\ref{scats}(a)]
the linearized Boltzmann equation can be solved exactly (Sec.~\ref{dis-only-2}). 
Mott's formula 
for the thermoelectric power $\a_\infty$ manifestly applies \cite{AMbook},
although the integral form must be employed away from Fermi degeneracy. 

The short-ranged-impurity only transport coefficients take simple expressions  
\be  \label{short-only}
	\s_\mathrm{imp}^{(\mathrm{s})} = N \frac{e^2}{h} \wtd{g}^{-1} , 
		\quad 
	\a_{\infty, \mathrm{imp}}^{(\mathrm{s})} =0, 
		\quad 
	\k_{\infty, \mathrm{imp}}^{(\mathrm{s})} = N \frac{\pi^2 k_\mathrm{B}^2 T}{3 h}\wtd{g}^{-1}.
\ee 
where $\wtd{g}$ is the effective dimensionless short-ranged disorder strength [Eq.~(\ref{imp-coll-M})]. 
Note that the Wiedemann-Franz law is manifestly satisfied.
The parameter $N$ is the number of independent 2-component Dirac species, equal to four in graphene.

Another analytically solvable limit is the long-ranged-impurity-only case in the 
absence of screening [see Eq.~(\ref{long-only})]. 
Especially, at the charge neutral point $\mu=0$ the Lorenz ratio is independent of temperature,
\be  \label{L-a-0}
	L _{\infty, \mathrm{imp}}^{(\mathrm{l})}
	( Q_\text{TF} \to 0, \mu=0) 	
	= 
	\frac{21}{5} \frac{\pi^2 k_\mathrm{B}^2}{3 e^2}.
\ee
Here $ Q_\text{TF} $ denotes the temperature and density-dependent Thomas-Fermi wavevector [Eq.~(\ref{q-tf})].	
Equation~(\ref{L-a-0}) violates the
Wiedemann-Franz law and enhances the Lorenz constant by a factor $L/L_0 = \frac{21}{5}=4.2$.  

\begin{figure}
\centering
\includegraphics[width=0.237\textwidth]{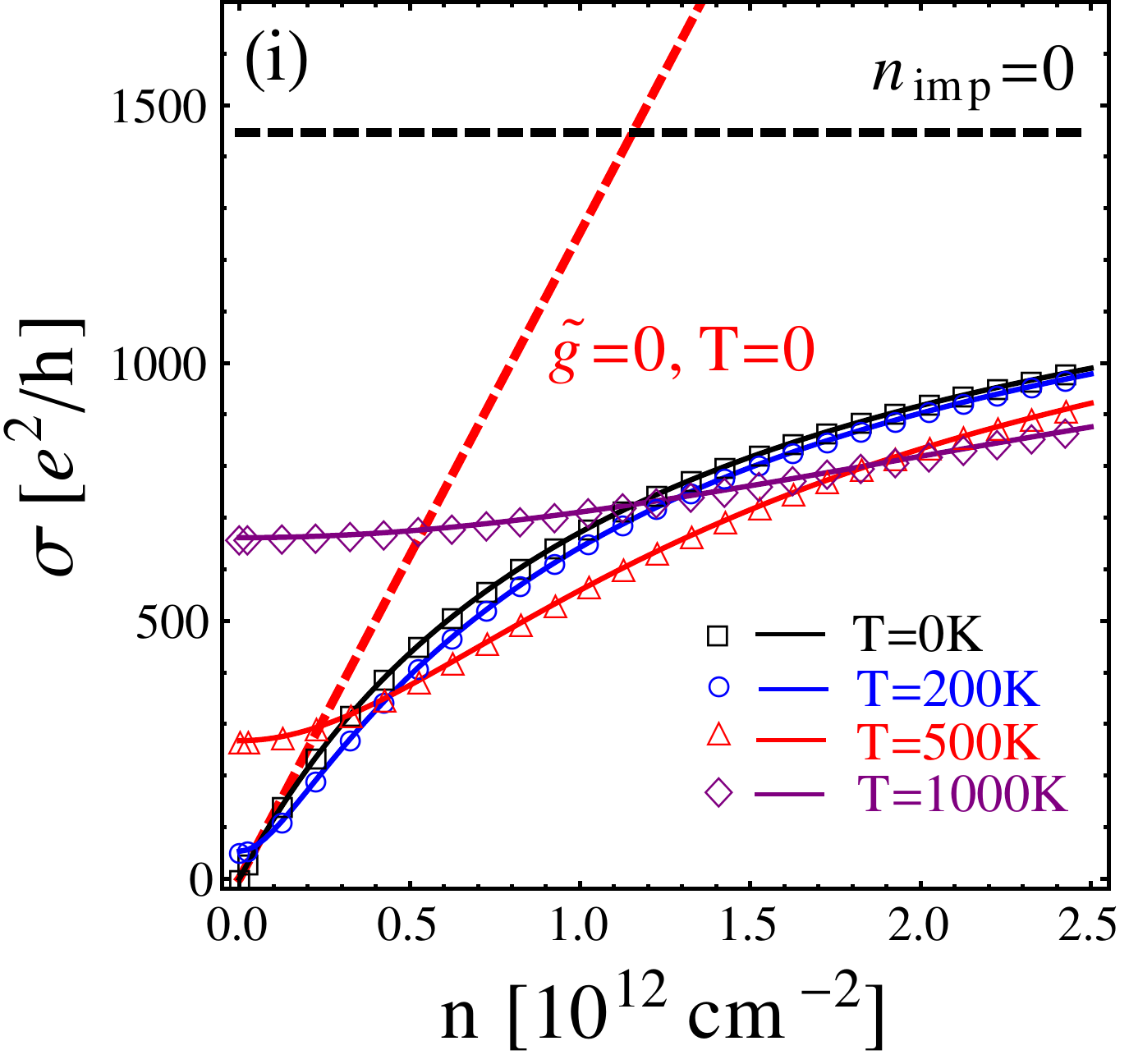} 
\includegraphics[width=0.23\textwidth]{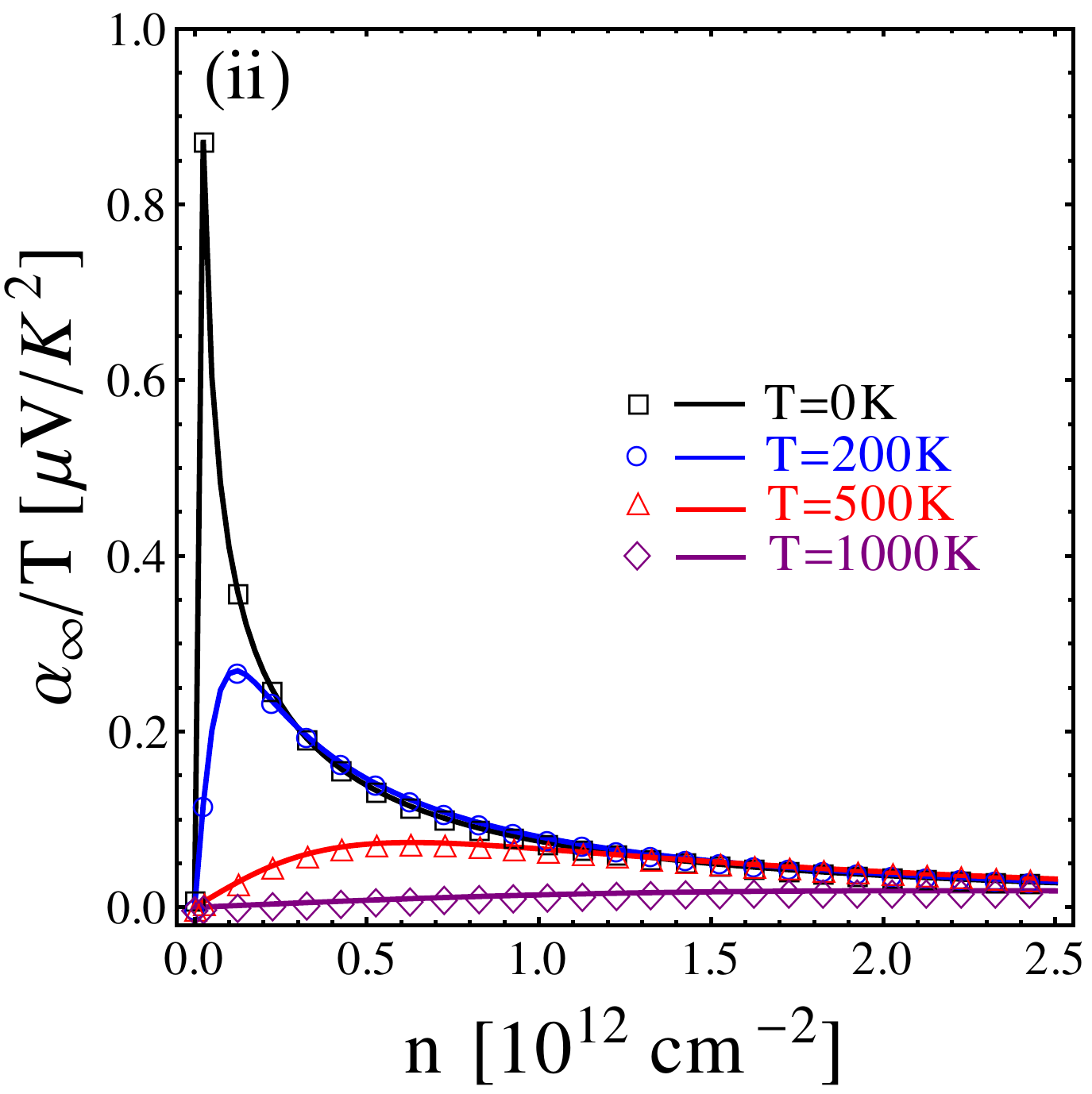} \\
\includegraphics[width=0.23\textwidth]{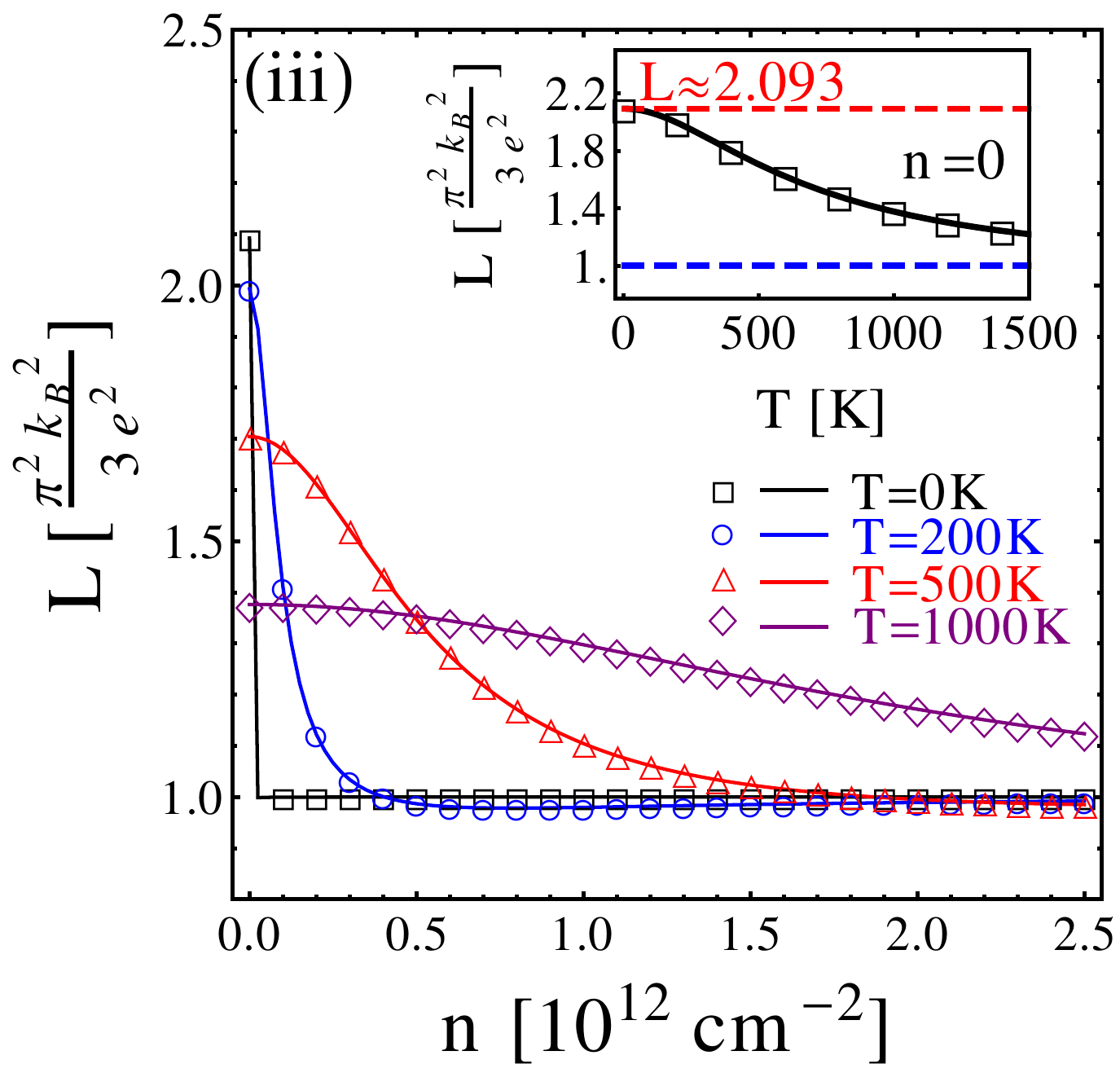}
\includegraphics[width=0.22\textwidth]{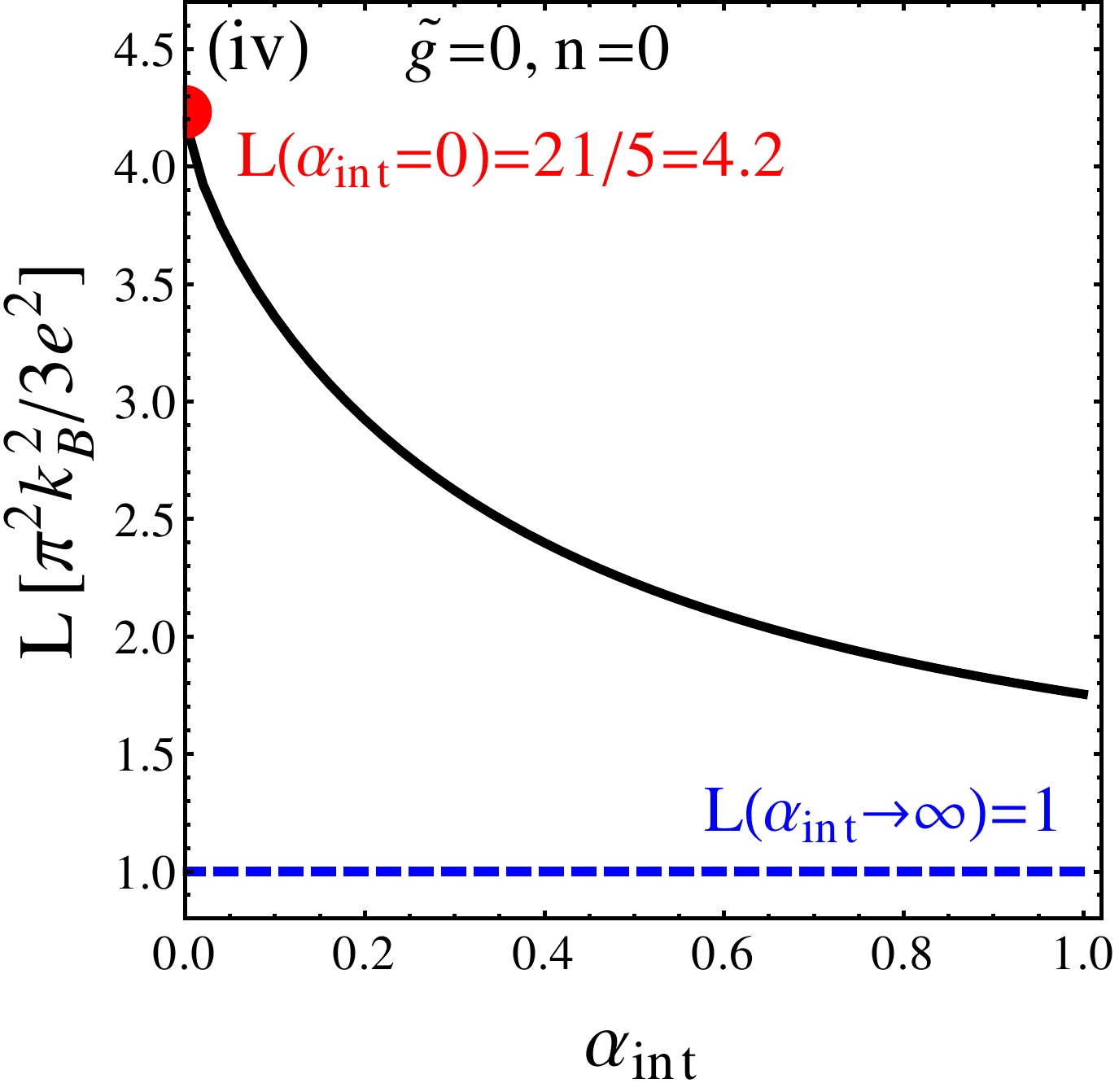}
\caption{Impurity-only transport coefficients as functions of 
the charge-carrier density $n$ for various temperatures. 
The symbols are the numerical result obtained by the orthogonal-polynomial method and the solid curves are 
the analytical result obtained by Eq.~(\ref{imp-only-L}). In our calculations we use the parameters in Ref.~\onlinecite{GXFKexp}: 
the effective short-ranged impurity strength $\wtd{g} \sim 1.1 \times 10^{-4}$, 
the long-ranged impurity concentration $n_\mathrm{imp} = 2.4 \times 10^{9} \mathrm{cm}^{-2}$, 
and the fine structure constant $\a_\mathrm{int} =0.6$. 
(i) Electric conductivity. 
The horizontal black 
(diagonal red) 
dashed line indicates the conductivity in the absence of long-ranged impurities 
(in the absence of short-ranged impurities) at $T=0$. 
(ii) Thermoelectric power. (iii) Lorenz ratio. The insert panel shows the Lorenz ratio as a function of temperature 
at the charge neutral point $n=0$. 
(iv) Lorenz ratio as a function of the fine structure constant 
[which determines the Thomas-Fermi wavevector Eq.~(\ref{q-tf})]
at charge neutrality $n=0$ in the absence of short-ranged impurities 
$\wtd{g}=0$.}  
\label{imp-only-pic}
\end{figure}

The transport coefficients due to the combination of short-ranged disorder and screened Coulomb impurities are shown 
in Fig.~\ref{imp-only-pic}(i)--(iii).  
We compare results
obtained by the 
orthogonal-polynomial algorithm to the exact results evaluated by Eq.~(\ref{imp-only-L}). 
We used  
the dimensionless short-ranged impurity strength $\tilde{g}$ and Coulomb impurity density $n_{\mathrm{imp}}$ [Eq.~(\ref{limp-col})]
determined by fitting the 
low-temperature, density-dependent conductivity data in Ref.~\onlinecite{GXFKexp}. 
Thomas-Fermi screening is limited by the 
fine structure constant 
\begin{align}\label{aintDef}
	\a_\mathrm{int}
	=
	\frac{2 e^2}{(\k_1+\k_2)\hbar\vf},
\end{align}
where $\k_{1,2}$ denotes the permittivities of the media above and below the graphene sheet. 
Here we take
$\a_\mathrm{int} = 0.6$, 
appropriate for BN encapsulation \cite{GXFKexp}.
We keep the order of the polynomial basis 
up to $\mathcal{N} =16$ in order to recover the analytical result. We observe that in the presence of 
Coulomb impurities
the Wiedemann-Franz law is in general broken. As shown in Fig.~\ref{imp-only-pic}(iii), the Lorenz ratio $L$ is 
a function of the charge density $n$ and temperature $T$ for a fixed $\a_\mathrm{int}$.

There exist two interesting limits: 
When $T \to \infty$ the effective long-ranged impurity strength vanishes 
[Eq.~(\ref{st-imp-dless-l})], 
so that short-ranged impurity scattering dominates transport and Wiedemann-Franz law restores. 
When $T \to 0$ the long-ranged impurity becomes dominant. 
At the charge neutral point $n = 0$,
the Thomas-Fermi wavevector $Q_\text{TF}$ divided by the temperature becomes a constant [Eq.~(\ref{qTFdimless})].
The Lorenz ratio is enhanced 
relative to the Wiedemann-Franz law
by a numerical constant depending on the fine structure constant ($L/L_0 \approx 2.093$ for $\a_\mathrm{int} = 0.6$). 
The Wiedemann-Franz law is recovered at sufficiently high charge densities $n \neq 0$ and/or temperatures.  

In Fig.~\ref{imp-only-pic}(iv) we show the Lorenz ratio $L$ as a function of the fine structure constant 
$\a_\mathrm{int}$ 
[appearing in the Thomas-Fermi wavevector Eq.~(\ref{q-tf})]
at charge neutrality in the absence of short-ranged impurity scattering. 
It is clear that for any finite $\a_\mathrm{int}$ the Wiedemann-Franz law is broken. 
Especially, for $\a_\mathrm{int} \to 0$ we obtain $L/L_0 = \frac{21}{5}=4.2$ [Eq.~(\ref{L-a-0})], which 
provides an upper bound for the Lorenz ratio induced solely by impurities.

\subsection{Crossover from interaction-limited regime to disorder-limited regime}  \label{hydro-dym-sec}

\begin{figure}
\centering
\includegraphics[width=0.23\textwidth]{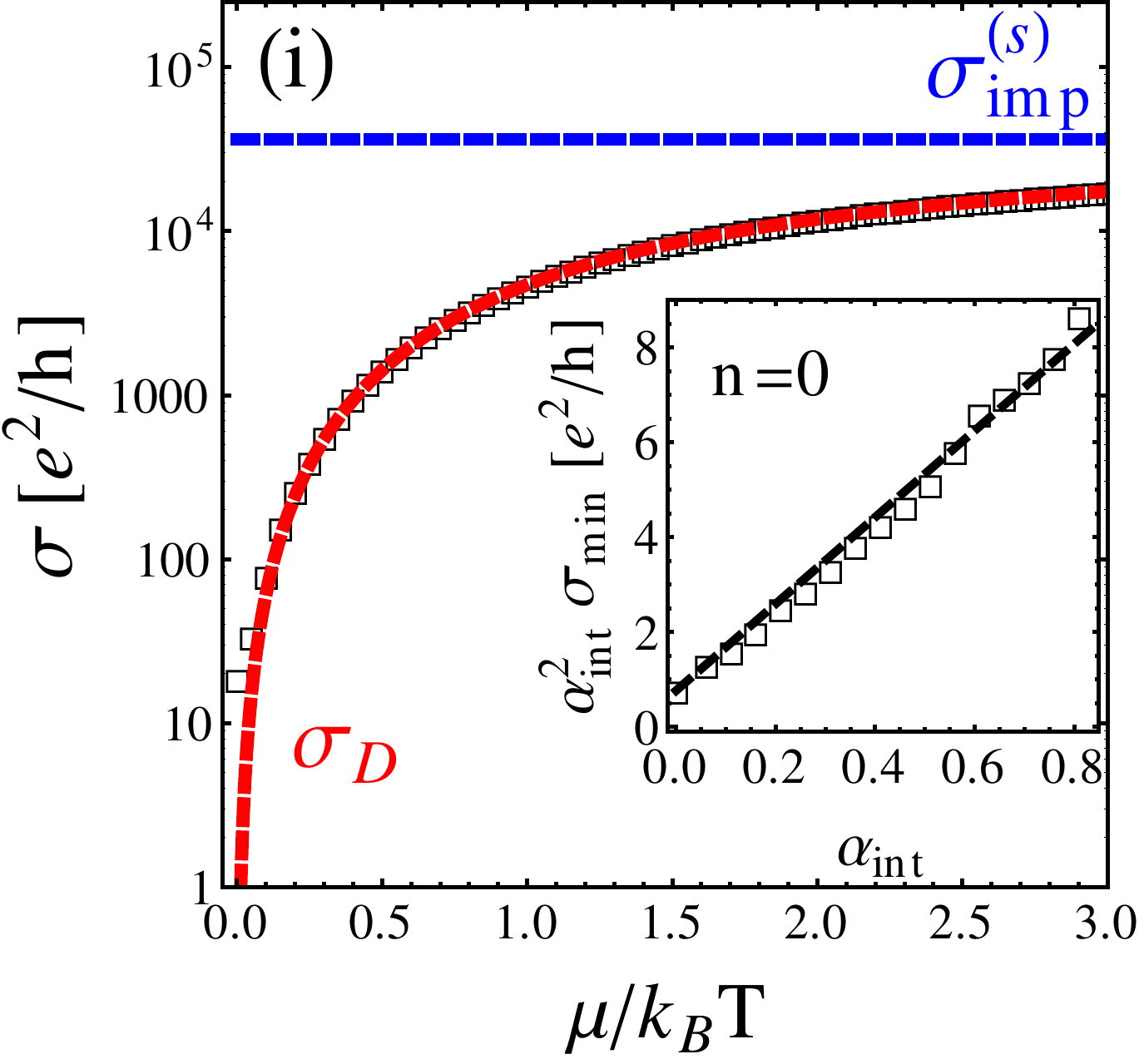} \,\,\,
\includegraphics[width=0.205\textwidth]{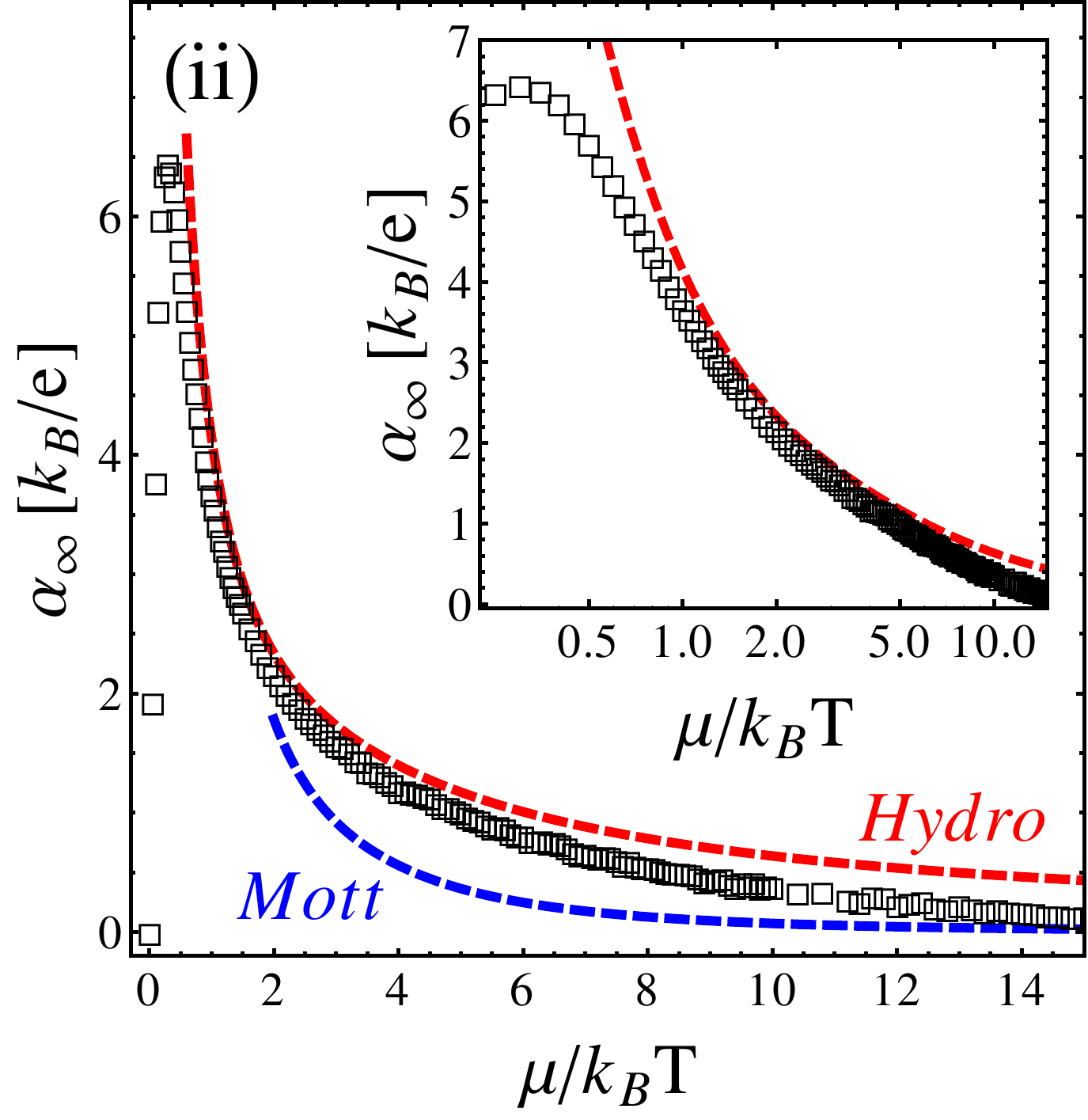} \\
\includegraphics[width=0.242\textwidth]{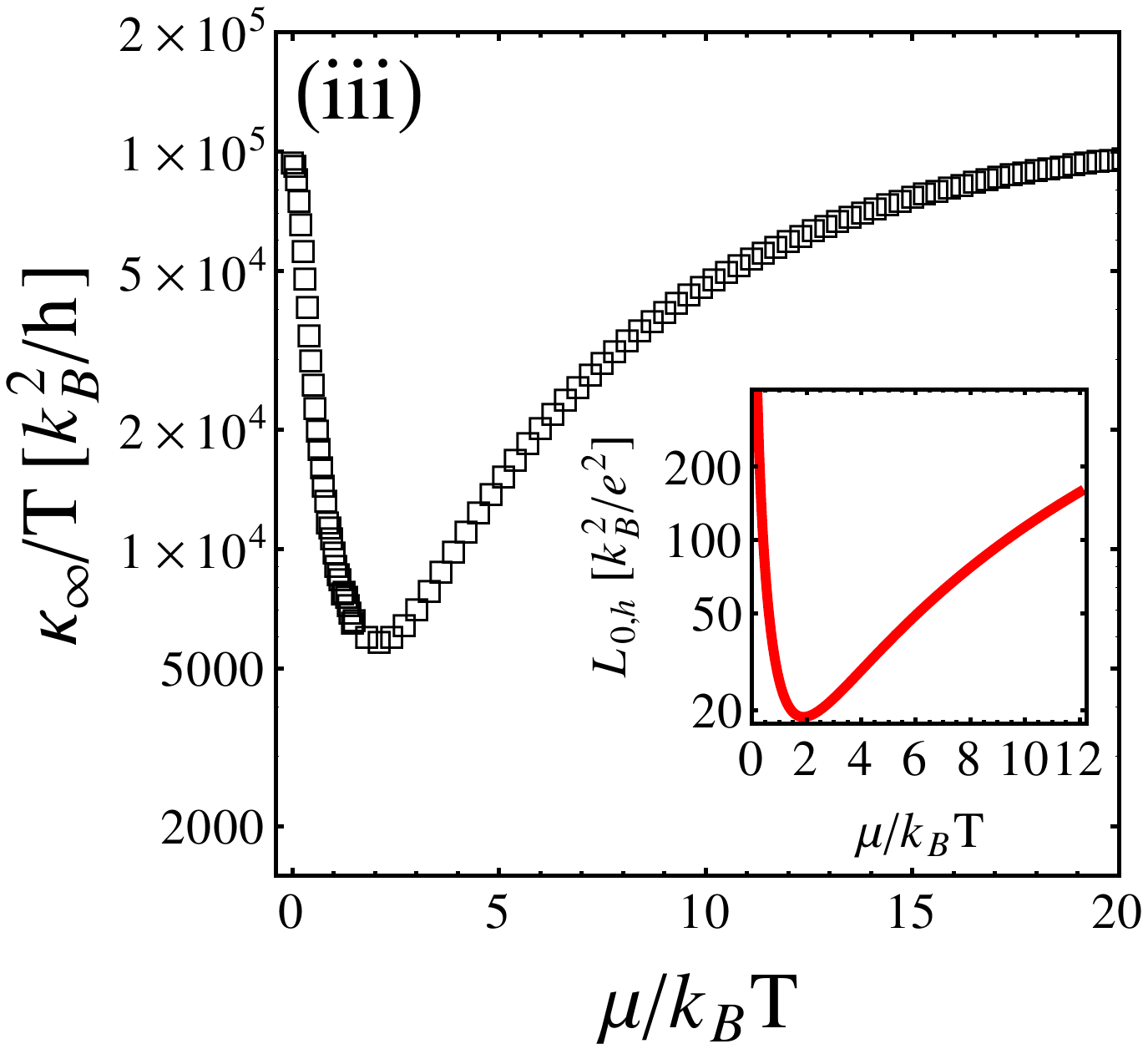}
\includegraphics[width=0.233\textwidth]{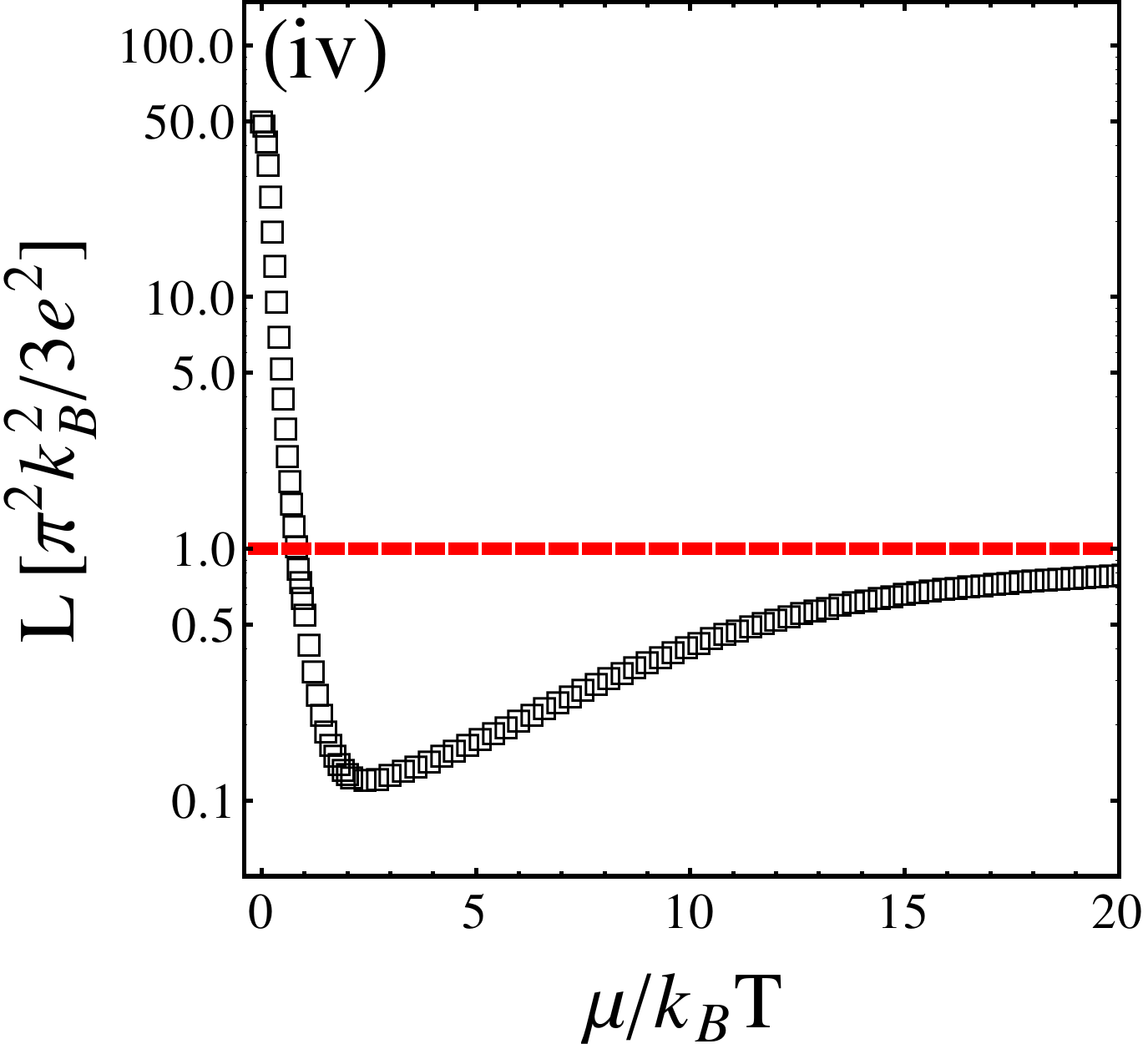}
\caption{Transport coefficients in the presence of Coulomb interactions and short-ranged disorder as  
functions of $\frac{\mu}{k_\mathrm{B} T}$. We take the short-ranged disorder strength, the fine structure constant, and 
the order of the polynomial basis the same as those in Fig.~\ref{imp-only-pic}. The black squares are the numerical result. 
(i) Conductivity. 
The horizontal blue 
dashed line indicates the disorder-only conductivity 
$\s_\mathrm{imp}^{(\mathrm{s})}$
in Eq.~(\ref{short-only}), 
while
the red dashed line is the 
``Drude'' component of the hydrodynamic conductivity
$\sigma_{D}$	
given by the first term of Eq.~(\ref{cond-hyd}). The inset panel shows the minimal conductivity at the charge neutrality 
as a function of the fine structure constant. A linear fit of the numerical result gives 
$	
	\a_\mathrm{int}^2 \s_\mathrm{min} \approx 0.79 + 9.13 \a_\mathrm{int}
$ (black dashed line).
This is consistent with the unscreened result in \cite{kashsmin,lars2007}.
(ii) Thermoelectric power. 
The top red dashed curve is the ideal clean hydrodynamic result in Eq.~(\ref{HydroTEP}) 
and the bottom blue dashed curve is the result obtained from Mott's formula. 
The insert panel is a semi-log plot for the hydrodynamic regime. 
(iii) Thermal conductivity. 
The insert panel shows the following ``synthetic'' Lorenz ratio:
This is a plot of the thermal conductivity for a hydrodynamic 
relativistic gas in the \emph{absence of impurities}, normalized to the minimal conductivity 
at charge neutrality, Eq.~(\ref{L-h-0}).  
(iv) Lorenz ratio for graphene with Coulomb interactions and short-ranged disorder only. 
The horizontal red dashed line indicates the Wiedemann-Franz law.
}    \label{hydro-coes}
\end{figure}

We combine Coulomb interactions 
[see Fig.~\ref{scats}(b)]
and short-ranged impurity to verify the predictions 
of the relativistic hydrodynamic theory \cite{mueller2008,matt09,narozhny15}. The transport coefficients obtained by the 
numerical solution of the Boltzmann equation are shown in Fig.~\ref{hydro-coes}. Close to charge neutrality 
$\mu \lesssim k_\mathrm{B}T$ the conductivity  [Fig.~\ref{hydro-coes}(i)] remains finite. This reflects the ``minimal'' 
conductivity due to the electron-hole collisions. 

In the hydrodynamic
(interaction-dominated) regime of primary interest, 
$\tau_\mathrm{in} \ll \tau_\mathrm{el}$ \cite{mueller2008,matt09}. 
Here $1/\tau_\mathrm{in}$ denotes the inelastic scattering rate due to electron-electron and electron-hole collisions,
while $1/\tau_\mathrm{el}$ is the scattering rate due to elastic electron-impurity and (quasi)elastic 
electron--acoustic-phonon collisions. 
(In this section we neglect optical phonons, which are dealt with below.) 
Strong inelastic scattering quickly relaxes fluctuations to local equilibrium. Intercarrier scattering is
special however, in that it preserves the total energy and momentum of the Dirac fluid \cite{mueller2008,matt09}.  
This means that the distribution function for electrons and holes is always close to Fermi-Dirac in some
co-moving reference frame, and this translates into strong constraints on kinetic coefficients. 

At charge neutrality, charge flow is decoupled from momentum flow, and can be relaxed by 
electron-hole collisions alone. In the interaction-dominated regime, the minimal conductance at the 
Dirac point is to a first approximation a function only of the dimensionless interaction strength $\a_\mathrm{int}$ 
[Eq.~(\ref{aintDef})] \cite{kashsmin,lars2007,MarkusSub2008,mueller2008,matt09},
and is therefore independent of temperature (ignoring logarithmic renormalization effects \cite{matt08}).
This is very different from the case of disorder-dominated transport due to Coulomb impurities. 
In a disorder-dominated sample, around charge neutrality scattering off Coulomb impurities leads 
to a decreasing resistance with temperature, as shown in Fig.~\ref{imp-only-pic}(i). 
This can be understood via dimensional analysis, since the resistivity is proportional to the
impurity density, and the only other length scale is the thermal de Broglie wavelength:
$\rho(T) \sim n_\mathrm{imp} (\hbar v_\mathrm{F}/k_\mathrm{B}T )^2$.

In the thermopower experiment \cite{GXFKexp}, no downturn in resistivity with increasing temperature 
was observed over the temperature range of interest (130--350 K). Instead, a superlinear rise was 
seen above 200 K that we attribute to electron--optical-phonon scattering, discussed below. 
This should be contrasted with earlier high-temperature experiments that observed 
a decreasing resistance \cite{Shao2008}; the latter can presumably be attributed
to disorder-dominated transport \cite{Vasko2007}.   

Away from charge neutrality and at intermediate temperatures, the thermopower shown in 
Fig.~\ref{hydro-coes}(ii) approaches the ideal clean hydrodynamic result
\begin{align}\label{HydroTEP}
	\a_{\infty} 
	= 
	\mathsf{s} / e n,
\end{align}
which is the thermodynamic entropy per charge; $\mathsf{s}$ denotes the entropy density.
At higher densities/lower temperatures, $\a_{\infty} \rightarrow 0$, consistent with the Mott relation
[Eq.~(\ref{short-only}) for short-ranged impurity scattering]. 
For $\mu \ll k_B T$ the Lorenz ratio [Fig.~\ref{hydro-coes}(iv)] is much larger than that of a Fermi liquid, 
$L/L_0 \gg 1$. Wiedemann-Franz recovers far away from the Dirac point $\mu \gg k_\mathrm{B}T$.

As shown in 
the inset of
Fig.~\ref{hydro-coes}(i), the rescaled minimal conductivity $\a_\mathrm{int}^2 \s_\mathrm{min}$ 
is almost linear in the fine structure constant $\a_\mathrm{int}$ for $\a_\mathrm{int} \lesssim 1$, 
which reflects the Coulomb screening effect. In the absence of screening $\a_\mathrm{int} \ll 1$, 
we recover the results of Refs.~\onlinecite{kashsmin} and \onlinecite{lars2007}. 
As shown in Fig.~\ref{hydro-coes}(iii), the non-monotonicity of the thermal conductivity $\k_\infty$ 
(or the Lorenz ratio $L$) as a function of $\mu/k_B T$ is simply a consequence of the ideal relativistic 
thermodynamics
[Eq.~(\ref{L-h-0})]. 
In the interaction-limited regime, 
the enhancement of the Lorenz ratio diverges as the 
strength of impurity scattering vanishes. 
The hydrodynamic enhancement will also dominate over that attributable to Coulomb impurities, Eq.~(\ref{L-a-0}). 
For sufficiently weak impurity scattering and in the absence of optical phonons, the hydrodynamic 
description should generally apply, regardless of the scattering mechanisms that lift the zero modes of 
the
Coulomb collision operator. 
In reality both short-ranged and long-ranged impurities are simultaneously present, and the resulting transport 
coefficients 
have similar features as shown in Fig.~\ref{hydro-coes}. 

\begin{figure}[h!]
\centering
\includegraphics[width=0.235\textwidth]{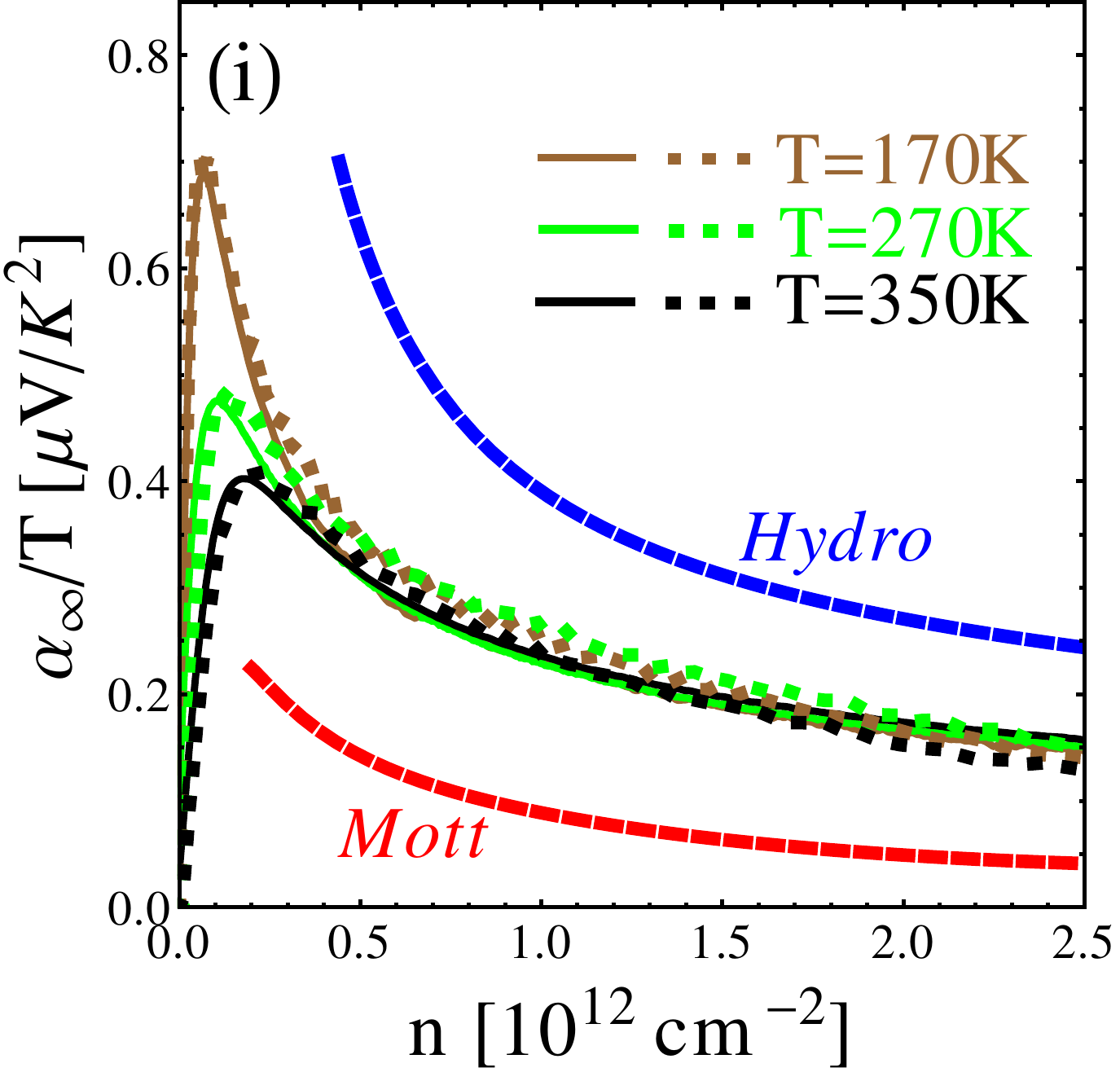}
\includegraphics[width=0.235\textwidth]{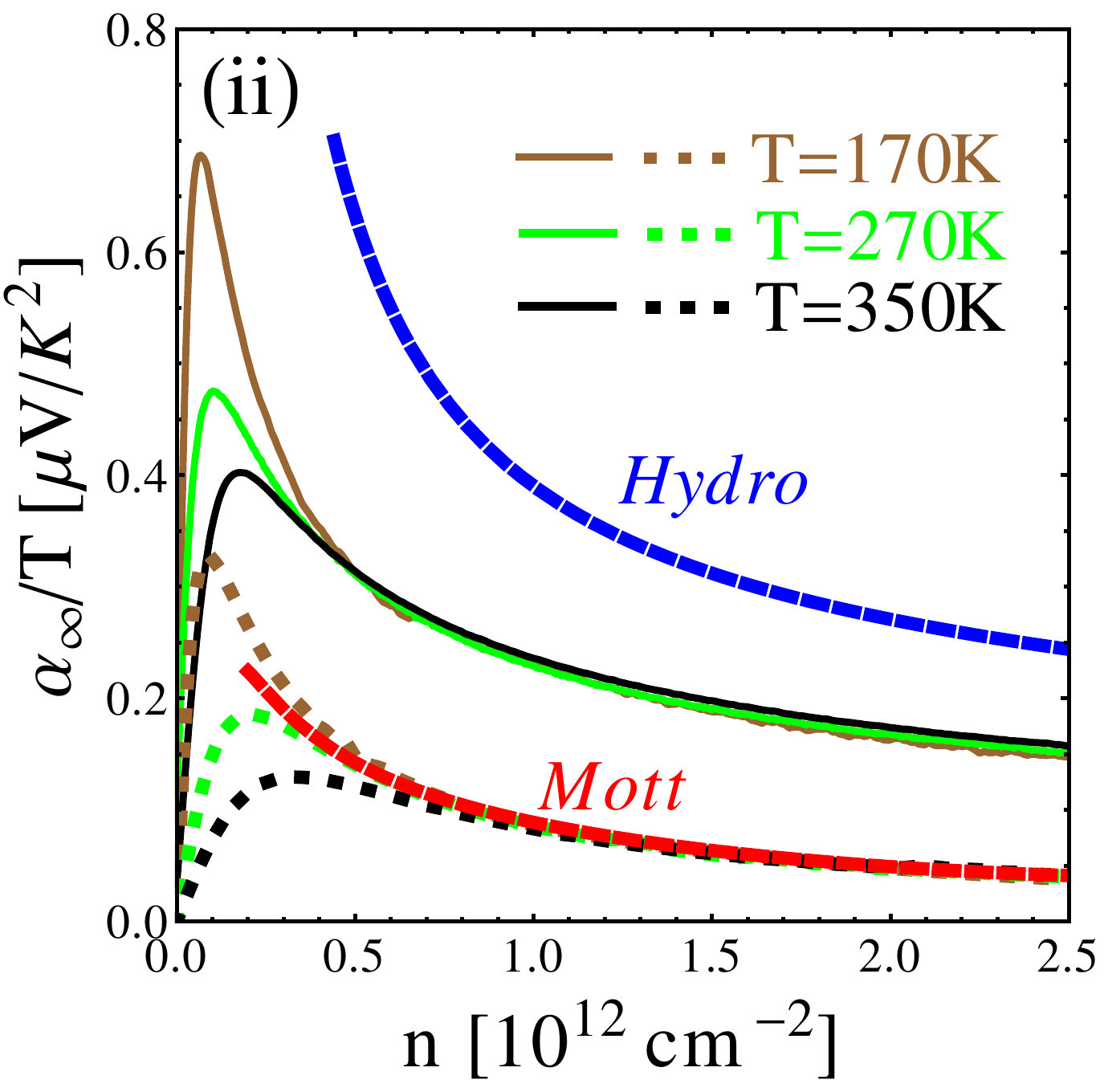} \\
\includegraphics[width=0.235\textwidth]{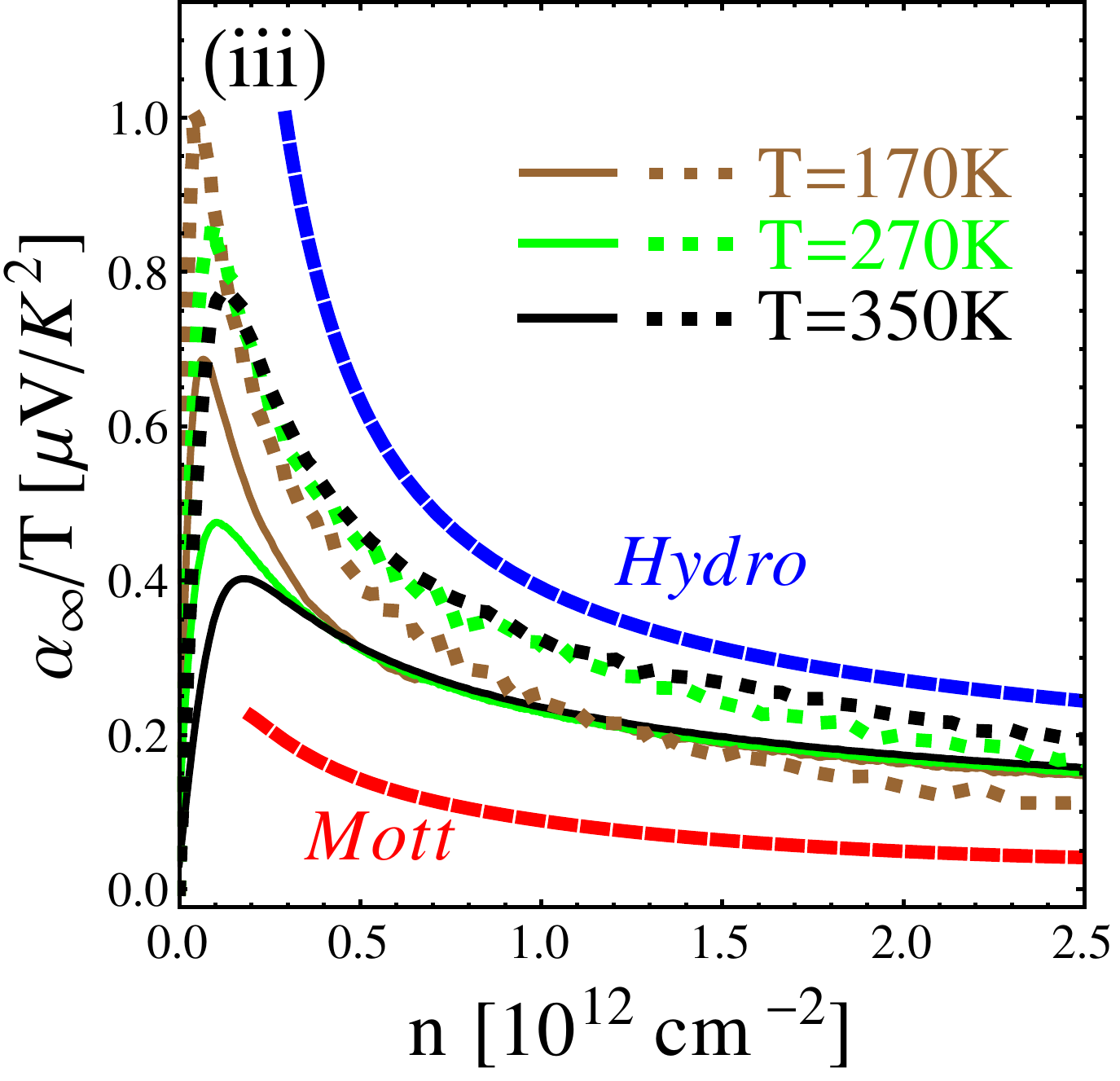}
\includegraphics[width=0.235\textwidth]{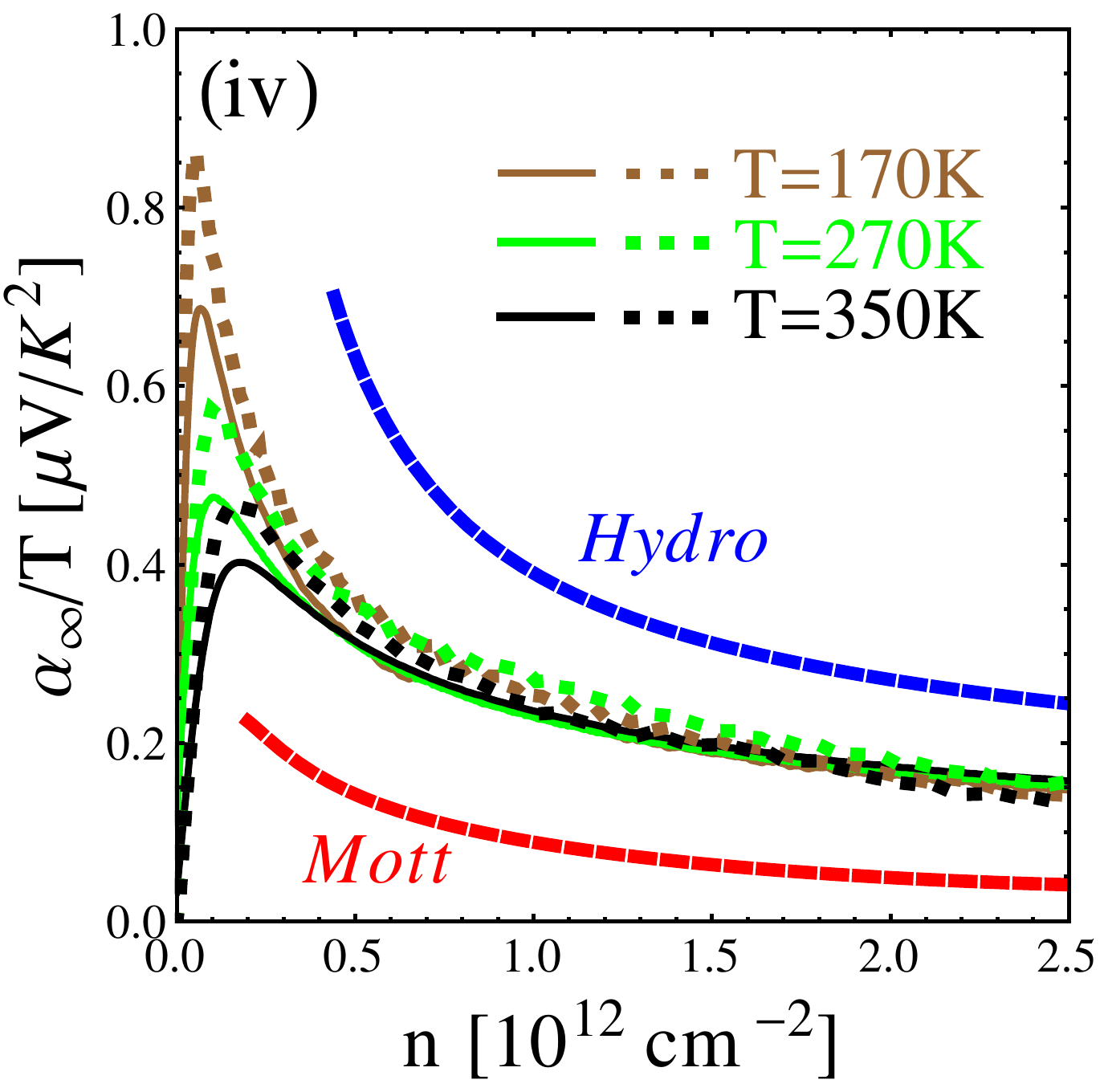}    \\
\includegraphics[width=0.235\textwidth]{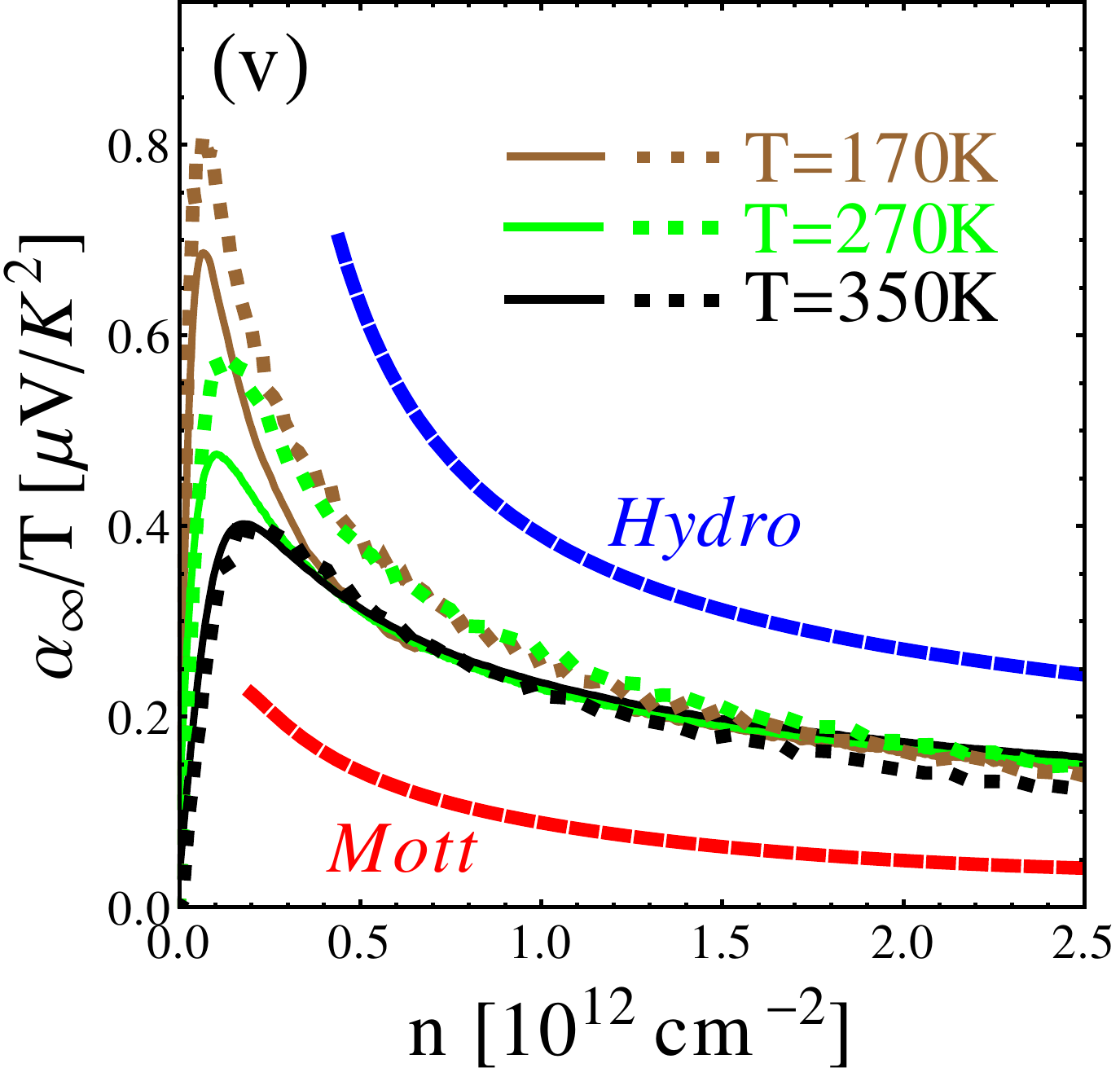}
\caption{
Thermopower as a function of doping and temperature including various scattering mechanisms,
and comparison to the experiment in \cite{GXFKexp}.	
The dotted (solid) curves are the result of theory (experiment). 
The bottom red and top blue dashed lines show the thermopower calculated from the experimental conductivity data
using 
Mott's formula~\cite{GXFKexp}
and the ideal hydrodynamic result [Eq.~(\ref{HydroTEP})], respectively.	
We use the same parameters and the 
temperature-dependent
optical-phonon-electron coupling strength as in Ref.~\onlinecite{GXFKexp}. 
(i) Thermopower incorporating impurities, Coulomb channels A and B, and 
all electron--optical-phonons scattering processes depicted in Fig.~\ref{scats}(c).
The ``optical'' electron-hole Coulomb scattering channel C [Fig.~\ref{scats}(b)$_{\msf{iii}}$], 
which shows a plasmon-enhancement in the RPA, is excluded by hand.
(ii) Thermopower incorporating only short- and long-ranged impurities. 
(iii) Thermopower incorporating impurities and Coulomb channels A, B, and C,
neglecting optical phonons.
(iv) Thermopower incorporating all scattering mechanisms, including the Coulomb channel C. 
(v) Thermopower incorporating disorder, Coulomb channels A and B, and optical phonons, 
but neglecting the optical-phonon mediated electron-hole imbalance relaxation processes
depicted in Figs.~\ref{scats}(c)$_{\msf{ii}}$ and \ref{scats}(c)$_{\msf{iv}}$.
}    
\label{hydro-coes-exp}
\end{figure}

Our result for the thermopower in the presence of both types of disorder and Coulomb carrier-carrier scattering,
but in the absence of optical phonons, is shown in Fig.~\ref{hydro-coes-exp}(iii). There it is compared to the experimental
results from~\cite{GXFKexp}. Our numerical results monotonically approach the ideal hydrodynamic limit 
[Eq.~(\ref{HydroTEP})] with increasing temperature, 
except near charge neutrality where a finite impurity density sends the thermopower to zero as $n \rightarrow 0$ 
[Eq.~(\ref{therp-hyd})]. 
The experimental results instead show a saturation of the
thermopower midway between the Mott and hydrodynamic bounds. Below we show that the additional inclusion of 
electron--optical-phonon scattering gives good agreement with the experiment, Fig.~\ref{hydro-coes-exp}(i).

\subsection{Optical-phonon-limited transport}

\begin{figure}
\centering
\includegraphics[width=0.237\textwidth]{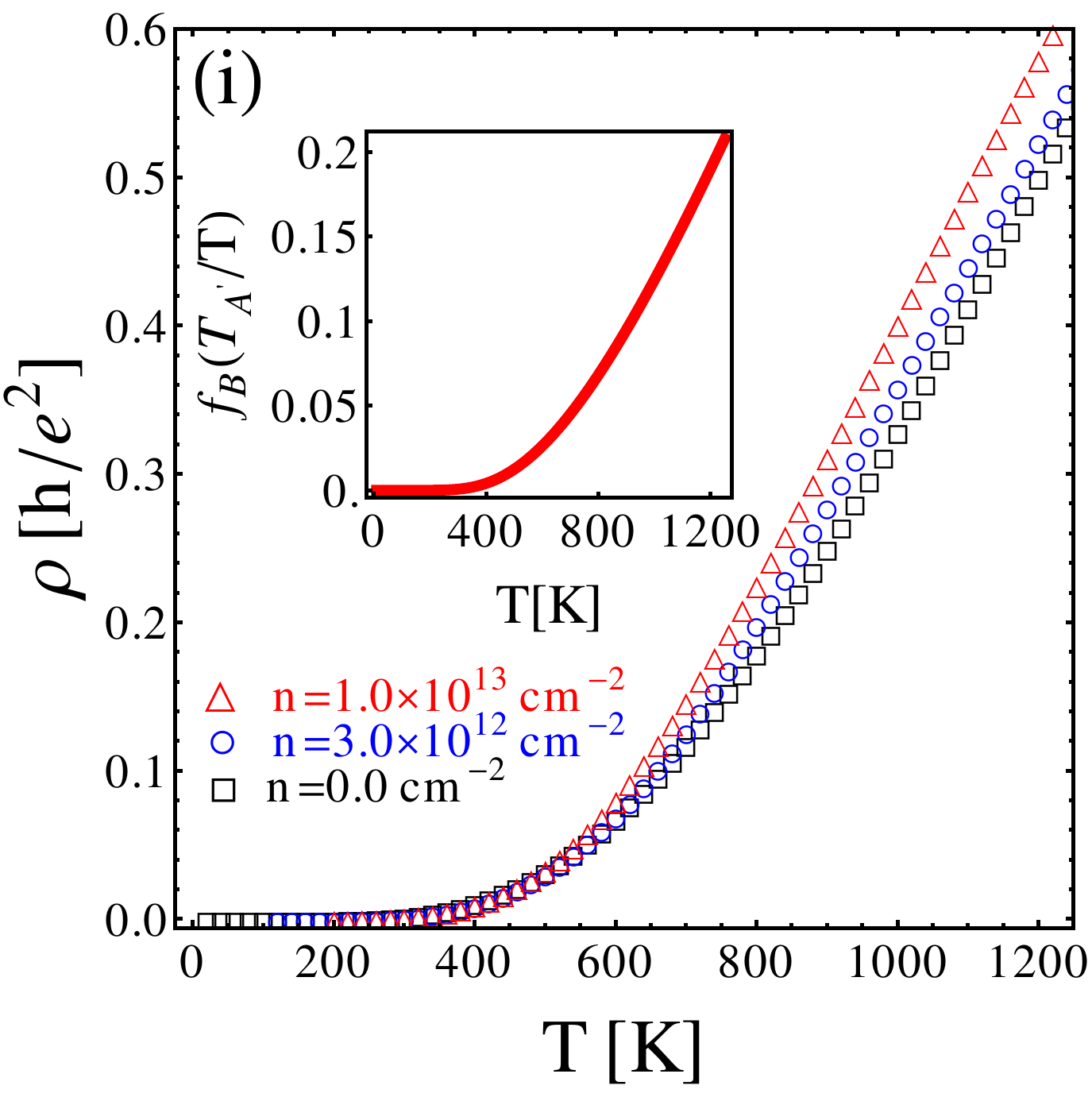} 
\includegraphics[width=0.235\textwidth]{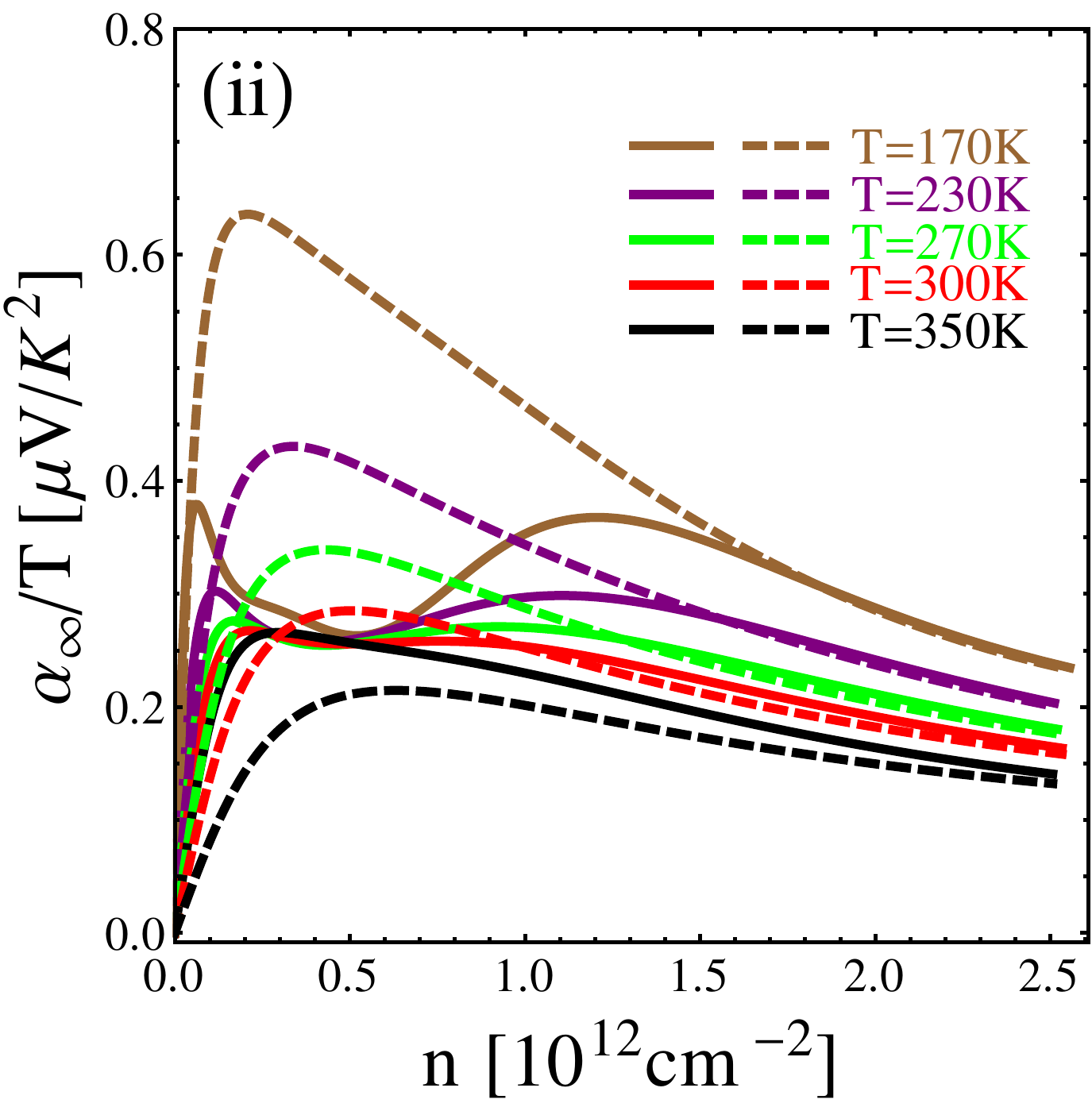}
\caption{Optical-phonon--limited transport coefficients as functions of temperature and charge density. 
We take the effective
dimensionless
electron--optical-phonon coupling strength $\wtd{\a}_{\mathrm{opt}}^2 = 1$ 
[c.f.~Eq.~(\ref{Oph-Params}); $\wtd{\a}_{\mathrm{opt}}^2 = {\a}_{\mathrm{opt}}^2/(16 \pi^2)$]	
and the optical-phonon 
temperature $T_{A^\prime} \equiv \hbar \, \w_{A^\prime}/ k_\mathrm{B} \approx 2200 \, \mathrm{K}$ \cite{GXFKexp}. 
(i) Resistivity $\rho \equiv \s^{-1}$ as a function of temperature for various charge densities. For comparison 
the insert panel shows the Bose-Einstein distribution function of optical phonons [Eq.~(\ref{bose-fun})]. 
(ii) Thermoelectric power $\a_\infty$ as a function of density for various temperatures. The solid (dashed) 
curves show the result in the presence (absence) of the electron-hole imbalance relaxation processes 
[see the diagrams (c)$_{\msf{ii}}$ and (c)$_{\msf{iv}}$ in Fig.~\ref{scats}].}  
\label{opt-ct}
\end{figure}

The total energy of electrons and holes is no longer conserved in the presence of the optical-phonon 
bath 
[Fig.~\ref{scats}(c)]. 
Via Eqs.~(\ref{trans-coe-def}), (\ref{def-Lij}), (\ref{F-12-c}), and (\ref{opt-mat}), 
we calculate the electronic transport coefficients due only to electron--optical-phonon scattering processes. 
The results are shown in Fig.~\ref{opt-ct}. 

We obtain the resistivity as a function of temperature and charge-carrier density that qualitatively coincides 
with the result in Ref.~\onlinecite{sohier2014}. The resistivity weakly depends on the charge carrier density, 
and, moreover, the optical phonons are thermally activated at the temperature about one order of magnitude 
lower than their frequency. Three temperature regimes can be observed.
(i) Collisionless regime ($T \lesssim 150 \, \mathrm{K}$). The resistivity is almost zero since the population of thermally activated phonons 
is
exponentially small when $T \ll T_{A^\prime}$. 
(ii) Crossover regime ($150 \, \mathrm{K} \lesssim T \lesssim 400 \, K$). The resistance increases superlinearly in temperature. 
(iii) High-temperature regime ($T \gtrsim 400 \mathrm{K}$). The resistance increases linearly in temperature. For high enough temperatures, 
the optical phonons play a similar role as impurities, 
yet the scattering amplitude is enhanced by the Bose-Einstein 
distribution function $f_\mathrm{B}(T_{A^\prime}/T) \sim T /T_{A^\prime}$.
The temperature dependence of the resistivity qualitatively follows the Bose-Einstein distribution function of the optical phonons. 

In the crossover regime, the electron--optical-phonon scattering is strongly inelastic. 
The thermopower [Fig.~\ref{opt-ct}(ii)] due to electron--optical-phonon scattering alone 
does not follow Mott's formula~\cite{non-mon-opt}. 
Furthermore, the electron-hole imbalance 
relaxation processes
[Figs.~\ref{scats}(c)$_{\msf{ii}}$ and \ref{scats}(c)$_{\msf{iv}}$] 
have significant effects at low doping.

\subsection{All scattering mechanisms; comparison to thermopower measurements} \label{comp-exp}

Finally we combine all scattering mechanisms to model the data of the experiment in Ref.~\onlinecite{GXFKexp}.
In order to interpret the data we need first to estimate all the effective parameters. 
Since the graphene sample is encapsulated between two hexagonal-boron-nitride substrates we estimate the fine structure 
constant as 
$
	\a_\mathrm{int}
	=
	2e^2/(\k_1+\k_2)\hbar\vf \approx 0.6
$ 
where $\k_1 = \k_2 \approx 3.8$ is the dielectric constant of boron nitride~\cite{jang08}. 
The 
dimensionless
short-ranged impurity strength $\wtd{g}$ and the Coulomb impurity concentration $n_\mathrm{imp}$ are determined by the 
conductivity data at low temperature and high doping, where inelastic scattering is negligible. According to this 
analysis we have 
$\wtd{g} \approx 1.1 \times 10^{-4}$ 
and 
$n_\mathrm{imp} \approx 3 \times 10^9 \, \mathrm{cm}^{-2}$. 
Finally, the electron--optical-phonon coupling is attained by fitting the 
electrical
conductivity data at high temperatures. Note that to reach a \emph{quantitative} agreement to 
the experimental data, we have tuned the optical-phonon frequency to $T_{\mathrm{A}'}=2200\,\mathrm{K}$, 
which is a little bit higher than the values reported in Refs.~\onlinecite{sohier2014} and \onlinecite{basko2008}. 
The reason for this enhancement might be that $\mathrm{A}'$ phonons are more rigid due to substrate encapsulation 
or 
that
higher-frequency optical-phonon branches are also involved.

The fitting procedure described above gives an electron--optical-phonon 
coupling that increases with decreasing temperature, see \cite{GXFKexp} for details.  
This is presumably due to a combination of ultraviolet renormalization \cite{baskoaleiner2008,basko2008} 
and the temperature-dependent Coulomb screening~\cite{sohier2014,basko2008}. 
We leave the theoretical study of the electron--optical-phonon vertex for 
\emph{deeply inelastic} energy and momentum transfers to future work. 

We have calculated the thermopower for every combination of the scattering sources in Fig.~\ref{scats} and present the most 
informative results in Fig.~\ref{hydro-coes-exp}. 
As shown in Fig.~\ref{hydro-coes-exp}(i), 
our theoretical result coincides quantitatively well with the experimental data if we take into account impurities, 
optical phonons, and Coulomb interactions, yet  
\emph{neglect} the electron-hole optical scattering channel C. 
Figure \ref{hydro-coes-exp}(ii) indicates that the result of the Mott's relation (red dashed line) merely reflects the 
impurity-only (both short- and long-ranged) thermopower at high doping.  

Fig.~\ref{hydro-coes-exp}(iii) shows the results in the absence of optical phonons, 
but including short-ranged and Coulomb-impurity
scattering, as well as carrier-carrier channels A, B, and C [Fig.~\ref{scats}(b)].
Although graphene is relatively degenerate for $n \geq 10^{12}$ cm${}^{-2}$ 
[$T_F = 1350$ K], the Mott relation is not recovered for the measured temperatures. 
At these high densities, this is due to the pure intraband electron-electron scattering in channel A.
The disorder is so weak in the experiment that we would need very high densities to observe Fermi liquid behavior;
in other words, it is possible to be both degenerate and hydrodynamic in a very clean sample.
We estimate that at $T = 170$ K, the Mott relation would be recovered only at densities above $n \sim 10^{13}$ cm${}^{-2}$. 
Comparing Fig.~\ref{hydro-coes-exp}(iii) (results in the absence of optical phonons) to Fig.~\ref{hydro-coes-exp}(i), we observe 
that the optical phonons significantly suppress the thermopower at higher temperatures
and drive the system further away from an ideal hydrodynamic fluid.

The thermopower $\alpha$ is well-defined and given by Eq.~(\ref{HydroTEP}) in the absence of a mechanism for momentum relaxation.
Away from charge neutrality however, even within the hydrodynamic regime \emph{some} such mechanism is 
necessary to separately define
$\sigma$ and $\sigma \alpha$ in Eq.~(\ref{Jlin-resp}). 
In general the ratio $\alpha$ is also sensitive to this mechanism.  
Here this role is filled by either disorder or optical phonon scattering. 
In particular, Coulomb impurities are poorly screened at low temperatures for charge-carrier densities not too large,
while optical phonons become important at higher temperatures. 
As discussed in Sec.~\ref{hydro-sec-2}, the optical-phonon scattering becomes nonnegligible when the collision matrix 
elements [Eqs.~(\ref{imp-coll-M}) and (\ref{opt-mat})] satisfy 
$
	\left( \mathcal{M}_{\mathrm{opt}} \right)_{00} 
	\gtrsim 
	\left( \mathcal{M}_{\mathrm{imp}} \right)_{00}
$. 
For a charge-carrier density $n \sim 10^{12} \, \mathrm{cm}^{-2}$ ($T_\mathrm{F} \approx T_{A^\prime} \approx 2000\,\mathrm{K}$) and 
temperature $T < 350 \, \mathrm{K}$, this leads to 
$
	T \gtrsim T^\ast \sim T_\mathrm{F}/\ln{(10^5 \,\wtd{\a}_\mathrm{opt}^2)}  \sim 200 \, \mathrm{K}
$ 
via a simple estimation~\cite{est-alpha}, based on the parameters in the experiments \cite{GXFKexp}. 

The plasmon pole in the dynamically-screened Coulomb interaction can enhance the electron-hole scattering in the Coulomb channel C 
[Fig.~\ref{scats}(b)${}_{\msf{iii}}$]. This mechanism could strengthen the hydrodynamic response. 
Comparing Fig.~\ref{hydro-coes-exp}(iv) (which includes channel C) to Fig.~\ref{hydro-coes-exp}(i) (which neglects it), 
we conclude that the associated plasmon enhancement~\cite{CatAleiner,Principi15,flensberg95,levitov} is somehow suppressed in 
the experiments. We propose that this suppression may be due to 
additional screening by metallic gates that soften the plasmon dispersion,
or damping induced by the plasmon--optical-phonon coupling~\cite{Abajo2014},
which is not accounted for in our treatment. 
Comparing 
Fig.~\ref{hydro-coes-exp}(v) [results in the absence of electron-hole imbalance relaxation processes due to optical phonons, Figs.~\ref{scats}(ii) and \ref{scats}(iv)] to Fig.~\ref{hydro-coes-exp}(i), we observe that these processes also significantly affect the thermopower at lower charge densities.

\section{Boltzmann equation in the presence of impurities, Coulomb interaction, and optical phonons}   \label{tech-sec}

\subsection{Collision integrals} \label{c-integrals}


The elastic collision integral in Eq.~(\ref{collisions}) 
gives Fermi's golden rule amplitudes associated to the diagram in Fig.~\ref{scats}(a), and reads
\begin{widetext}
\begin{subequations} \label{imp-col}
\be
	\St_{\mathrm{imp},\l}[f_\l] 
	= 
	\St_{\mathrm{imp},\l}^{(\mathrm{s})}[f_\l]
	+
	\St_{\mathrm{imp},\l}^{(\mathrm{l})}[f_\l],
\ee
where $\St_{\mathrm{imp},\l}^{(\mathrm{s})}$ and $\St_{\mathrm{imp},\l}^{(\mathrm{l})}$ describe the short- and long-ranged impurity scattering, respectively,
\begin{align}
	\St_{\mathrm{imp},\l}^{(\mathrm{s})}[f_\l] 
	= & \, 
	\int_{\bq} \,\delta(\ep_{\bq}-\ep_{\bp})  
		\left[ G_0 + G_\mathrm{f} 
		\left(\frac{1 + \hat{\bp}\cdot\hat{\bq}}{2}\right) + G_\mathrm{b} 
		\left(\frac{1 - \hat{\bp}\cdot\hat{\bq}}{2}\right)
		\right] 
		\left[ f_\l(\bq, \bfr) - f_\l(\bp,\bfr) \right], \label{simp-col} \\
	\St_{\mathrm{imp},\l}^{(\mathrm{l})}[f_\l] 
	= &\, 
		\frac{2  \pi n_\mathrm{imp}}{\hbar}
		\int_{\bq} \,\delta(\ep_{\bq}-\ep_{\bp}) 
		\left(\frac{1 + \hat{\bp}\cdot\hat{\bq}}{2}\right) 
		\left|U_\mathrm{eff}(\w=0,|\bp-\bq|)\right|^2 
		\left[ f_\l(\bq, \bfr) - f_\l(\bp,\bfr) \right]. 
	\label{limp-col}
\end{align}
\end{subequations}
In Eq.~(\ref{simp-col}) the effective short-ranged impurity strengths are 
$G_0 = (2\pi)^2 (2 g_A + g_{A3})$, 
$G_\mathrm{f}=(2\pi)^2g_u$, and 
$G_\mathrm{b} = (2\pi)^2(2 g_m + g_v) $ \cite{matt08}. 
In Eq.~(\ref{limp-col}) the long-ranged impurity scattering is characterized by the Coulomb impurity number per unit area 
$n_\mathrm{imp}$ and the static RPA Coulomb interaction $|U_\mathrm{eff}(\w=0,k)|^2$ [see Appendix~\ref{ftsc}]. 
The Dirac delta function $\delta(\ep_{\bq}-\ep_{\bp})$ enforces energy conservation. 
The terms associated to the factors $(1 \pm \hat{\bp}\cdot\hat{\bq} )/2$ describe the enhancement of forward ($+$) and backward ($-$) scattering.
In Eq.~(\ref{imp-col}), we have introduced the shorthand notation
\[
	\int_{\bq} \equiv \int \frac{d^2 \vex{q}}{(2 \pi)^2}.
\]

The Coulomb collision integral is evaluated at the RPA level associated to the three scattering processes depicted in Fig.~\ref{scats}(b), 
\begin{subequations}  \label{coulomb-col}
\begin{align} 
	\St_{\mathrm{int}, \l}[\{f_\l\}] 
	= & \, 
	\frac{N}{\hbar} 
	\, \int_{\bp_2, \bp_3, \bp_4} 
	\! \frac{1+\hat{\bp}\cdot\hat{\bp}_2}{2} 
	\, \frac{1+\hat{\bp}_3\cdot\hat{\bp}_4}{2} 
	\left( 2 \pi \right)^3   \nn \\
	& \times \big[ \delta^{(3)}{(\sfp+\sfp_4 - \sfp_2 - \sfp_3)} \, |U_\mathrm{eff}(\sfp-\sfp_2)|^2 \, \big\{ \left[1-f_\l(\bp,\bfr)\right] f_\l(\bp_2,\bfr) \, [1-f_\l(\bp_4,\bfr)] \, f_\l(\bp_3,\bfr) \nn \\
	&  \hspace{5.95cm}  \left.          
		- \left[1-f_\l(\bp_2,\bfr)\right] f_\l(\bp,\bfr) \, [1-f_\l(\bp_3,\bfr)] \, f_\l(\bp_4,\bfr) \big\} \right. \label{col-1} \\ 
	&  \,\, + \delta^{(3)}(\sfp-\sfp_4-\sfp_2+\sfp_3) \, |U_\mathrm{eff}(\sfp-\sfp_2)|^2 \, \big\{ \left[1-f_\l(\bp,\bfr)\right] f_\l(\bp_2,\bfr) \, f_{-\l}(\bp_4,\bfr) \, [1-f_{-\l}(\bp_3,\bfr)]\nn \\
	& \hspace{5.95cm} 
		- \left[1-f_\l(\bp_2)\right] f_\l(\bp) \, f_{-\l}(\bp_3) \, [1-f_{-\l}(\bp_4)]\big\} \label{col-2} \\
	&  \left.  \,\,  + \delta^{(3)}(\sfp-\sfp_4-\sfp_3+\sfp_2) \, |U_\mathrm{eff}(-\sfp-\sfp_2)|^2 \, \big\{ \left[1-f_\l(\bp,\bfr)\right] 
		\left[ 1-f_{-\l}(\bp_2,\bfr)\right] \, f_\l(\bp_3,\bfr) \, f_{-\l}(\bp_4,\bfr) \right. \nn \\
	& \hspace{6cm} 
		- f_\l(\bp,\bfr)\, f_{-\l}(\bp_2,\bfr) \, [1-f_\l(\bp_3,\bfr)] \, [1-f_{-\l}(\bp_4,\bfr)] \big\} \big]. \label{col-3}
\end{align}
\end{subequations}
Equations.~(\ref{col-1})--(\ref{col-3}) correspond to the 
Coulomb scattering channels A--C,
diagrams (b)${}_{\msf{i}}$--(b)${}_{\msf{iii}}$, 
respectively. 
The quasiparticle energy and momentum are written in the three-vector form $\sfp \equiv (\ep_\bp, \bp)$, and the three-dimensional Dirac delta functions $\delta^{(3)}(\cdots)$ describe energy and momentum conservation. 
Channel A, Eq.~(\ref{col-1})
is electron-electron scattering, while 
channels B and C, Eqs.~(\ref{col-2}) and (\ref{col-3})
are electron-hole scattering processes. The RPA screened Coulomb interaction takes the form as shown in Appendix~\ref{ftsc}. 
We emphasize that even at charge neutrality, dynamical screening is crucial at finite temperature due to the thermal activation
of electron-hole pairs.
Interaction-mediated ``Auger''
imbalance relaxation processes are suppressed because of the linear dispersion of electrons and holes \cite{matt09}. 
Due to kinematic constraints, channels A and B act in the ``quasi-static'' regime $|\omega| \leq v_F q$,  
while channel C acts in the ``optical'' regime $|\omega| \geq v_F q$
[Fig.~\ref{kinemitics}]. Here $\omega$ and $q$ are the frequency and momentum transferred across the Coulomb line.

The carrier--optical-phonon scattering is described by the diagrams in Fig.~\ref{scats}(c) and leads to the collision integral
\begin{subequations} \label{oph-e}
\begin{align} 
	& \St_{\mathrm{oph},\l} [\{f_\l\} ] 
	= 
	\frac{(2\pi)^2\,\b_{A^\prime}^2 \, s_0}{\w_{A^\prime} \, M}	 
	\int_{\bq} \, \left( \frac{1- \hat{\bp}\cdot\hat{\bq}}{2} \right) \nn \\
	& \times \Big( f_\mathrm{B}(\w_{A^\prime}) \big\{ \delta\left( \ep_\bp-\ep_\bq - \w_{A^\prime}\right)  
	\left[ 1- f_\l (\bp,\bfr) \right] f_\l(\bq,\bfr) - \delta\left( \ep_\bp-\ep_\bq 
		+ \w_{A^\prime} \right) f_\l (\bp,\bfr) \left[ 1- f_\l(\bq,\bfr) \right] \big\} \label{e-e-oph-abs} \\
	& \quad 
	+ 
	\left[ 1+f_\mathrm{B}(\w_{A^\prime}) \right] \big\{ \delta\left( \ep_\bp-\ep_\bq + \w_{A^\prime} \right) \left[ 1- f_\l (\bp,\bfr) \right] f_\l( \bq,\bfr) 
		- \delta\left( \ep_\bp-\ep_\bq - \w_{A^\prime} \right) f_\l (\bp,\bfr) \left[ 1- f_\l(\bq,\bfr) \right] \big\} \Big) \label{e-e-oph-emi} \\
	& + 
	\frac{(2\pi)^2\,\b_{A^\prime}^2 \, s_0}{\w_{A^\prime} \, M}	
	\int_{\bq} \, 
	\left( \frac{1 -  \hat{\bp}\cdot\hat{\bq}  	}{2} \right) 
	\delta\left( \ep_\bp  + \ep_\bq - \w_{A^\prime} 	\right) \, 
	\Big\{ f_\mathrm{B}(\w_{A^\prime}) \, \left[ 1- f_\l (\bp,\bfr) \right] \left[ 1 - f_{-\l}(	\bq	,\bfr) \right] \label{e-h-oph-abs} \\
	& \vspace{5cm} 
	- 
	\left[1+f_\mathrm{B}(\w_{A^\prime})\right] \, f_\l (\bp,\bfr) \, f_{-\l}( \bq	,\bfr) \Big\}, 
	\label{e-h-oph-emi}
\end{align} 
\end{subequations} 
\end{widetext}
where $M = 2.0 \times 10^{-23} \, \mathrm{g}$ 
is the carbon atom mass and $s_0 = 2.62 \,\text{\r{A}}^2$ the area per carbon atom.
Equations~(\ref{e-e-oph-abs},\ref{e-e-oph-emi}) 
[(\ref{e-h-oph-abs},\ref{e-h-oph-emi})] 
correspond to the diagrams in Figs.~\ref{scats}(c)$_{\msf{i,iii}}$ [\ref{scats}(c)$_{\msf{ii,iv}}$], respectively.	
We note that the processes 
(c)$_{\msf{ii}}$ and (c)$_{\msf{iv}}$
are absent for acoustic phonon scattering~\cite{kim2010} because the acoustic-phonon velocity is much smaller than the Fermi velocity. 
To compare to the experiment in \cite{GXFKexp}, 		
we take the $A^\prime$ phonon temperature 
$
	T_{A^\prime} \equiv \hbar \w_{A^\prime}/k_\mathrm{B} \approx 2200 \, \mathrm{K}
$,
larger than in some previous studies \cite{sohier2014}.	
The coupling strength $\beta_{A^\prime}$ has been suggested to be strongly 
energy
dependent due to renormalization and screening by the Coulomb interactions~\cite{baskoaleiner2008,basko2008,sohier2014}. 
We treat $\beta_{A^\prime}$ as a fitting parameter when interpreting the experimental data~\cite{GXFKexp}.


We separate the distribution function $f_{\l}( \bp,\bfr )$ into two parts,
\be   \label{distf-f-1}
f_{\l} ( \bp,\bfr ) \equiv f_{\l}^{(0)}( \bp, \bfr ) + \delta{f}_{\l} ( \bp, \bfr ),
\ee
where $f_{\l}^{(0)} \left( \bp,\bfr \right)$ is the local equilibrium Fermi-Dirac function ($\beta = 1/k_\mathrm{B}T$)
\begin{subequations}  
\be  \label{fd-0th}
f_{\l}^{(0)}(\bp, \bfr) = \frac{1}{e^{ \beta (\vep_\bp - \mu_\l ) } + 1}, \quad \vep_\bp = \hbar \vf |\bp|, \quad \mu_\l = \l \mu,
\ee
and $\delta{f}_{\l} \left( \bp,\bfr \right)$ is the deviation from the local equilibrium and can be conveniently cast into the form 
\be   \label{delf-0}
\begin{split}
\delta{f}_{\l} \left( \bp,\bfr \right) = \frac{1}{\beta} \left[ -\frac{\rmd f_{\l}^{(0)}}{\rmd \varepsilon_\bp } \right] \chi_{\l} \left( \bp,\bfr \right). 
\end{split}
\ee
\end{subequations}
Via the standard derivation~\cite{LPv10}, from Eq.~(\ref{f-bolt-eq}) we obtain the time-independent linearized Boltzmann's equation for $\chi_{\l}$,
\be  \label{l-bolt-eq}
\sff_{\l}^{\prime}(p,z) \, \velf \cdot \left( \l e \, \b \, \boldsymbol{\mathcal{E}}  -  \frac{p - \lambda \ln{z}}{T} \nabla_\bfr{T} \right) =\frac{1}{ \hbar \beta } \wtd{\St}_{\l} [\{ \chi_{\l^\prime} \}],
\ee
where we have introduced the electrochemical field $\boldsymbol{\mathcal{E}} \equiv \mathbf{E} + \frac{1}{e}\nabla_\bfr{\mu}$, the effective Fermi-Dirac distribution function, and its derivative 
\be  \label{dless-df}
\sff_{\l}(p,z) \equiv \frac{1}{z^{-\l}e^{p} + 1}, \quad \sff_{\l}^\prime(p,z) \equiv -\partial_p \sff_{\l}(p,z),  
\ee
which depends on the dimensionless momentum 
$
	p = \beta \hbar \vf |\bp|
$ 
and the ``fugacity'' 
$
	z = \exp(\beta \mu).
$
On the right hand side of Eq.~(\ref{l-bolt-eq}), the linearized collision integral reads    
\be
\wtd{\St}_{\l} [\{ \chi_{\l^\prime} \}] =  \wtd{\St}_{\mathrm{imp}, \l} [\chi_\l]  + \wtd{\St}_{\text{int}, \l} [\{ \chi_{\l^\prime} \}]+ \wtd{\St}_{\text{oph}, \l}[\{ \chi_{\l^\prime} \}],
\ee
where the impurity collision integral is
\begin{widetext}
\begin{subequations} \label{st-imp-dless}
\be  
\wtd{\St}_{\mathrm{imp}, \l} [\chi_\l] = \wtd{\St}_{\mathrm{imp}, \l}^{(\mathrm{s})} [\chi_\l] + \wtd{\St}_{\mathrm{imp}, \l}^{(\mathrm{l})} [\chi_\l]
\ee
with short- and long-ranged components
\begin{align} 
\wtd{\St}_{\mathrm{imp}, \l}^{(\mathrm{s})} [\chi_\l] = & \, \int_{\bq} \delta{\left(p - q \right)} \left[ g_0 + g_\mathrm{f} \left(\frac{1 + \hat{\bp}\cdot\hat{\bq}}{2}\right) + g_\mathrm{b} \left(\frac{1 - \hat{\bp}\cdot\hat{\bq}}{2}\right)\right] \big[ \sff_{\l}^\prime(q) \, \chi_\l(\bq)  - \sff_{\l}^\prime(p) \, \chi_\l(\bp) \big], \label{st-imp-dless-s} \\
\wtd{\St}_{\mathrm{imp}, \l}^{(\mathrm{l})} [\chi_\l] = & \, \gamma^2 \int_{\bq}  \delta{\left(p -q \right)}\left( \frac{1+\hat{\bp}\cdot\hat{\bq}}{2} \right) \left|\wtd{U}_\mathrm{eff}(\w=0,|\bp-\bq|)\right|^2 \big[ \sff_{\l}^\prime(q) \, \chi_\l(\bq)  - \sff_{\l}^\prime(p) \, \chi_\l(\bp) \big], \label{st-imp-dless-l}
\end{align}
\end{subequations}
the Coulomb collision integral is
\begin{subequations}
\be
	\wtd{\St}_{\mathrm{int}, \l} [\{ \chi_\l\}] 
	= 
	\wtd{\St}_{\mathrm{int}, \l}^{\mathrm{(i)}} [\{ \chi_\l\}]  
	+ 
	\wtd{\St}_{\mathrm{int}, \l}^{\mathrm{(ii)}} [\{ \chi_\l\}]  
	+ 
	\wtd{\St}_{\mathrm{int}, \l}^{\mathrm{(iii)}} [\{ \chi_\l\}], 
\ee
with components corresponding to 
channels A--C in Fig.~\ref{scats}(b)	
\begin{align}
	\wtd{\St}_{\mathrm{int}, \l}^{\mathrm{(i)}}[\{ \chi_\l\}]   
	=  
	& \, 2\pi N \int_{\bk,\bq} \! 
	\frac{1+\hat{\bp}\cdot \what{\bp-\bq}}{2}\frac{1+\hat{\bk} \cdot \what{\bk-\bq}}{2} 
	\, \delta(p-|\bp-\bq|-k+|\bk-\bq|) \nn \\ & \, 
	\times |\wtd{U}_\mathrm{eff}(p-|\bp-\bq|,q)|^2 \, 
	\Xi_{p, |\bp-\bq|;k,|\bk-\bq|}^{\l,\l; \l,\l} 
	\big[ -\chi_{\l}(\bp) + \chi_{\l}(\bp-\bq) + \chi_{\l}(\bk) - \chi_{\l}(\bk-\bq) \big], 
		\label{channel-A} \\
	\wtd{\St}_{\mathrm{int}, \l}^{\mathrm{(ii)}}[\{ \chi_\l\}]  
	=  
	& \, 2\pi N \int_{\bk,\bq} \! 
	\frac{1+\hat{\bp}\cdot \what{\bp+\bq}}{2}\frac{1+\hat{\bk} \cdot \what{\bk-\bq}}{2} 
	\,  \delta(p-|\bp+\bq|+k-|\bk-\bq|) \nn \\ & \, 
	\times |\wtd{U}_\mathrm{eff}(p-|\bp+\bq|,q)|^2 \, 
	\Xi_{p, |\bp+\bq|;k,|\bk-\bq|}^{\l,\l ; -\l ,-\l} 
	\big[ -\chi_{\l}(\bp) + \chi_{\l}(\bp+\bq) - \chi_{-\l}(\bk) + \chi_{-\l}(\bk-\bq) \big], 
		\label{channel-B} \\
	\wtd{\St}_{\mathrm{int}, \l}^{\mathrm{(iii)}}[\{ \chi_\l\}] 
	=  
	& \, 2\pi N \int_{\bk,\bq} \! 
	\frac{1-\hat{\bp}\cdot \what{\bp-\bq}}{2}\frac{1-\hat{\bk} \cdot \what{\bk-\bq}}{2} 
	\, \delta(p+|\bp-\bq|-k-|\bk-\bq|) \nn \\ & \, 
	\times |\wtd{U}_\mathrm{eff}(p+|\bp-\bq|,q)|^2 \, 
	\Xi_{p, |\bp-\bq|;k,|\bk-\bq|}^{\l,-\l;\l,-\l} 
	\big[ -\chi_{\l}(\bp) - \chi_{-\l}(-\bp+\bq) + \chi_{\l}(\bk) + \chi_{-\l}(-\bk+\bq) \big], 
		\label{channel-C}
\end{align}
\end{subequations}
and the carrier--optical-phonon collision integral is
\begin{subequations} \label{oph-col-dl}
\begin{align}
	\wtd{\St}_{\mathrm{oph},\l} [\{\chi_{\l^\prime}\}] 
	= 
	& \, \a_\mathrm{oph}^2 
	\int_\bq \frac{1-\hat{\bp}\cdot\hat{\bq}  }{2} \, \nn \\
	& \times 
	\Big[ 
		\sff_{\l}^\prime(q) \,\big\{ \delta(p - q - \W_{\mathrm{A}^\prime}) \left[ \sff_\mathrm{B}(\wA) + \sff_{\l}(p) \right] 
		- 
		\delta(p - q + \W_{A^\prime}) \left[ \sff_\mathrm{B}(-\wA) + f_{\l}(p) \right] \big\} \chi_{\l}(\bq) \nn 	
		\\
                &   \,\,\, 
		- \sff_{\l}^\prime(p) \, \big\{ \delta(p - q + \W_{A^\prime}) \left[ \sff_\mathrm{B}(\wA) + \sff_{\l}(q) \right] 
		- 
		\delta(p - q - \W_{A^\prime}) \left[ \sff_\mathrm{B}(-\wA)+ \sff_{\l}(q) \right] \big\} \chi_{\l}(\bp)  
	\Big] \\
	-&\, 
	\a_\mathrm{oph}^2 
	\int_\bq \frac{1  - \hat{\bp}\cdot\hat{\bq}  }{2} \, 
	\delta(p + q - \W_{A^\prime}) 
	\Big\{ 
		\sff_{\l}^\prime(p) \big[ \sff_\mathrm{B}(\wA) + \sff_{-\l}(q) \big] \chi_{\l}(\bp) 
		+ 
		\sff_{\l}^\prime(q) \big[ \sff_\mathrm{B}(\wA) + \sff_{\l}(p) \big]\chi_{-\l}(\bq) 
	\Big\}.
\end{align}
\end{subequations}
\end{widetext}
In Eq.~(\ref{st-imp-dless-s}) we have introduced the dimensionless short-ranged impurity strengths 
	$(g_0,g_\mathrm{f},g_\mathrm{b}) = (G_0,G_\mathrm{f},G_\mathrm{b})/\hbar \vf^2$. 
In Eq.~(\ref{st-imp-dless-l}), we define the dimensionless long-ranged impurity strength 
	$\gamma^2 = 2\pi n_{\mathrm{imp}} (\b \hbar \vf)^2$. 
The dimensionless screened Coulomb interaction $\wtd{U}_\mathrm{eff}(\w,q)$ is presented in Appendix~\ref{ftsc},
Eq.~(\ref{UeffDIMLESS}). 
In Eqs.~(\ref{channel-A})--(\ref{channel-C}) the integrand kernel reads 
\be
	\Xi_{p_1,p_2;p_3,p_4}^{\l_1, \l_2 ; \l_3 , \l_4} 
	= 
	\frac{1}{8} \prod_{j=1}^{4} \sech{\left(\frac{p_j - \l_j \ln{z}}{2}\right)}.
\ee
In Eq.~(\ref{oph-col-dl}) the Bose-Einstein distribution function is
\be \label{bose-fun}
\sff_\mathrm{B}(\W) = \frac{1}{e^\W -1}, 
\ee
the effective optical-phonon frequency and coupling constant are
\be
	\W_{A^\prime} =\b \hbar \w_{A^\prime}, 
		\quad 
	\a_\mathrm{oph}^2  \equiv \frac{4 \pi^2 s_0 \beta_{A^\prime}^2}{\hbar \w_{A^\prime} M \vf^2},
	\label{Oph-Params}
\ee 
respectively.

The charge current $\mathbf{J}$ and heat current $\mathbf{J}_\mathrm{Q}$ are determined by the distribution function $f_\l (\bp,\bfr)$ as 
\begin{subequations}  \label{currents}
\begin{align}
& \mathbf{J} =  \, -e \sum_{\l= \pm 1} \l \int_\bp \! \velf \, \Tr{f_\l (\bp)}, \\
& \mathbf{J}_\mathrm{Q} = \sum_{\l= \pm 1} \int_\bp \! \velf \, \left( \ep 
_{\bp} - \mu_\l \right) \Tr{f_\l (\bp)}. 
\end{align}
\end{subequations}

\subsection{Solution of linearized Boltzmann equation}  \label{solution-subsec}

The collision integral of the Boltzmann equation (\ref{l-bolt-eq}) is a linear 
operator
acting on the solution $\{\chi_{\l} (\bp)\}$. It is convenient to expand the solution $\chi_{\l} (\bp)$ as~\cite{LPv10, allen1978, comp-mb},
\begin{align} \label{chi-exp}
	& 
	\chi_\l(\bp) 
	=  
	\sum_{J=-\infty}^{\infty}
	\sum_{n=0}^{\infty} 
	\eta_{n}(\l, p) \, e^{i J \varphi_\bp} \, \phi_{n J},
\end{align}
where $\bp \equiv p \, (\cos{\varphi_\bp}, \sin{\varphi_\bp})$, 
$J$ is the rank of the two-dimensional spherical harmonics $\{ e^{i J \varphi_\bp} \}$ supporting the angular variable 
$\varphi_\bp$, and $n$ is the rank of some basis supporting the radial (energy) variable $p$.
The coefficients  $\{\phi_{n J}\}$ determine the solution.

In order to compute the longitudinal transport coefficients, we assume that both the temperature gradient $\nabla_\bfr T$ and the electrochemical field $\boldsymbol{\mathcal{E}}$ are along the $x$-direction. Consequently, Eq.~(\ref{l-bolt-eq}) takes the form 
\begin{align} \label{l-boltz-x}
	\vf \, \sff_\l^\prime(p,z) \left( \l e \, \b \, \mathcal{E}_x  -  \frac{p - \lambda \ln{z}}{T} \p_x{T} \right) \cos{\varphi_\bp} & \nn \\
	=   \frac{1}{\b \hbar} \sum_{\l^\prime = \pm 1} \! \int_\bq \mathcal{R}_{\l \l^\prime} (\bp , \bq) \, \chi_{\l^\prime} (\bq) &,  
\end{align}
where the linear operator $\mathcal{R}_{\l \l^\prime} (\bp , \bq)$ is determined by the collision integrals 
(\ref{st-imp-dless})--(\ref{oph-col-dl}). 
Due to the p-wave form of the driving fields, in Eq.~(\ref{chi-exp}) the solution $\chi_{\l} (\bp)$ can be simplified to
\be \label{chi-exp-simp}
	\chi_\l(\boldsymbol{\bp}) =  \boldsymbol{\eta}(\l, p)  \cdot \boldsymbol{\Phi} \, \cos{\varphi_\bp} .
\ee 
where the vectors $\boldsymbol{\Phi}$ and $\boldsymbol{\eta}(\l,p)$ 
determine	
the solution $\{\Phi_n\}$ 
within the assumed	
radial basis $\{\eta_n(\l, p)\}$. Substituting Eq.~(\ref{chi-exp-simp}) into the right-hand side 
of Eq.~(\ref{l-boltz-x}), multiplying both sides of Eq.~(\ref{l-boltz-x}) by 
$ 
	\boldsymbol{\eta}(\l, p) \, \cos{\varphi_p}	
$, 
and integrating over $\bp$ and summing over $\l$, we obtain the Boltzmann equation for $\boldsymbol{\Phi}$,
\be  \label{l-boltz-simp}
\frac{\vf}{\sqrt{\pi}} \left\{e \b \, \boldsymbol{\mathcal{F}}^{(1)} \mathcal{E}_x + \frac{1}{T}\left[ \boldsymbol{\mathcal{F}}^{(1)} \ln{z} -\boldsymbol{\mathcal{F}}^{(2)} \right] \partial_x{T} \right\} = \frac{1}{\b\hbar} \hat{\mathcal{M}} \boldsymbol{\Phi} ,
\ee 
where the force vectors are given by
\begin{subequations} \label{liou-ab}
\begin{align}
\boldsymbol{\mathcal{F}}^{(1)} \equiv &\, \frac{1}{\sqrt{4 \pi}} \sum_{\l=\pm 1} \! \l \int_0^{\infty} \!\! \rmd{p} \, \sff_\l^\prime(p,z)  \, p \, \boldsymbol{\eta}(\l, p) , \label{liou-a} \\
\boldsymbol{\mathcal{F}}^{(2)} \equiv &\, \frac{1}{\sqrt{4 \pi}} \sum_{\l=\pm 1} \! \int_0^{\infty} \!\! \rmd{p} \, \sff_\l^\prime(p,z)   p^2 \boldsymbol{\eta}(\l, p), \label{liou-b}
\end{align}
\end{subequations}
and the collision matrix $\hat{\mathcal{M}}$ is determined by 
\be \label{def-M-mat}
	\hat{\mathcal{M}} 
	\equiv 
	\sum_{\l,\l^\prime =\pm 1} 
	\iint_{\bp,\bk} 
	\hat{\bp} \cdot \hat{\bk} \, 
	\mathcal{R}_{\l \l^\prime} (\bp , \bk) \, 
	\boldsymbol{\eta} (\l , p) \otimes \boldsymbol{\eta} (\l^\prime , k), 
\ee
where ``$\otimes$'' is the Kronecker product. 
We show the form of $\hat{\mathcal{M}}$ in Sec.~\ref{collision-mats}. Finally, formally inverting $\hat{\mathcal{M}}$ in Eq.~(\ref{l-boltz-simp}) we obtain the solution
\be \label{phi-0} 
\boldsymbol{\Phi} =\frac{\b \hbar \vf}{\sqrt{\pi}}\hat{\mathcal{M}}^{-1} \left\{ e \b \boldsymbol{\mathcal{F}}^{(1)} \mathcal{E}_x + \frac{1}{T}\left[ \boldsymbol{\mathcal{F}}^{(1)} \ln{z} - \boldsymbol{\mathcal{F}}^{(2)} \right] \partial_x T \right\}.
\ee

Inserting Eqs.~(\ref{distf-f-1}), (\ref{delf-0}), and (\ref{chi-exp-simp}) into the definition (\ref{currents}), we obtain the electric and thermal current along the $x$-direction in terms of $\boldsymbol{\Phi}$,
\begin{align}  \label{e-current-l}
& J^x =  \frac{N e}{\sqrt{4 \pi} \b^2  \hbar^2  \vf}  \boldsymbol{\mathcal{F}}^{(1)} \cdot \boldsymbol{\Phi}, \\
& J_\mathrm{Q}^x = \frac{N}{ \sqrt{4 \pi} \b^3  \hbar^2  \vf} \left( \boldsymbol{\mathcal{F}}^{(2)}  - \boldsymbol{\mathcal{F}}^{(1)} \ln{z} \right) \cdot \boldsymbol{\Phi}. 
\end{align}
Inserting Eq.~(\ref{phi-0}) into Eq.~(\ref{e-current-l}) and comparing to Eq.~(\ref{linear-resp}), we obtain the transport coefficients
\begin{subequations} \label{trans-coe-def}
\begin{align} 
\s = & \, N \frac{e^2}{h} \mathsf{L}_{11}, \\
\a_\infty = & \, \frac{k_\mathrm{B}}{e}  \left( \frac{ \mathsf{L}_{12} }{ \mathsf{L}_{11} } -\ln{z} \right), \\
\k_\infty = & \, N \frac{k_\mathrm{B}^2 T}{h}  \left( \mathsf{L}_{22} - \frac{ \mathsf{L}_{21}^2 }{ \mathsf{L}_{11} } \right),
\end{align}
\end{subequations}
where 
\be  \label{def-Lij}
\mathsf{L}_{ij} \equiv \boldsymbol{\mathcal{F}}^{(i)} \cdot \hat{\mathcal{M}}^{-1} \boldsymbol{\mathcal{F}}^{(j)}, \quad i,j \in \{1,2\}.
\ee

\subsubsection{Collision matrix}  \label{collision-mats}

The collision matrix $\hat{\mathcal{M}}$ defined by Eq.~(\ref{def-M-mat}) has three parts 
\be \label{all-coll-mat}
\hat{\mathcal{M}} = \hat{\mathcal{M}}_\mathrm{imp} + \hat{\mathcal{M}}_{\mathrm{int}} +  \hat{\mathcal{M}}_{\mathrm{oph}}, 
\ee
corresponding to the collision integrals~(\ref{st-imp-dless})--(\ref{oph-col-dl}), respectively. 

The impurity collision matrix elements read
\begin{subequations}   \label{imp-coll-M}
\be
\begin{split}
\left( \mathcal{M}_{\mathrm{imp}} \right)_{mn}=  &\, \frac{1}{4\pi}\sum_\l \int_{0}^{\infty} \! \rmd p \, \sff_{\l}^\prime(p,z) p^2  \, \wtd{G}(p)  \\ 
&\, \times \eta_m(\l, p) \eta_n(\l, p),
\end{split}
\ee
with
\be   \label{wtdG}
	\wtd{G}(p) =  \td{g} + \td{\gamma}^2 \,p^{-2} \, F\left(\frac{q_\mathrm{TF}}{2p}\right),
\ee
\end{subequations}
where the effective short-ranged disorder strength is $\td{g} \equiv  \left(4 g_0+ g_\mathrm{f} + 3 g_\mathrm{b}\right)/(4\pi)$, the effective long-ranged disorder strength is 
	$\td{\gamma}^2 \equiv \gamma^2 \a_\mathrm{int}^2/2 = \pi n_\mathrm{imp} \a_\mathrm{int}^2 (\b \hbar  \vf)^2$, 
and the function
\be  \label{f1-fun}
\begin{split}
 F(x) = &\, \int_{0}^{\pi} \! \rmd \theta \, \frac{\sin^2\theta}{\left(\sin{\frac{\theta}{2}}+x\right)^2} ,
\end{split}
\ee
with $q_\mathrm{TF}$ the dimensionless Thomas-Fermi wavevector [Eq.~(\ref{qTFdimless})]. We note that in the strong-interaction limit 
	$\lim_{\a_\mathrm{int} \to \infty} \, F(q_\mathrm{TF}/2p) = 2\pi(p/q_\mathrm{TF})^2$, 
so that the Coulomb impurity becomes short-ranged. 

The Coulomb interaction collision matrix is
\be  \label{mat-ABC}
\hat{\mathcal{M}}_{\mathrm{int}} = \hat{\mathcal{M}}_{\mathrm{int}}^{\mathrm{A}}+\hat{\mathcal{M}}_{\mathrm{int}}^{\mathrm{B}}+\hat{\mathcal{M}}_{\mathrm{int}}^{\mathrm{C}},
\ee
where the elements of each component are given by
\begin{widetext}
\begin{subequations}    \label{mat-int}
\begin{align}
	\left(\mathcal{M}_{\mathrm{int}}^{\mathrm{A}}\right)_{mn} 
	= 
	& \, 
	\frac{\pi \, N}{2}	
	\sum_{\l=\pm 1} 
	\int_{\bp,\bk,\bq} \! 
	\frac{1+\hat{\bp}\cdot \what{\bp-\bq}}{2}\frac{1+\hat{\bk} \cdot \what{\bk-\bq}}{2} \, 
	\delta(p-|\bp-\bq|-k+|\bk-\bq|)  
	\, |\wtd{U}_\mathrm{eff}(p-|\bp-\bq|),q|^2 \nn 	\\ 
	& \times \Xi_{p, |\bp-\bq|;k,|\bk-\bq|}^{\l,\l; \l,\l}  
	\left[ \eta_m(\l,p) \, \hat{\bp} - \eta_m(\l,|\bp-\bq|) \, \what{\bp-\bq} - \eta_m(\l,k) \, \hat{\bk} + \eta_m(\l, |\bk-\bq|) \, \what{\bk-\bq} \right] \nn \\
	& \, \cdot  
	\left[ \eta_n(\l,p) \, \hat{\bp} - \eta_n(\l,|\bp-\bq|) \, \what{\bp-\bq} - \eta_n(\l,k) \, \hat{\bk} + \eta_n(\l, |\bk-\bq|) \, \what{\bk-\bq} \right], \label{mat-int-a} \\
	\left(\mathcal{M}_{\mathrm{int}}^{\mathrm{B}}\right)_{mn} 
	=
	& \, 
	\frac{\pi \, N}{2}	
	\sum_{\l=\pm 1} 
	\int_{\bp, \bk,\bq} \! 
	\frac{1+\hat{\bp}\cdot \what{\bp+\bq}}{2}\frac{1+\hat{\bk} \cdot \what{\bk-\bq}}{2} \,  
	\delta(p-|\bp+\bq|+k-|\bk-\bq|) 
	\, |\wtd{U}_\mathrm{eff}(p-|\bp+\bq|,q)|^2 \nn 	\\
	&  \times \Xi_{p, |\bp+\bq|;k,|\bk-\bq|}^{\l,\l ; -\l,-\l}  
	\left[ \eta_m(\l,p) \, \hat{\bp} - \eta_m(\l,|\bp+\bq|) \, \what{\bp+\bq} + \eta_m(-\l,|\bk|) \, \hat{\bk} - \eta_m(-\l,|\bk-\bq|) \, \what{\bk-\bq} \right] \nn \\
	& \,  \cdot 
	\left[ \eta_n(\l,p) \, \hat{\bp} - \eta_n(\l,|\bp+\bq|) \, \what{\bp+\bq} + \eta_n(-\l,|\bk|) \, \hat{\bk} - \eta_n(-\l,|\bk-\bq|) \, \what{\bk-\bq} \right], \label{mat-int-b} 
\end{align}
\begin{align}
	\left(\mathcal{M}_{\mathrm{int}}^{\mathrm{C}}\right)_{mn} = 
	& \, 
	\frac{\pi \, N}{2}	
	\sum_{\l=\pm 1} 
	\int_{\bp, \bk,\bq} \! 
	\frac{1-\hat{\bp}\cdot \what{\bp-\bq}}{2}\frac{1-\hat{\bk} \cdot \what{\bk-\bq}}{2} \, 
	\delta(p+|\bp-\bq|-k-|\bk-\bq|) 
	\, |\wtd{U}_\mathrm{eff}(p+|\bp-\bq|,q)|^2  \nn 	\\
	&  \times\Xi_{p, |\bp-\bq|;k,|\bk-\bq|}^{\l, -\l;\l, -\l} 
	\left[ \eta_m(\l,p) \, \hat{\bp} - \eta_m(-\l,|\bp-\bq|) \, \what{\bp-\bq}  - \eta_m(\l,\bk) \, \hat{\bk} + \eta_m(-\l,|\bk-\bq|) \, \what{\bk-\bq}  \right] \nn \\
	& \, \cdot 
	\left[ \eta_n(\l,p) \, \hat{\bp} - \eta_n(-\l,|\bp-\bq|) \, \what{\bp-\bq}  - \eta_n(\l,\bk) \, \hat{\bk} + \eta_n(-\l,|\bk-\bq|) \, \what{\bk-\bq}  \right]. \label{mat-int-c}
\end{align}
\end{subequations}

Finally, the carrier--optical-phonon collision matrix reads
\be  \label{opt-mat}
\begin{split}
	\left( \mathcal{M}_{\mathrm{opt}}\right)_{mn}  
	= & \,
	\wtd{\a}_\mathrm{opt}^2 
	\sum_\l \int_{0}^{\infty} \! \rmd p \, 
	\sff_{\l}^\prime(p,z) \, p \left( p + \W_{A^\prime} \right) 
		\left[2 \, \eta_{m}(\l, p) + \eta_{m}(\l, p+\W_{A^\prime}) \right]  
		\eta_{n}(\l, p) \left[ \sff_\mathrm{B}(\W_{A^\prime}) + \sff_{\l}(p+\W_{A^\prime}) \right] \\ 
	- & \,
	\wtd{\a}_\mathrm{opt}^2 
	\sum_\l \int_{\W_{A^\prime}}^{\infty} \! \rmd p \, 
	\sff_{\l}^\prime(p,z) \, p \left( p - \W_{A^\prime} \right) 
	\left[2 \, \eta_{m}(\l, p) + \eta_{m}(\l, p-\W_{A^\prime}) \right] 
	\eta_{n}(\l, p) \left[\sff_\mathrm{B}(-\W_{A^\prime}) + \sff_{\l}(p-\W_{A^\prime}) \right]  \\
	+& \,
	\wtd{\a}_\mathrm{opt}^2 
	\sum_\l \int_{0}^{\W_{A^\prime}} \! \! \! \! \rmd p  \, 
	\sff_{\l}^\prime(p,z) \, p \left( \W_{A^\prime}-p \right) 
	\left[ 2 \, \eta_{m}(\l, p) - \eta_{m}(-\l, \W_{A^\prime}-p) \right] 
	\eta_{n}(\l, p) \left[ \sff_\mathrm{B}(\W_{A^\prime}) + \sff_{-\l}(\W_{A^\prime}-p) \right],
\end{split}
\ee
where the effective electron--optical-phonon coupling $\wtd{\a}_\mathrm{opt}^2 = \a_\mathrm{opt}^2/(16\pi^2)	$.

\end{widetext}

\subsubsection{Orthogonal polynomials}

A
computationally
efficient method is to choose the basis $\{\eta_{n}(\l,p)\}$ as a set of orthogonal polynomials~\cite{allen1978,comp-mb}
in two variables $\l$ and $p$, 
taking into account the
symmetry properties of the collision matrix and the force vectors. 
We define the orthonormal condition as
\be \label{orthnorm}
\langle \eta_m, \, \eta_n \rangle \equiv \sum_{\l=\pm}\int_0^{\infty} \! \rmd p \, \mathcal{K}_\l(p,z) \, \eta_m(\l,p) \, \eta_n(\l,p) = \delta_{m,n},
\ee
where the kernel function depends on the fugacity $z$,
\be \label{kernel}
\mathcal{K}_\l(p,z) \equiv p \, \sff_{\l}^\prime(p,z),
\ee
and the function $\sff_{\l}^\prime(p,z)$ is defined in Eq.~(\ref{dless-df}). We note that in general the transport coefficients obtained via Eq.~(\ref{trans-coe-def}) are independent of the choice of basis, and moreover, the normalization condition in Eq.~(\ref{orthnorm}) can be relaxed.  

Our 
objective
is to orthonormalize the monomial system $\{ p^n, \l p^n\}_{n \ge -1}$ by the orthogonal condition (\ref{orthnorm}). 
Note that we have included the negative power $n=-1$ because $\delta f_\l \sim 1/p$ is the lowest power of $p$ that 
leads to finite charge and thermal currents in two spatial dimensions according to Eq.~(\ref{currents}). 
Via the Gram-Schmidt orthogonalization process we recursively generate the polynomials in the form
\begin{subequations} \label{orth-ele}
\be \label{orth-ele-1}
	\eta_{-2}(\l,p) = \l/p, \quad \eta_{-1}(\l,p) = 1/p,
\ee
and for $n \ge 0$
\be \label{orth-ele-2}
	\eta_n(\l,p) = \sum_{m=0}^{n} a_{nm} u_m,
\ee
where $\{u_m\}$ 
are the
monomials $\{ p^n, \l p^n\}_{n \ge 0}$ 
ordered as follows:
\begin{align} \label{orth-ele-2c}
\begin{array}{cccccccc}
 \, u_0  & \,  u_1   & \,  u_2  & \,  u_3 &  \, u_4 & \,  u_5 & \, u_6 & \cdots \\
\hline \\
 \, p    &  \,  \l   & \, 1 & \, p^2 & \, \l p &  p^3 &  \, \l p^2 & \cdots \\
\hline
\end{array}
\end{align}
\end{subequations} 
As discussed in Sec.~\ref{sec-results}, the negative-power basis ``$\l/p$'' [see Eq.~(\ref{orth-ele-1})] 
is crucial for solving the Boltzmann equation with only disorder or acoustic-phonon scattering processes. 
The leading positive-power basis ``$p$'', ``$\l$'', and ``$1$'' [see Eq.~(\ref{orth-ele-2c})] multiplied by 
the Fermi velocity $\mathbf{v}_\mathrm{F}$ correspond to the momentum, charge velocity, and energy velocity, 
respectively, and play the key role in the hydrodynamic description~\cite{fordunlenbook, mueller2008, matt09, mirlin2011, narozhny15}. 

In order to calculate the coefficients $\{a_{mn}\}$ in Eq.~(\ref{orth-ele-2}), we introduce the functions
\be \label{f-Omega} \Omega_{n,\pm}(z) = - \Gamma(n+1) \, \left[ \mathrm{Li}_n(-z) \pm \mathrm{Li}_n(-z^{-1})\right], \ee
where $\Gamma(n)$ is the gamma function and $\mathrm{Li}_n(-z)$ the polylogarithm defined by
\be
\mathrm{Li}_n(-z^\l) \equiv -\frac{1}{\Gamma(n)}\int_0^{\infty} \! \rmd p \, p^{n-1} \, \sff_{\l}(p,z), \quad n \ge 0. 
\ee
The leading coefficients read
\be  \label{a-some}
\begin{split}
a_{00} = &\, \frac{1}{\sqrt{\W_{3,+}(z)}}, \quad a_{10} =  -a_{11} \frac{\W_{2,-}(z)}{\W_{3,+}(z)}, \\
a_{11} = &\, \sqrt{\frac{\W_{3,+}(z)}{\W_{1,+}(z)\,\W_{3,+}(z)- \W_{2,-}^2(z)}}, \\
\end{split}
\ee 
which are important for
writing
down the force vectors [see Eq.~(\ref{F-12-c})]. 
Higher order coefficients can be generated numerically.

Substituting Eq.~(\ref{orth-ele}) into Eq.~(\ref{liou-ab}) we obtain the force vectors
\be  \label{F-12-c}
\begin{split}
\boldsymbol{\mathcal{F}}^{(1)} = & \, \frac{1}{\sqrt{4\pi}}\begin{bmatrix} 1 & \Omega_{0,-}(z) & -\frac{a_{10}}{a_{00}a_{11}} & \frac{1}{a_{11}} & 0 & \cdots \end{bmatrix}^\trasp, \\
\boldsymbol{\mathcal{F}}^{(2)} = & \, \frac{1}{\sqrt{4\pi}}\begin{bmatrix} \Omega_{1,-}(z) & \Omega_{1,+}(z) & \frac{1}{a_{00}} & 0 & 0 & \cdots \end{bmatrix}^\trasp,
\end{split}
\ee     
where we have used $\Omega_{0,+}(z) = 1$ and the coefficients $a_{00,10,11}$ are given in Eq.~(\ref{a-some}). In Eq.~(\ref{F-12-c}) only the leading four (three) components of the force vectors $\boldsymbol{\mathcal{F}}^{(1)}$ [$\boldsymbol{\mathcal{F}}^{(2)}$] are nonzero so that we only need the $ 4 \times 4$ block of the inverse collision matrix $\hat{\mathcal{M}}^{-1}$ to evaluate the transport coefficients in Eq.~(\ref{trans-coe-def}).

\subsubsection{Thermodynamics}

We present some useful thermodynamic relations for 
the ideal two-component
relativistic gas. 
The charge-carrier density $n$ and the internal energy density $\vep$ are fixed by the Fermi-Dirac function [Eq.~(\ref{fd-0th})] as 
\be  \label{fix-z-ep}
 n =  \sum_{\l=\pm 1} \l \int_\bp \! f_\l^{(0)}(\bp), \quad \vep =  \sum_{\l=\pm 1} \int_\bp \! \vep_\bp f_\l^{(0)}(\bp), 
\ee
which leads to
\begin{subequations}  \label{hyd-relts}
\begin{align} 
& \, n = \frac{N k_\mathrm{B}^2 T^2}{4\pi \hbar^2 \vf^2}\W_{2,-}(z), \label{nc-fix} \\
& \, \varepsilon = \frac{N  k_\mathrm{B}^3  T^3}{6  \pi \hbar^2  \vf^2} \W_{3,+}(z),  \label{vep-fix}
\end{align}
\end{subequations}
where $\W_{n,\pm}(z)$ are defined in Eq.~(\ref{f-Omega}). One can use the charge carrier density (\ref{nc-fix}) to determine the 
fugacity 
$z = z(n,T)$. 

Moreover, the enthalpy density $\mathsf{h}$ and entropy density $\mathsf{s}$ obey the thermodynamic relations $\mathsf{h} = \varepsilon + \mathsf{P}$ and  $T  \mathsf{s} = \mathsf{h} - n \mu$, where $\mathsf{P}$ is the pressure. For 
the
ideal relativistic gas we exploit
scale invariance
$\mathsf{h}= 3 \, \mathsf{P}$~\cite{LLbook5} so that
\be  \label{entha-entro}
\mathsf{h} = \frac{3}{2} \vep, \quad \mathsf{s} = \frac{1}{T}\left( \frac{3}{2} \vep - n \mu \right).
\ee
Explicit formulae for all thermodynamic potentials in terms of $n$ and $T$ are useful for analyzing the
hydrodynamic description [Sec.~\ref{hydro-dym-sec}]. Transport coefficients in the interaction-limited regime 
are expressed in terms of these (irrespective of Fermi degeneracy), see Eq.~(\ref{trans-coe-int}).

\subsection{Transport coefficients \label{Sec: Transport Coeffs}}

\subsubsection{Impurity-only transport} \label{dis-only-2}

In the presence of only elastic scattering the linearized Boltzmann equation can be solved by
\begin{subequations}  \label{imp-ex-solu}
\be
\chi_\l(\bp) = \Phi_\l(p) \cos{\varphi_\bp},
\ee
with 
\be \label{sol-imp-only}
\Phi_\l(p)=\frac{2 \hbar \b \vf}{\wtd{G}(p)} \left[\frac{\l}{p}\left( e \b \mathcal{E}_x + \frac{\p_x{T}}{T} \ln{z}\right) - \frac{\p_x{T}}{T} \right].
\ee
\end{subequations}
The dimensionless 
``scattering rate''
$\wtd{G}(p)$ is defined in Eq.~(\ref{wtdG}). Substituting Eqs.~(\ref{distf-f-1}), (\ref{delf-0}), (\ref{imp-ex-solu}), and (\ref{sol-imp-only}) into the currents (\ref{currents}) we obtain the transport coefficients in the form of Eq.~(\ref{trans-coe-def})  
\begin{subequations}  \label{imp-only-L}
\begin{align}
	\mathsf{L}_{11}^{\mathrm{(imp)}} = &\, \sum_{\l=\pm 1}\int_{0}^{\infty} \! \rmd p \, \sff_\l^\prime(p,z) \, \left[\wtd{G}(p)\right]^{-1},    	\label{L11-imp}  	\\
	\mathsf{L}_{12}^{\mathrm{(imp)}} = &\, \sum_{\l=\pm 1}\int_{0}^{\infty} \! \rmd p \, \sff_\l^\prime(p,z) \, \l p \left[\wtd{G}(p)\right]^{-1},  	  		\\
	\mathsf{L}_{22}^{\mathrm{(imp)}} = &\, \sum_{\l=\pm 1}\int_{0}^{\infty} \! \rmd p \, \sff_\l^\prime(p,z) \, p^2 \left[\wtd{G}(p)\right]^{-1}.
\end{align}
\end{subequations}

For the short-ranged-impurity-only case $\wtd{G}(p)= \mathrm{const.}$, we observe that the basis 
``$\l/p$'' and ``$1$'' in Eq.~(\ref{imp-ex-solu}) are complete to cover the solution, and, 
furthermore, the integrals in Eq.~(\ref{imp-only-L}) can be evaluated analytically. 
The transport coefficients take simple expressions (\ref{short-only}). 

For the long-ranged-impurity-only case in the absence of screening, where $\wtd{g}=0$ and $\a_\mathrm{int} \to 0$ so that $q_\mathrm{TF} \to 0$, using $\lim_{x\to0} F(x) = 2 \pi$, we readily obtain
\begin{subequations}  \label{long-only}
\begin{align}
& \s_{\mathrm{imp}}^{(\mathrm{l}),\a_\mathrm{int}\to 0} = N \frac{e^2}{h}\frac{1}{2\pi \wtd{\g}^{2}} \Omega_{2,+}(z), \\
& \a_{\infty, \mathrm{imp}}^{(\mathrm{l}),\a_\mathrm{int}\to 0}  = \frac{k_\mathrm{B}}{e}\left[ \frac{\Omega_{3,-}(z)}{\Omega_{2,+}(z)} -\ln{z} \right], \\
& \k_{\infty, \mathrm{imp}}^{(\mathrm{l}),\a_\mathrm{int}\to 0}  = N \frac{k_\mathrm{B}^2 T}{h} \frac{1}{2\pi \wtd{\g}^{2}} \left[ \Omega_{4,+}(z) -\frac{\Omega_{3,-}^2(z)}{\Omega_{2,+}(z)} \right].
\end{align}
\end{subequations}   
In Fig.~\ref{imp-only-pic}(i)--(iii) we compare the transport coefficients obtained by the orthogonal-polynomial algorithm to the exact result evaluated by Eq.~(\ref{imp-only-L}). In practice we keep the order of the polynomial basis up to $\mathcal{N} =16$ to recover the analytical result.

\subsubsection{Interaction-limited transport}  \label{hydro-sec-2}
In the presence of Coulomb interactions we write the collision matrix (\ref{all-coll-mat}) as 
$
	\hat{\mathcal{M}} = \hat{\mathcal{M}}_{\mathrm{int}} + \delta{\hat{\mathcal{M}}},
$ 
where 
$\delta{\hat{\mathcal{M}}}$ 
can be any combination of $\hat{\mathcal{M}}_{\mathrm{imp},\mathrm{oph}}$. 
Due to momentum conservation the basis 
element
$u_0 = p$ [Eq.~(\ref{orth-ele-2c})] 
is a zero mode
of the Coulomb collision matrix $\hat{\mathcal{M}}_{\mathrm{int}}$, that is, 
$ 
	\left( \mathcal{M}_{\mathrm{int}} \right)_{0 n}
	=
	\left( \mathcal{M}_{\mathrm{int}} \right)_{n 0} = 0
$ for any $n \ge -2$ 
[negative $n$ is defined via Eq.~(\ref{orth-ele-1})]. 
Therefore, the perturbation $\delta{\hat{\mathcal{M}}}$ breaking the translation invariance 
regularizes the collision matrix and yields finite transport coefficients. 
Here 
we choose $\delta{\hat{\mathcal{M}}} = \hat{\mathcal{M}}_{\mathrm{imp}}$ with only short-ranged impurity 
scattering 
($n_\mathrm{imp}=0$),
\be  \label{model-hyd}
	\hat{\mathcal{M}} = \hat{\mathcal{M}}_{\mathrm{int}} + \hat{\mathcal{M}}_{\mathrm{imp}}^{(\mathrm{s})},
\ee
and we study the transport coefficients in the interaction-limited regime 
$	
	\wtd{g} \ll \a_\mathrm{int}^2 \sim \mathcal{O}(1)
$. 
We note that the discussion applies to any scattering mechanism that lifts the zero modes of Coulomb collision operator.

Via Eqs.~(\ref{imp-coll-M}), (\ref{mat-ABC}), and (\ref{F-12-c}) we expand the coefficients $\mathsf{L}_{ij}$ in Eq.~(\ref{def-Lij}) in $\wtd{g}$ up to order of ``1'', 
\begin{subequations}  \label{Lij-hyd}
\begin{align}
\mathsf{L}_{11}^{(\mathrm{h})} & = \frac{\W_{2,-}^2(z)}{\W_{4,+}(z)} \wtd{g}^{-1} + \del{\mathsf{L}}_{11}^{(\mathrm{h})}, \label{hyd-L11} \\
\mathsf{L}_{12}^{(\mathrm{h})} & = \frac{\W_{2,-}(z) \, \W_{3,+}(z)}{\W_{4,+}(z)} \wtd{g}^{-1} + \del{\mathsf{L}}_{12}^{(\mathrm{h})}, \label{hyd-L12}\\
\mathsf{L}_{22}^{(\mathrm{h})} & = \frac{\W_{3,+}^2(z)}{\W_{4,+}(z)} \wtd{g}^{-1} + \del{\mathsf{L}}_{22}^{(\mathrm{h})}, \label{hyd-L22}
\end{align}
\end{subequations}
where $\del{\mathsf{L}}_{ij}^{(\mathrm{h})} \sim \a_\mathrm{int}^{-2} \sim \mathcal{O}(1)$ ($i,j \in \{1,2\}$) 
encode
the contributions of 
polynomial modes $\eta_{n \geq 1}(\lambda,p)$ orthogonal to $u_0 = p$; 
these can be evaluated numerically. 
Substituting Eq.~(\ref{Lij-hyd}) into Eq.~(\ref{trans-coe-def}) and exploiting the thermodynamic relations 
(\ref{hyd-relts}) and (\ref{entha-entro}), 
we obtain the interaction-limited transport coefficients 
\begin{subequations}  \label{trans-coe-int}
\begin{align}
	\s_{\mathrm{h}} & =\frac{ (e n \vf)^2 \tau_\mathrm{el} }{\mathsf{h}} + \s_{\mathrm{min}}, \label{cond-hyd} \\
	\a_{\infty,\mathrm{h}} & = \frac{e n \vf^2 \tau_\mathrm{el} }{T \s_{\mathrm{h}}} - \frac{\mu}{eT}, \label{therp-hyd} \\
	\k_{\infty,\mathrm{h}} & = \frac{\mathsf{h} \vf^2 \tau_\mathrm{el} }{T \s_{\mathrm{h}}} \, \overline{\s}_{\mathrm{min}}. \label{tcond-hyd}
\end{align} 
\end{subequations}
Except for a slight discrepancy in the form of the thermal conductivity (discussed below), these match 
the results of relativistic hydrodynamics~\cite{mueller2008,matt09}.	

In Eq.~(\ref{trans-coe-int}) $\tau_\mathrm{el}^{-1}$ is the elastic scattering rate 
induced by short-ranged impurities. It is defined by 
\be \label{tau-el}
\tau_\mathrm{el}^{-1}  \equiv \frac{k_\mathrm{B} T}{2 \hbar}\frac{\W_{4,+}(z)}{\W_{3,+}(z)}\wtd{g}
\approx 
\begin{cases}
\vspace{0.1cm}
\frac{7 \pi^4}{270 \, \zeta(3)} \frac{k_\mathrm{B} T}{\hbar} \wtd{g}, & \quad |\mu| \ll k_\mathrm{B} T, \\
\frac{1}{2}\frac{|\mu|}{\hbar} \wtd{g}, & \quad |\mu| \gg k_\mathrm{B} T, 
\end{cases}
\ee 
where $\zeta(n)$ is the Riemann zeta function. For comparison we estimate the inelastic-scattering rate due to the Coulomb interactions~\cite{mueller2008}
\be  \label{tau-ee}
\tau_\mathrm{ee}^{-1} \sim \begin{cases}
\vspace{0.1cm}
\frac{k_\mathrm{B} T}{\hbar} \a_\mathrm{int}^2 , & \quad |\mu| \ll k_\mathrm{B} T, \\
\frac{(k_\mathrm{B} T)^2}{\hbar |\mu|} \a_\mathrm{int}^2 , & \quad |\mu| \gg k_\mathrm{B} T. 
\end{cases}
\ee
Here we note that the expression for $|\mu| \gg k_\mathrm{B} T$ is the standard Fermi liquid behavior arising from channel A~\cite{com-tau-ee}.

The minimal conductivity $\s_{\mathrm{min}}$, and the related parameter $\overline{\s}_{\mathrm{min}}$ appearing in the thermal conductivity
[Eq.~(\ref{tcond-hyd})] take the form	
\begin{subequations}
\be
\s_{\mathrm{min}}  = N\frac{e^2}{h} \del{\mathsf{L}}_{11}^{(\mathrm{h})}, \quad \overline{\s}_{\mathrm{min}}  = \s_{\mathrm{min}} \left(1+ \varrho \right),
\ee
where $\varrho$ is an enhancement factor 
\be   \label{varrho}
\varrho = \left(\frac{n}{\beta \mathsf{h}}\right)^2  \frac{\del{\mathsf{L}}_{22}^{(\mathrm{h})}}{\del{\mathsf{L}}_{11}^{(\mathrm{h})}} -2\left(\frac{n}{\beta \mathsf{h}}\right) \frac{\del{\mathsf{L}}_{12}^{(\mathrm{h})}}{\del{\mathsf{L}}_{11}^{(\mathrm{h})}}.
\ee
\end{subequations}
This factor was not taken into account in previous works.

We emphasize four points. 
(i) 
As long as the Coulomb interactions dominate the collisions of electrons and/or holes
so that $\tau_{\mathrm{ee}}$ is the shortest scattering time,
the hydrodynamic description 
(\ref{trans-coe-int}) applies, where, however, the expression for the scattering rate $\tau_\mathrm{el}^{-1}$ 
and the values of $\del{\mathsf{L}}_{ij}^{(\mathrm{h})}$ should be determined by the mechanism that lifts the 
zero modes of the Coulomb collision integrals. 
(ii) 
As a simplification, if considering the effect of the
impurity collision matrix $ \hat{\mathcal{M}}_{\mathrm{imp}}^{(\mathrm{s})}$ 
\emph{only by its projection} 
onto the zero modes of the Coulomb collision $\hat{\mathcal{M}}_{\mathrm{int}}$~\cite{mueller2008,matt09}, 
one can show that 
$
	\del{\mathsf{L}}_{12}^{(\mathrm{h})} = \del{\mathsf{L}}_{22}^{(\mathrm{h})}=0,
$ 
so that $\varrho=0$  
and $\overline{\s}_{\mathrm{min}} = \s_{\mathrm{min}}$ 
[Eqs.~(\ref{cond-hyd}) and (\ref{tcond-hyd})].
In this case the 
thermal conductivity (\ref{tcond-hyd}) precisely recovers the expression in Refs.~\onlinecite{mueller2008} 
and \onlinecite{matt09}. 
(iii) 
The minimal conductivity $\s_{\mathrm{min}}$ dominates the charge conductivity [Eq.~(\ref{cond-hyd})] 
only in the vicinity of the charge neutrality,
i.e.\ for
$|\mu| / k_\mathrm{B} T \lesssim \sqrt{ \wtd{g} / \a_\mathrm{int}^2 }$ [see Fig.~\ref{hydro-coes}(i)].
In this regime we can show that the enhancement factor $\varrho \lesssim \mathcal{O}(\wtd{g})$ can be neglected, 
which is consistent with the conclusion in Refs.~\onlinecite{mueller2008} and \onlinecite{matt09}. 
(iv) 
At high density $|\mu| \gg k_\mathrm{B} T$ the impurity scattering starts to dominate when 
$\tau_\mathrm{el} \lesssim \tau_\mathrm{ee}$ [Eqs.~(\ref{tau-el}) and (\ref{tau-ee})], 
which leads to 
$|\mu| / k_\mathrm{B} T \gtrsim \sqrt{\a_\mathrm{int}^2 / \wtd{g}}$,  
and the expansion (\ref{Lij-hyd}) is no longer justified.

In the ideal hydrodynamic regime \cite{mueller2008,matt09}
\[
	\sqrt{\frac{\wtd{g}}{\a_\mathrm{int}^2}}  
	\ll 
	\frac{|\mu|}{k_\mathrm{B} T} 
	\ll 
	\sqrt{\frac{\a_\mathrm{int}^2}{\wtd{g}}},
\]
the thermoelectric power [Fig.~\ref{hydro-coes}(ii)] 
approaches
the 
thermodynamic
expression
\be  \label{hyd-tep}
	\a_{\infty,\mathrm{h}} \to \frac{\mathsf{s}}{e n}.
\ee     
The thermal conductivity [Fig.~\ref{hydro-coes}(iii)] takes the form
\be
	\frac{\k_{\infty,\mathrm{h}}}{T} \to L_{0,h} \, \overline{\s}_\mathrm{min},
\ee
where we define Lorenz ratio of an idea relativistic gas [see the panel in Fig.~\ref{hydro-coes}(iii)] as
\be  \label{L-h-0}
	L_{0,h}  \equiv \left( \frac{\mathsf{h}}{e n T} \right)^2.
\ee
Moreover, the Lorenz ratio $L$ [Fig.~\ref{hydro-coes}(iv)] tends to diverge as 
$|\mu| / k_\mathrm{B} T$ approaches the lower bound $\sqrt{ \wtd{g} / \a_\mathrm{int}^2}$,
\be
	L \to C\frac{ \a_{\mathrm{int}}^4 }{\wtd{g}} \frac{h \s_\mathrm{min}}{N e^2 } \left( \frac{k_\mathrm{B}}{e} \right)^2, 
\ee
where the constant $C = \frac{ 189 }{1280} \left[ \frac{\pi^2 \zeta(3)}{\ln(2)}\right]^2 \approx 90.03$. 
Manifestly, both Mott's law and the Wiedemann-Franz law are violated in the ideal hydrodynamic regime. 
By contrast, for
$|\mu| / k_\mathrm{B} T \gg \sqrt{ \a_\mathrm{int}^2 / \wtd{g} }$ 
we recover the disorder-limited behavior in Eq.~(\ref{short-only}) 
[see Figs.~\ref{hydro-coes}(ii) and ~\ref{hydro-coes}(iv)].


\begin{acknowledgments}

We are grateful to Fereshte Ghahari and Philip Kim for sharing their 
experimental thermopower data before publication and to Kin Chung Fong, Jesse Crossno,
and Markus Mueller 
for stimulating discussions. This research was supported by the Welch Foundation under 
Grant No.~C-1809 and by an Alfred P. Sloan Research Fellowship (No.~BR2014-035).

\end{acknowledgments}

\appendix

\section{Elliptic coordinate system for the Coulomb collisions (\ref{mat-int})}  \label{collinear-singularity}

To evaluate the Coulomb collision matrix [Eq.~(\ref{mat-int})] we first perform the integration over the momentum 
transfer $\bq$ and for the moment keep the incoming and outgoing momenta $\bp$ and $\bk$ constant. 
It is convenient to solve the energy conservation constraint by parameterizing $\bq$ in the 
elliptic (or hyperbolic) coordinate system~\cite{sachdev1998}, where the collinear scattering 
singularity~\cite{sachdev1998, polini2014} is shown explicitly.  However, as discussed in Appendix~\ref{ftsc}, 
this singularity is compensated by 
a line of zeroes in the
RPA screened Coulomb interaction along 
the forward scattering direction $\omega = v_F q$.

\begin{figure}
\centering
\includegraphics[width=0.237\textwidth]{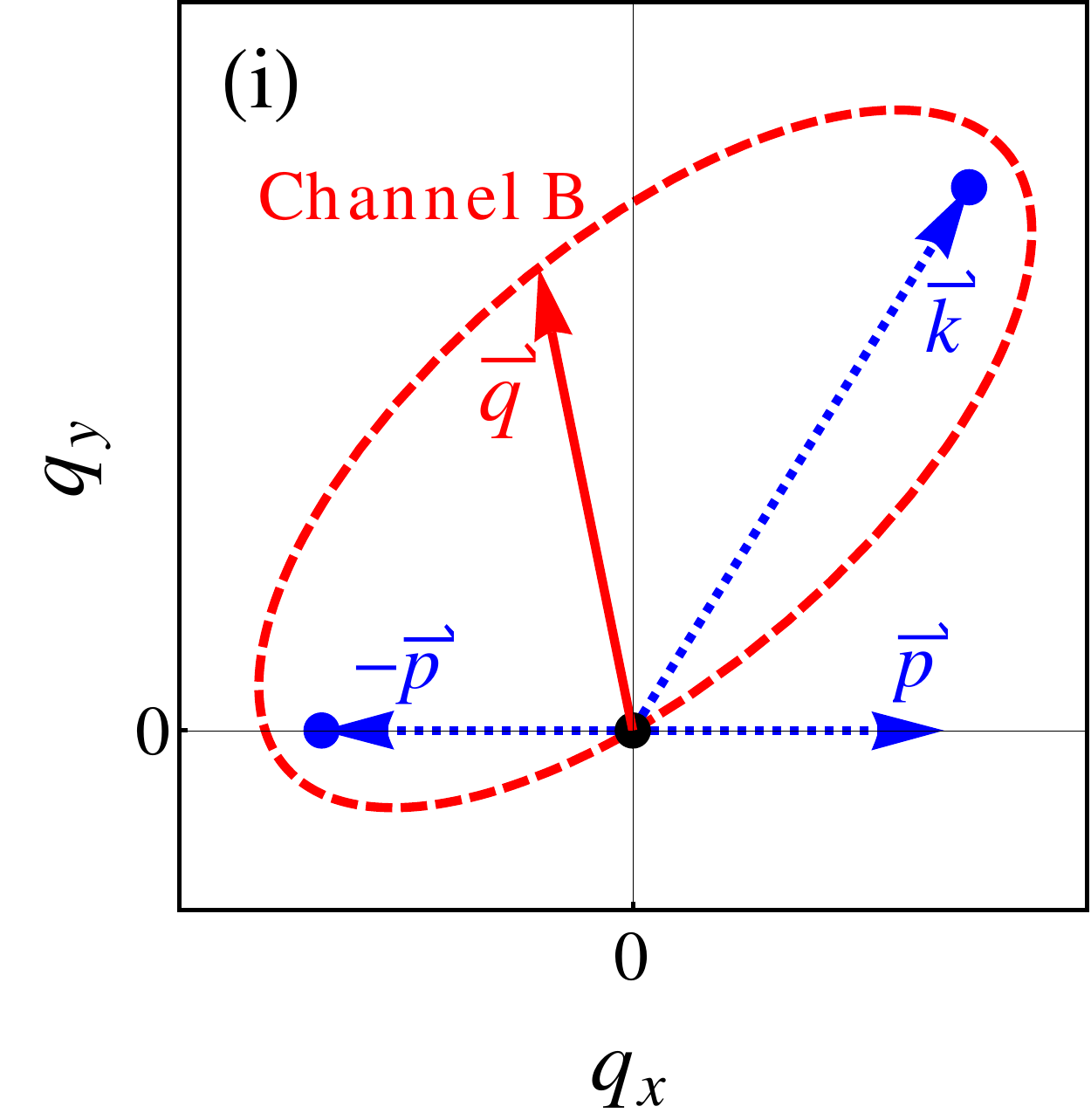}
\includegraphics[width=0.237\textwidth]{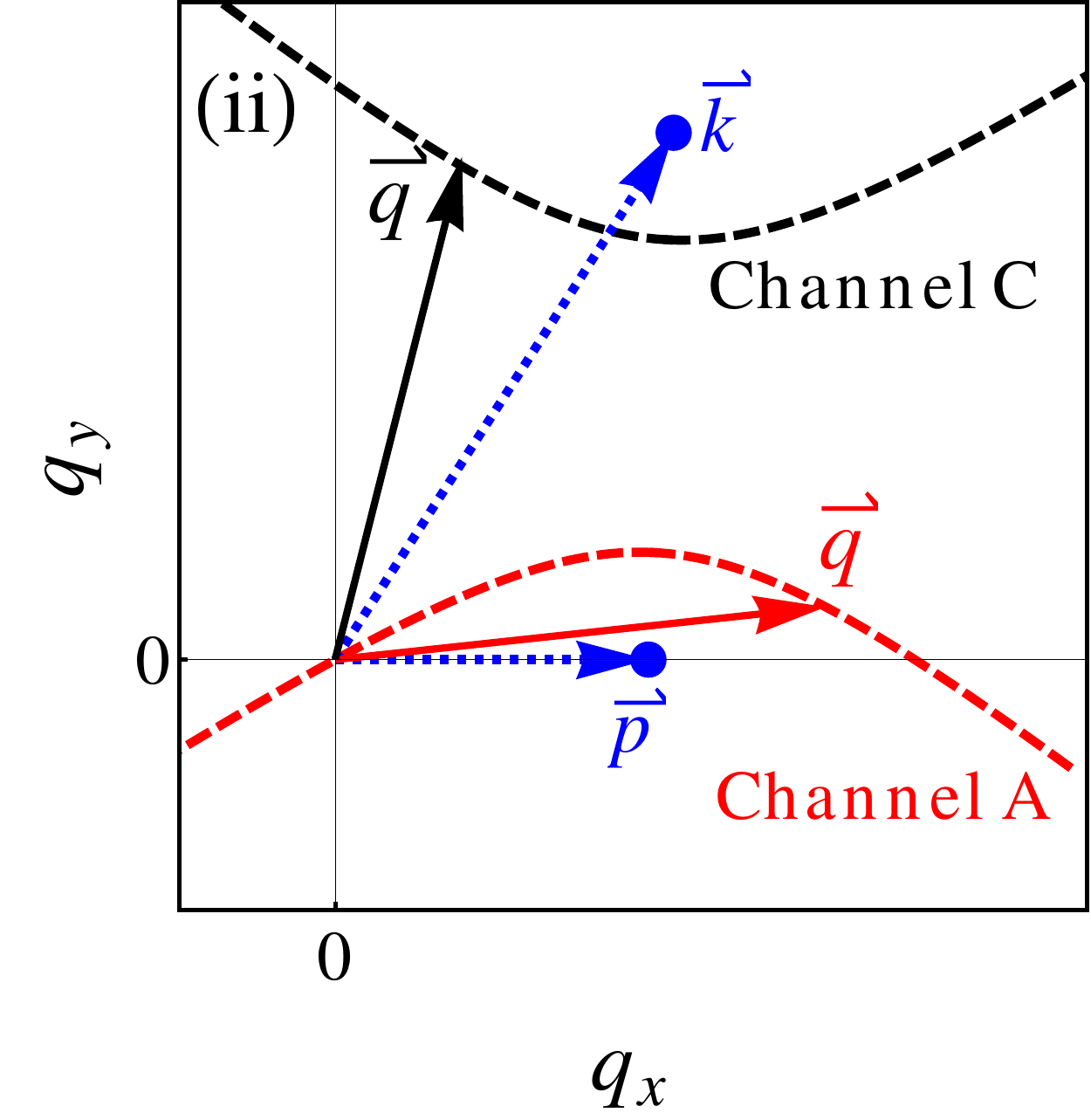}
\caption{Elliptic and hyperbolic coordinates of the momentum transfer for Coulomb interactions. 
(i) Channel B. The momentum transfer $\bq$ lays on a ellipse (red dashed curve) 
with two foci at the incident momenta $-\bp$ and $\bk$ (blue dotted lines).
(ii) Channels A and C. The momentum transfer $\bq$ lays on the two branches of a hyperbola (dashed curves) 
with two foci at the incident momenta $\bp$ and $\bk$ (blue dotted lines). 
The lower branch (red) corresponds to channel A and the upper branch (black) to channel C.}  
\label{collision}
\end{figure}

For channel B [Eq.~(\ref{mat-int-b})] one needs to evaluate an integral in the form
\be  \label{int-delta}
\mathcal{I}_\mathrm{B}(\bp,\bk) = \int_\bq (2\pi) \, \delta(p-|\bp+\bq|+k-|\bk-\bq|) \,	G(\bp,\bk ; \bq),
\ee 
where 
$G(\bp,\bk;\bq)$	
is a general function of $\bp$, $\bk$ and $\bq$. The elliptic coordinates $(\rho, \theta)$ of $\bq$ are defined by
\be  \label{ellip-coord}
\begin{split}
\begin{pmatrix} q_x \\ q_y \end{pmatrix} = \frac{1}{2}\begin{pmatrix}  k_x - p_x \\ k_y - p_y \end{pmatrix}                                       
+  &\, \frac{1}{2}\begin{pmatrix} 
k_x + p_x  &  -(k_y +p_y) \\ k_y + p_y  &  k_x + p_x 
\end{pmatrix} \\
&\,  \times  \begin{pmatrix} \cosh{\rho} \, \cos{\theta} \\ \sinh{\rho} \, \sin{\theta} \end{pmatrix},  
\end{split}
\ee
where
\be  \label{rho-the-int}
0 \le \rho < \infty, \quad -\pi \le \theta < \pi. 
\ee
As shown in Fig.~\ref{collision}(i), the momentum transfer $\bq$ lays on a ellipse with two foci at the incident momenta $-\bp$ and $\bk$. Via Eq.~(\ref{ellip-coord}) we readily obtain the relations
\begin{subequations} \label{use-rel-B}
\begin{align}
& |\bp+\bq|+|\bk-\bq| = |\bk+\bp|\cosh{\rho}, \\
& \int_\bq = \int_{-\pi}^{\pi} \!\frac{\rmd \theta}{(2 \pi)^2}\int_{0}^{\infty}\!\!\rmd\rho \,\frac{|\bk+\bp|^2}{4} \left( \cosh^2{\rho} -\cos^2{\theta} \right).
\end{align}
\end{subequations}
Substituting Eq.~(\ref{use-rel-B}) into Eq.~(\ref{int-delta}) we obtain 
\be\label{IBEval}
\begin{split}
\mathcal{I}_\mathrm{B}(\bp,\bk) = \frac{|\bk+\bp|}{4  \sinh{\rho_0}} \int_{-\pi}^{\pi}\!  &\, \frac{\rmd \theta}{2 \pi} \, \left( \cosh^2{\rho_0} -\cos^2{\theta} \right) \\
                                  &\, \times G[\bp,\bk;\bq(\rho_0,\theta)],
\end{split}
\ee
where $\rho_0$ is fixed by
\be \label{cosh0}
\cosh{\rho_0} = \frac{k+p}{|\bk+\bp|}.
\ee
The phase space of the collinear collision 
is manifestly divergent since $\rho_0 = 0$ when $k+p = |\bk+\bp|$.

For channels A and C [Eqs.~(\ref{mat-int-a}) and (\ref{mat-int-c})] we need to evaluate 
\be  \label{int-delta-2}
\mathcal{I}_\mathrm{C/A}(\bp,\bk) =\int_\bq \! \! (2\pi)\delta(p-k \pm |\bp-\bq| \mp |\bk-\bq|) \, G(\bp,\bk;\bq),
\ee 
where the upper (lower) signs are for channel C (A). 
The hyperbolic coordinates are defined by
\be  \label{hyper-coord}
\begin{split}
\begin{pmatrix} q_x \\ q_y \end{pmatrix} = \frac{1}{2}\begin{pmatrix}  k_x + p_x \\ k_y + p_y \end{pmatrix} 
                                         + \frac{1}{2}  &\, \begin{pmatrix} k_x - p_x & -k_y + p_y \\ k_y - p_y  & k_x - p_x \end{pmatrix} \\ 
                                           \times &\, \begin{pmatrix} \cosh{\rho} \, \cos{\theta} \\ \sinh{\rho} \, \sin{\theta} \end{pmatrix},
\end{split}  
\ee
where $\rho$ and $\theta$ are defined in the intervals in Eq.~(\ref{rho-the-int}). As shown in Fig.~\ref{collision}(ii), the momentum transfer $\bq$ lays on the two branches of a hyperbola (dashed curves) with two foci at the incident momenta $\bp$ and $\bk$.  Via Eq.~(\ref{hyper-coord}) we obtain
\begin{subequations}  \label{use-rel-AC}
\begin{align}
& |\bp-\bq|-|\bk-\bq| = |\bk-\bp|\cos{\theta}, \label{use-rel-AC-A}\\
& \int_\bq = \int_{-\pi}^{\pi}\!\frac{\rmd \theta}{(2 \pi)^2}\int_{0}^{\infty}\!\!\rmd\rho \,\frac{|\bk-\bp|^2}{4} \left( \cosh^2{\rho} -\cos^2{\theta} \right).
\end{align}
\end{subequations}
Substituting Eq.~(\ref{use-rel-AC}) into Eq.~(\ref{int-delta-2}) gives
\be\label{ICAEval}
\begin{split}
	\mathcal{I}_\mathrm{C/A}(\bp,\bk) = 
	\frac{|\bk-\bp|}{8\pi \, |\sin{\theta_0}|} \int_{0}^{\infty}\!\!\rmd \rho \, \left( \cosh^2{\rho} -\cos^2{\theta_0} \right) \\
	\times 
	\left\{ G[\bp,\bk;\bq(\rho,\theta_0)]+ G[\bp,\bk;\bq(\rho,-\theta_0)]		\right\},
\end{split}
\ee
where $\theta_0$ is fixed by
\be \label{cost0}
	\cos{\theta_0} = \pm\frac{k-p}{|\bk-\bp|}, \quad 0 \le \theta_0 < \pi,
\ee
and the sign ``$+$'' (``$-$'') is for channel C (A). The phase space of the collinear collision is divergent since $\theta_0=0$ when $k-p = \pm|\bk-\bp|$.

We note that for channel A, Eq.~(\ref{cost0}) is consistent with Eq.~(\ref{use-rel-AC-A}), so that 
the zero momentum transfer condition $\vex{q} = 0$ resides on the corresponding branch of the hyperbola shown in Fig.~\ref{collision}(ii).
For channel C, Eq.~(\ref{cost0}) is in general inconsistent with Eq.~(\ref{use-rel-AC-A}), so that $\vex{q} \neq 0$ does not reside
on this branch. The exception has $p = k$.

\begin{figure}[b]
\centering
\includegraphics[width=0.28\textwidth]{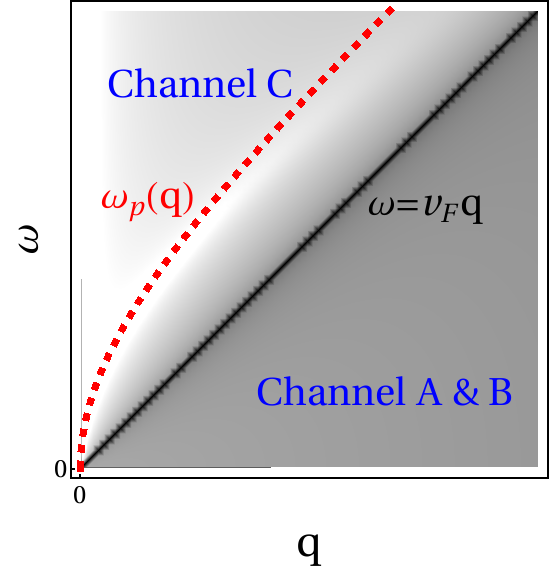}
\caption{Schematic density plot for the 
modulus-squared of the
RPA screened Coulomb interaction (\ref{rpa-int-def}). 
The screening is 
perfect along the forward scattering direction
$\w = \vf q$. 
This 
compensates the collinear singularity 
in the Coulomb collision integrals due to the linear dispersion \cite{kashsmin,lars2007},
see Eq.~(\ref{ForwardCancel}).
The red dashed line indicates the plasmon dispersion [Eq.~(\ref{plasmon})]. 
Due to kinematic constraints channels A and B 
[Figs.~\ref{scats}(b)${}_{\msf{i,ii}}$]
act in the 
``quasi-static''
regime $\vf q \geq |\w|$, 
while channel C 
[Fig.~\ref{scats}(b)${}_{\msf{iii}}$]
acts the ``optical'' 
regime $\vf q \leq |\w|$. } \label{kinemitics}
\end{figure}

\section{RPA screening of Coulomb interaction, cancellation of the collinear collision singularity, and plasmon enhancement in channel C}  \label{ftsc} 

\begin{figure}[t]
\centering
\includegraphics[width=0.235\textwidth]{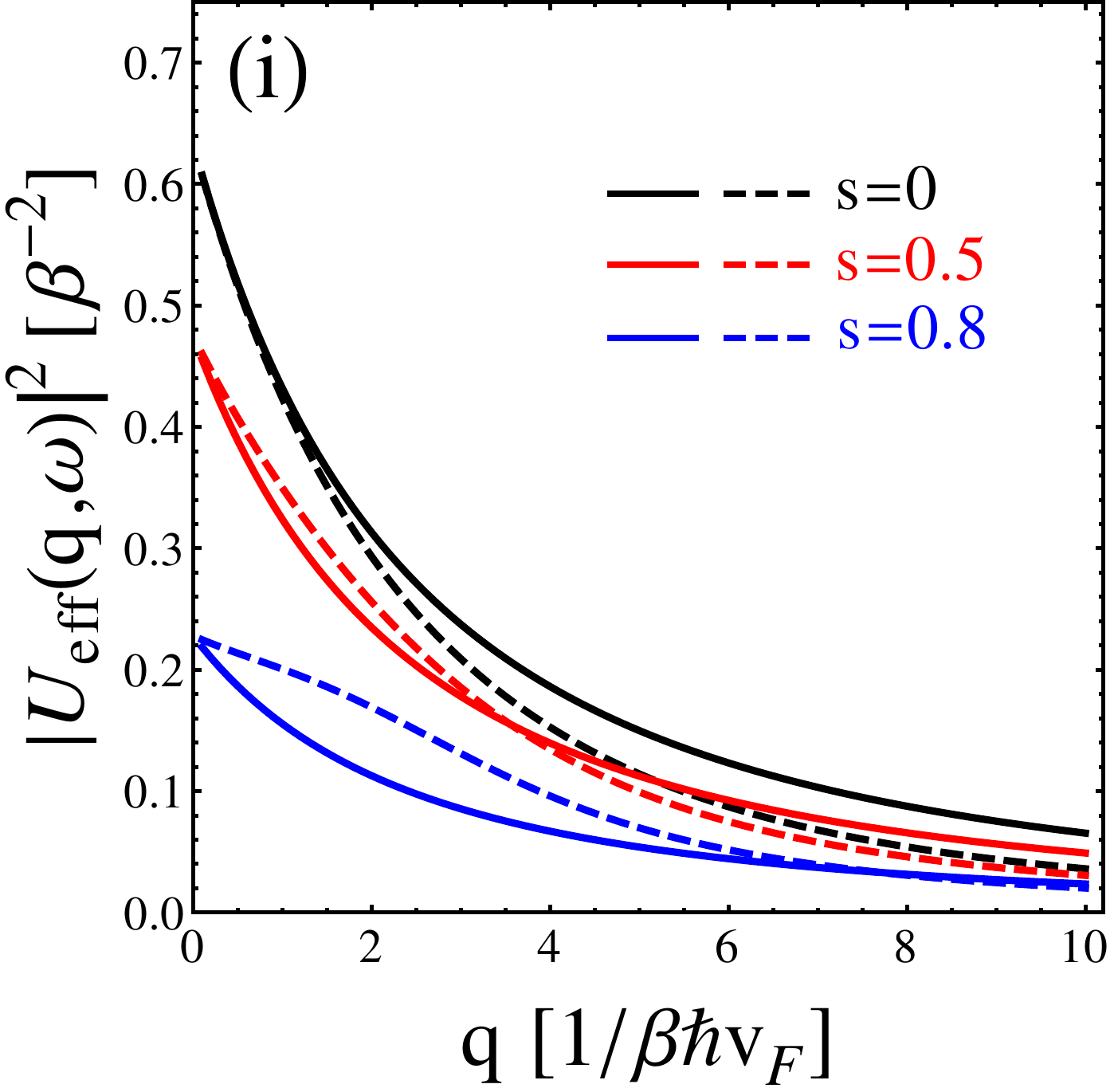}
\includegraphics[width=0.24\textwidth]{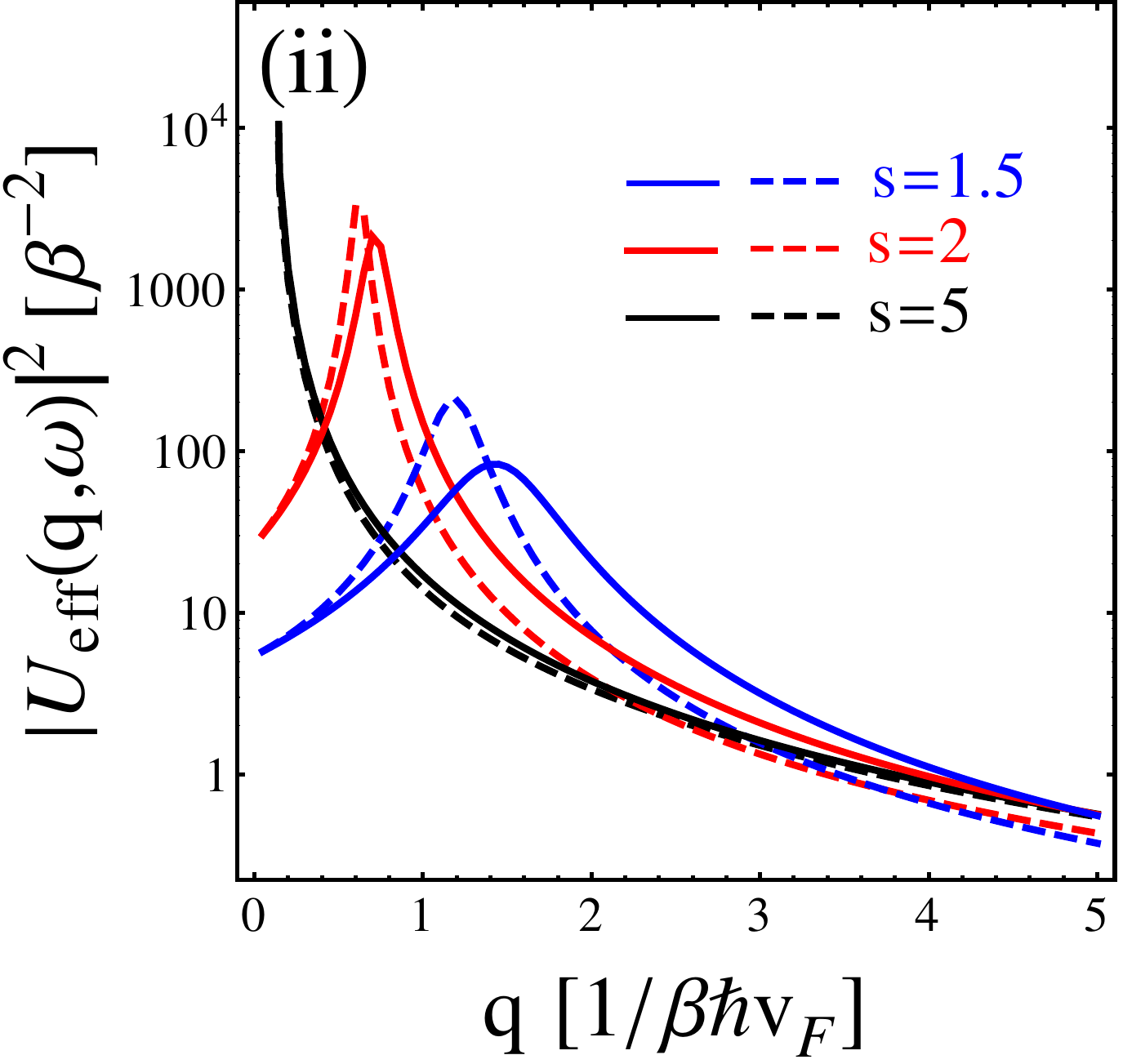}
\caption{Comparison between the approximate Coulomb interaction (\ref{app-ueff}) 
and the exact RPA result (\ref{rpa-int-def}). We take $\a_\mathrm{int} =0.6$ and 
$z= \exp(\beta \mu) = 5$
and define $s \equiv |\w|/\vf q$. The solid and dashed curves are the results of 
Eq.~(\ref{app-ueff}) and Eq.~(\ref{rpa-int-def}), respectively. 
(i) Quasi-static regime $\vf q \geq |\w|$. (ii) Optical regime $\vf q \leq |\w|$.} \label{comp-rpa}
\end{figure}

\begin{figure}[b]
\includegraphics[width=0.245\textwidth]{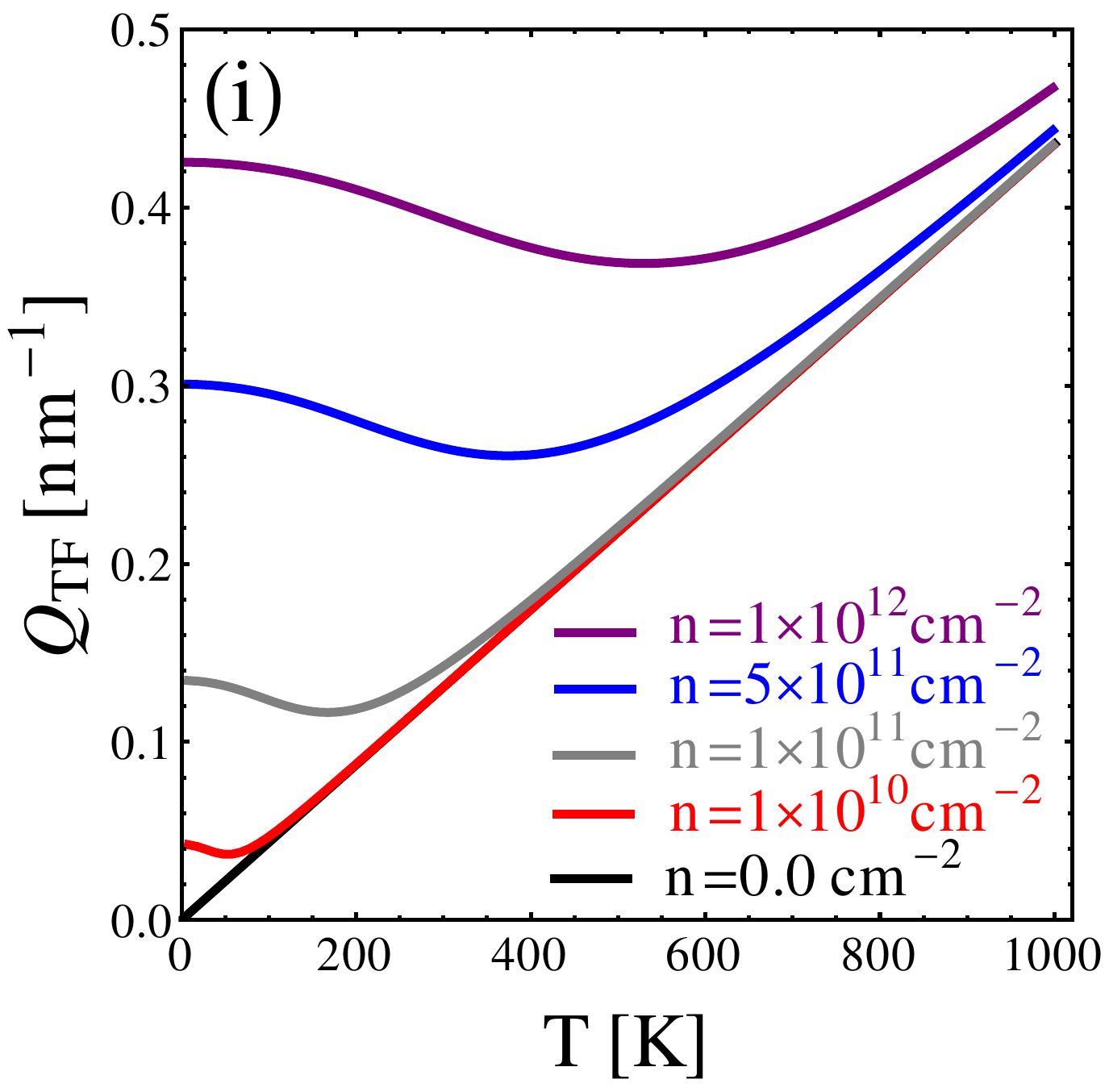}
\includegraphics[width=0.23\textwidth]{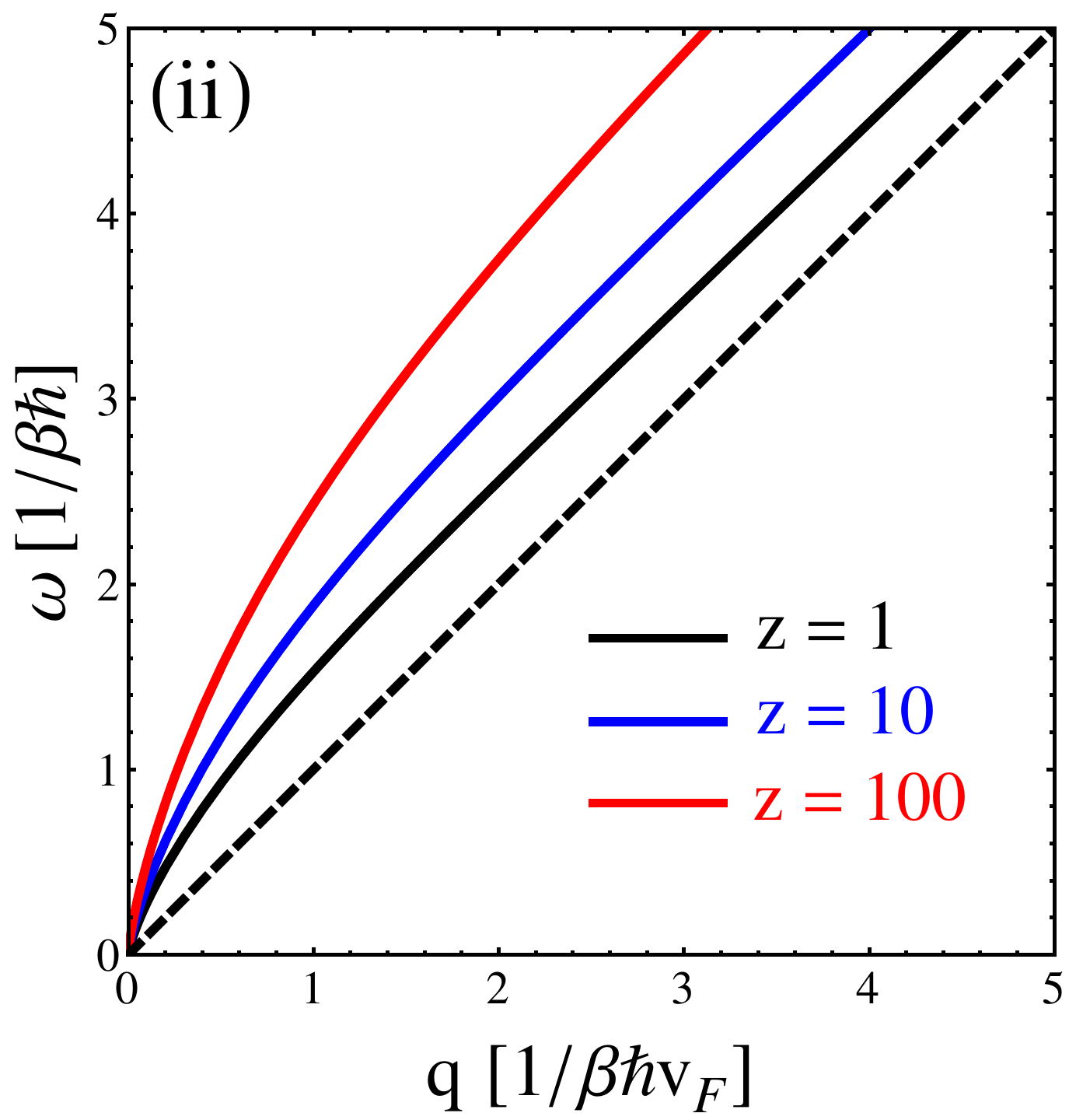}
\caption{Thomas-Fermi wavevector and plasmon dispersion as functions of density and temperature. 
We take $\a_\mathrm{int}=0.6$. 
(i) Thomas-Fermi wavevector [Eq.~(\ref{q-tf})]. 
(ii) Plasmon dispersion [Eq.~(\ref{plasmon})]. 
The dashed line depicts $\w =\vf q$ for guiding the eyes. }  \label{Qtf-plasmon}
\end{figure}

In this appendix we employ physical units, but we set $\hbar = 1$ unless noted.
At finite temperature the screened Coulomb interaction takes the form [see Fig.~\ref{kinemitics}]
\be  \label{rpa-int-def}
	U_{\mathrm{eff}}(\w,q)	
	= 
	\frac{V_q}{\ep(q,\w)}, \quad V_q = \frac{2 \pi \a_\mathrm{int}\vf}{q},
\ee
where $\a_\mathrm{int}$ is the fine structure constant 
and $\ep(q,\w)$ is the dynamical screening function. In the RPA approximation, 
$\ep(q,\w) = 1 - V_q \chi_0(q,\w)$, where $\chi_0(q,\w)$ is the Lindhard function. The real and imaginary parts of $\chi_0(q,\w)$ take the forms~\cite{macdonald2009}
\begin{subequations} \label{linhard-def}
\begin{align}
\textrm{Re}  \chi_0(q,\w) = &\, -\Xi(T,z) -\frac{q^2}{\sqrt{\left|\vf^2 q^2 - \w^2\right|}} \nn \\
                            &\, \times 
\begin{cases} 
\vspace{0.2cm}
\Lambda_{-}(q,\w), \quad & \,  |\w| \le \vf q, \\
\Lambda(q,\w), \quad & \, |\w| \ge \vf q, 
\end{cases} \\
\textrm{Im} \chi_0(q,\w) = &\,  \frac{q^2}{\sqrt{\left| \vf^2 q^2-\w^2 \right|}} 
\begin{cases} 
\vspace{0.2cm}
\Lambda(q,\w), \quad & |\w| \le \vf q,  \\ 
-\Lambda_{+}(q,\w), \quad & |\w| \ge \vf q, \end{cases}
\end{align}
\end{subequations}
where
\begin{subequations}
\begin{align}
\Xi(T,z) = & \, \frac{N k_\mathrm{B} T}{2\pi \vf^2}\ln\left[(1+z)(1+z^{-1})\right], \\
\Lambda_{\pm}(q,\w) = & \, \frac{N}{8 \pi} \left[ \frac{\pi}{2} -  \sum_{\l}H_{\pm,\l}(q,\w) \right], \\
\Lambda(q,\w)= & \, \frac{N}{8 \pi} \sum_{\l,\l^\prime=\pm}{\l G_{\l,\l^\prime}(q,\w)}.
\end{align}
\end{subequations}
Here $H_{\pm,\l}(q,\w)$ and $G_{\l,\l^\prime}(q,\w)$ are defined by
\begin{subequations}  \label{f-G-H}
\begin{align}
G_{\l,\l^\prime}(q,\w) = &\, \int_{1}^{+\infty} \! \rmd u \, \frac{\sqrt{u^2-1}}{z^{-\l^\prime} \exp \left( \frac{|\vf q u + \l \w|}{2 k_\mathrm{B} T} \right) +1}, \\
H_{\l,\l^\prime}(q,\w) = & \,\int_{-1}^{1} \! \rmd u \, \frac{\sqrt{1-u^2}}{z^{-\l^\prime} \exp \left(  \frac{|\vf q u + \l \w|}{2 k_\mathrm{B} T} \right) +1}.
\end{align}
\end{subequations}
In numerical calculations we use the dimensionless form of the Coulomb interaction 
\begin{align}\label{UeffDIMLESS}
	\wtd{U}_{\mathrm{eff}}(\w,q) 
	\equiv 
	\frac{1}{\b (\hbar \vf)^2} 
	U_{\mathrm{eff}}\left(	\frac{1}{\b \hbar} \w, \frac{1}{\b \hbar \vf} q	\right),	
\end{align} 
where $\w$ and $q$ are dimensionless. As shown in Fig.~\ref{kinemitics}, due to energy conservation one has the 
following kinematic constraints. For channel A and B, 
$
	|\omega| = \vf |p - |\bp \mp \bq|| 
$ 
so that $\vf q \geq |\w|$ (``quasi-static'' regime), while for channel C, $|\w| = \vf (p+|\bp-\bq|
)$ and $\vf q \leq |\w|$ (``optical'' regime).

Combining Eqs.~(\ref{rpa-int-def}) and (\ref{linhard-def}) we obtain
\begin{widetext}
\be \label{ueff2}
\begin{split}
	& 	|U_{\mathrm{eff}}(\w,q)|^2 		= \\
	& \, \left( 2 \pi \a_\mathrm{int} v_F \right)^2 \times
	\begin{cases} 
	\displaystyle 
		\frac{
		\left| (\vf q)^2-\w^2 \right|
		}{
		\left\{ \sqrt{\left| (\vf q)^2-\w^2 \right|}\left[ q + Q_\mathrm{TF}(T,z)  \right]+ 2\pi\a_\mathrm{int} v_F q^2 \Lambda_{-}(q,\w)  \right\}^2 
		+ 
		\left[ 2 \pi \a_\mathrm{int} v_F q^2 \Lambda(q,\w) \right]^2
		}, 
			\quad |\w| \le \vf q \\
	\displaystyle 
		\frac{
		\left| (\vf q)^2-\w^2 \right|
		}{
		\left\{ \sqrt{\left| (\vf q)^2-\w^2 \right|}\left[ q + Q_\mathrm{TF}(T,z)  \right]+ 2\pi\a_\mathrm{int} v_F q^2 \Lambda(q,\w)  \right\}^2 
		+ 
		\left[2 \pi \a_\mathrm{int} v_F q^2 \Lambda_{+}(q,\w) \right]^2}, \quad |\w| \ge \vf q \\
\end{cases}
\end{split}
\ee
where the Thomas-Fermi wavevector $Q_\text{TF}(T,z)$ [see Fig.~\ref{Qtf-plasmon}(i)] is given by 
\be \label{q-tf}
	Q_\text{TF}(T,z) = \frac{N \a_\mathrm{int}}{\hbar \b  \vf} \ln\left[(1+z)(1+z^{-1})\right].
\ee
We also introduce the dimensionless Thomas-Fermi wavevector 
\begin{align}\label{qTFdimless}
	q_\mathrm{TF} \equiv \hbar \b \vf Q_\text{TF}=N \a_\mathrm{int}\ln\left[(1+z)(1+z^{-1})\right]. 
\end{align}
We emphasize that the factor $\left| (\vf q)^2-\w^2 \right|$ in Eq.~(\ref{ueff2}) leads to the cancellation of 
the collinear collision singularity that occurs along $|\w|=\vf q$ 
[Eqs.~(\ref{IBEval},\ref{cosh0}) and (\ref{ICAEval},\ref{cost0})],	
\begin{subequations}\label{ForwardCancel}
\begin{align}
	\frac{\vf^2q^2 - \w^2}{\vf^2|\sin{\theta}|} \bigg|_{\theta=\theta_0} 
	= & 
	\, \sqrt{k\,p} \, \left[ -|\bk - \bp| + ( k+p) \, \cosh{\rho} \right] 
	\, \left| \sin{\left( \frac{\varphi_\bk - \varphi_\bp}{2} \right) }\right| - \mathrm{sgn}(\theta_0) k \, p \, \sinh{\rho} \, \sin{(\varphi_\bk - \varphi_\bp)}, 
\\
	\frac{\vf^2q^2 - \w^2}{\vf^2\sinh{\rho}} \bigg|_{\rho=\rho_0} 
	= & 
	\, \sqrt{k\,p} \, \left[ |\bk + \bp| + ( k-p) \, \cos{\theta} \right] 
	\, \left| \sin{\left( \frac{\varphi_\bk - \varphi_\bp}{2} \right) }\right| + k \, p \, \sin{\theta} \, \sin{(\varphi_\bk - \varphi_\bp)}, 
\\
	\frac{\w^2-\vf^2 q^2}{\vf^2|\sin{\theta}|} \bigg|_{\theta=\theta_0} 
	= & 
	\, \sqrt{k\,p} \, \left[ |\bk - \bp| + ( k+p) \, \cosh{\rho} \right] 
	\, \left| \sin{\left( \frac{\varphi_\bk - \varphi_\bp}{2} \right) }\right| + \mathrm{sgn}(\theta_0) k \, p \, \sinh{\rho} \, \sin{(\varphi_\bk - \varphi_\bp)},
\end{align}
\end{subequations}
which correspond to channels A, B, and C, respectively.

In our calculation we further simplify the RPA screened interaction (\ref{ueff2}). 
For $\vf q > |\w|$ (channels A and B) we take the static limit $\w \to 0$ of the denominator  
and neglect the 
dielectric
enhancement arising from the residue $\Lambda_{-}(q,\w=0)$,
\begin{subequations} \label{app-ueff}
\be 
	|U_{\mathrm{eff}}(\w,q)|^2 
	\approx 
	\left( \frac{2 \pi \a_\mathrm{int}}{q} \right)^2 \frac{\vf^2 q^2-\w^2}{\left( q+  Q_\mathrm{TF} \right)^2}, \quad \vf q > |\w|.
\ee
For $\vf q < |\w|$ (channel C), up to the leading order in $ \vf q/|\w| \ll 1$ we obtain
\be
	|U_{\mathrm{eff}}(\w,q)|^2 
	\approx 
	\frac{(2 \pi \a_\mathrm{int})^2 (\w^2 - \vf^2 q^2)}
	{
	\frac{1}{v_F^2}	
	\left[ \sqrt{ \w^2-\vf^2 q^2 } \left( q+ Q_\mathrm{TF} \right) - \w \, Q_\mathrm{TF} \right]^2
	+ 
	\frac{1}{4} 
	\left[ 
	\frac{ 
		\pi \a_\mathrm{int} \vf q^2 \sinh \left(\frac{\beta \w}{2}\right)
	}{
		\cosh \left(\frac{\beta \w}{2} \right)+ \cosh \left(\b \mu \right)
	} \right]^2 }, 
	\quad \vf q < |\w|.
\ee
\end{subequations}
\end{widetext} 
As compared in Fig.~\ref{comp-rpa}, Eq.~(\ref{app-ueff}) provides a good approximation for the RPA interaction (\ref{rpa-int-def}).

The plasmon dispersion [see Fig.~\ref{Qtf-plasmon}(ii)] is determined by $1 - V_q \, \re \chi_0 =0$, which gives for $\vf q/|\w| \ll 1$,
\be \label{plasmon}
	\w_\mathrm{p}(q) = 
	v_F
	\sqrt{
	\frac{q \left( q + Q_\mathrm{TF} \right)^2}{q + 2 Q_\mathrm{TF}}
	}.			
\ee
Channel C is strongly enhanced along this plasmon dispersion. However, the presence of metallic gates or plasmon-phonon 
coupling~\cite{Abajo2014} may significantly broaden the plasmon 
peak
so that the enhancement of channel C is suppressed.


\begin{thebibliography}{99}
\bibitem{why-inplane} We assume that the out-of-plane phonons are suppressed since the graphene sample in Ref.~\onlinecite{GXFKexp} is encapsulated between two boron-nitride substrates.
\bibitem{ph-drag} The acoustic-phonon drag effect should be also negligible compared to the direct electronic component due to phonon-phonon interactions [see Refs.~\onlinecite{Kubakaddi09}~and~\onlinecite{Bao}].
\bibitem{negoffdiag} {We argue that no valley or spin imbalance develops, so that the distribution functions 
			remain diagonal in valley and spin spaces.  
			Both elastic impurity scattering and inelastic Coulomb scattering preserve the valley and spin symmetry. 
			Although optical phonons involve intervalley scattering, a non--valley-diagonal component of the distribution function
			is not expected.}
\bibitem{non-mon-opt} The non-monotonicity in doping at low temperature could be an artifact of the semi-classical Boltzmann equation and might be smeared out by multiple scattering processes that are not captured by the Boltzmann equation approach.
\bibitem{est-alpha} We estimate $\left( \mathcal{M}_{\mathrm{opt}}\right)_{00} \sim \wtd{\a}_\mathrm{opt}^2 \, \left(T_\mathrm{F}/T\right)^4 e^{-T_\mathrm{F}/T}$ and $\left( \mathcal{M}_{\mathrm{imp}}\right)_{00} \sim \left( \wtd{g} + \pi \a_\mathrm{int}^2 \hbar^2 \vf^2 n_\mathrm{imp} / T_\mathrm{F}^2 \right)\left(T_\mathrm{F}/T\right)^4$, where $\wtd{g}$, $\a_\mathrm{int}$, and $n_\mathrm{imp}$ take the values extracted from experiment~\cite{GXFKexp}.
\bibitem{com-tau-ee} See the diagram 
(b)${}_{\msf{i}}$
in Fig.~\ref{scats}. 
If we artificially suppress
channel A, one should obtain $\tau_\mathrm{ee}^{-1} \sim e^{-\frac{\mu}{k_\mathrm{B} T}} \frac{k_\mathrm{B} T}{\hbar} \a_\mathrm{int}^2$ for $\mu \gg k_\mathrm{B} T$. Here we assume finite screening.
\end{thebibliography}
\end{document}